\documentclass[12pt]{article}
\usepackage{amssymb}
\usepackage{amsmath}
\usepackage{bm}
\usepackage{a4}
\usepackage{hyperref}
\usepackage{graphicx}
\usepackage{comment}
\usepackage{color}
\usepackage[normalem]{ulem}

\DeclareMathSymbol{\mrq}{\mathord}{operators}{`'}

\evensidemargin \oddsidemargin
\newcommand{\NN}{\mathbb{N}}

\def\1{{\bf 1}}
\def\0{{\bf 0}}

\def \C {{\cal C}}
\def \D {{\cal D}}

\newcommand{\E}{{\cal E}}

\def \I {{\cal I}}

\newcommand{\K}{{\cal K}}

\def \U {{\cal U}}

\def \xip {\xi_{{\scriptscriptstyle +}}}
\newcommand{\uxi}{\underline{\xi}}
\newcommand{\ouxi}{\bar{\underline{\xi}}}
\newcommand{\sm}{s_{{\scriptscriptstyle -}}}
\def \sp {s_{{\scriptscriptstyle +}}}
\newcommand{\spm}{s_{{\scriptscriptstyle \pm}}}

\newcommand{\smm}{s_{{\scriptscriptstyle m}}}
\newcommand{\sM}{s_{{\scriptscriptstyle M}}}
\newcommand{\ospm}{\bar s_{{\scriptscriptstyle \pm}}}
\newcommand{\osmm}{\bar s_{{\scriptscriptstyle m}}}
\newcommand{\osM}{\bar s_{{\scriptscriptstyle M}}}

\newcommand{\Be}{{\bm \epsilon}}
\newcommand{\Bep}{{\bm \epsilon}^{\scriptscriptstyle \perp}}

\newcommand{\bi}{\mathbf{i}}
\newcommand{\bj}{\mathbf{j}}
\newcommand{\bk}{\mathbf{k}}
\newcommand{\bjc}{{\bm j}}

\newcommand{\hs}{\hat s}

\newcommand{\hze}{\hat  z_e}

\def\nn{\nonumber \\}

\newcommand{\bx}{{\bm x}}
\newcommand{\bxp}{{\bm x}^{\scriptscriptstyle \perp}}
\newcommand{\bX}{{\bm X}}
\newcommand{\bXp}{{\bm X}^{\scriptscriptstyle \perp}}
\newcommand{\bu}{{\bm u}}
\newcommand{\bup}{{\bm u}^{\scriptscriptstyle \perp}}

\newcommand{\cbUp}{\check{\bm U}^{\scriptscriptstyle \perp}}

\newcommand{\hbup}{\hat {\bf u}^{\scriptscriptstyle \perp}}
\newcommand{\bv}{{\bm v}}

\newcommand{\Bap}{{\bm \alpha}^{\scriptscriptstyle \perp}}
\newcommand{\bb}{{\bm \beta}}

\newcommand{\BD}{{\bm \Delta}}

\newcommand{\bE}{{\bm E}}
\newcommand{\bEp}{{\bm E}^{\scriptscriptstyle \perp}}
\newcommand{\bB}{{\bm B}}
\newcommand{\bBp}{{\bm B}^{\scriptscriptstyle \perp}}
\newcommand{\bA}{{\bm A}}
\newcommand{\bAp}{{\bm A}\!^{\scriptscriptstyle \perp}}
\newcommand{\bjp}{{\bm j}^{\scriptscriptstyle \perp}}

\newcommand{\Bp}{{\bm p}}

\newcommand{\DM}{\Delta_{{\scriptscriptstyle M}}}

\newcommand{\Dm}{\Delta_{{\scriptscriptstyle m}}}
\newcommand{\oDm}{\bar\Delta_{{\scriptscriptstyle m}}}
\newcommand{\oDM}{\bar\Delta_{{\scriptscriptstyle M}}}

\newcommand{\xiH}{\xi_{{\scriptscriptstyle H}}}
\newcommand{\bxiH}{\bar\xi_{{\scriptscriptstyle H}}}

\newcommand{\gammaM}{\gamma^{{\scriptscriptstyle M}}}
\newcommand{\znd}{z_{{\scriptscriptstyle dp}}}
\newcommand{\zM}{z_{{\scriptscriptstyle M}}}

\newcommand{\be}{\begin{equation}}
\newcommand{\ee}{\end{equation}}
\newcommand{\bea}{\begin{eqnarray}}
\newcommand{\eea}{\end{eqnarray}}
\newcommand{\ba}{\begin{array}}
\newcommand{\ea}{\end{array}}
\newtheorem{prop}{Proposition}

\def\sq{\mbox{\rlap{$\sqcap$}$\sqcup$}}
\newenvironment{proof}[1]{\vspace{5pt}\noindent{\bf Proof #1}\hspace{6pt}}%
{\hfill\sq}
\newcommand{\bp}{\begin{proof}}
\newcommand{\ep}{\end{proof}\par\vspace{10pt}\noindent}
\begin{document}

\title{
 An analytical optimization of plasma density profiles 
for downramp injection in laser wake-field acceleration
}

\author{Gaetano Fiore$^{1,2}$\footnote{Corresponding author. Email: gaetano.fiore@na.infn.it}, \  Paolo Tomassini$^{3}$  \\    
$^{1}$ Dip. di Matematica e Applicazioni, Universit\`a di Napoli ``Federico II'', \\
Complesso Universitario  M. S. Angelo, Via Cintia, 80126 Napoli, Italy\\         
$^{2}$         INFN, Sez. di Napoli, Complesso  MSA,  Via Cintia, 80126 Napoli, Italy\\
$^{3}$ Extreme Light Infrastructure - Nuclear Physics, IFIN-HH, 
\\ 30 Reactorului Street, 077125 Magurele, Romania
}

\date{}

\maketitle

\begin{abstract}
We propose and detail a multi-step analytical procedure, based on an improved fully relativistic plane model for Laser Wake Field Acceleration, to tailor the initial density of a cold diluted plasma to the laser pulse profile, so as to control 
the spacetime localization and features of wave-breakings of the plasma wave and maximize the early stage acceleration of small bunches of electrons self-injected by the first wave-breaking at the density down-ramp. We find an excellent agreement with the results of 1D Particle In Cell simulations obtained with the same input data.
\end{abstract}



\section{Introduction}  

\begin{figure}
\includegraphics[width=16cm]{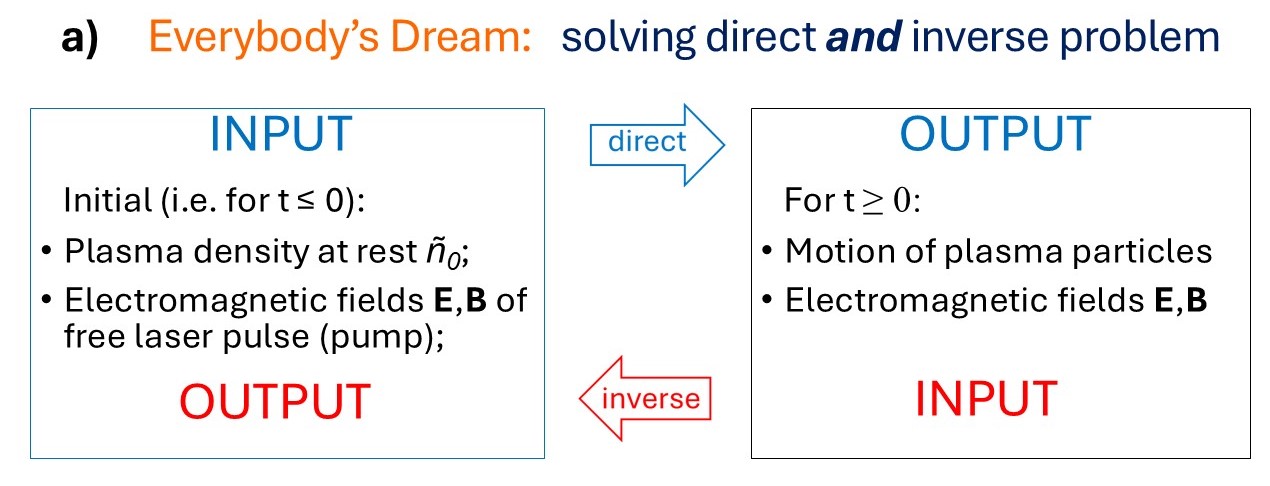}
\includegraphics[width=16cm]{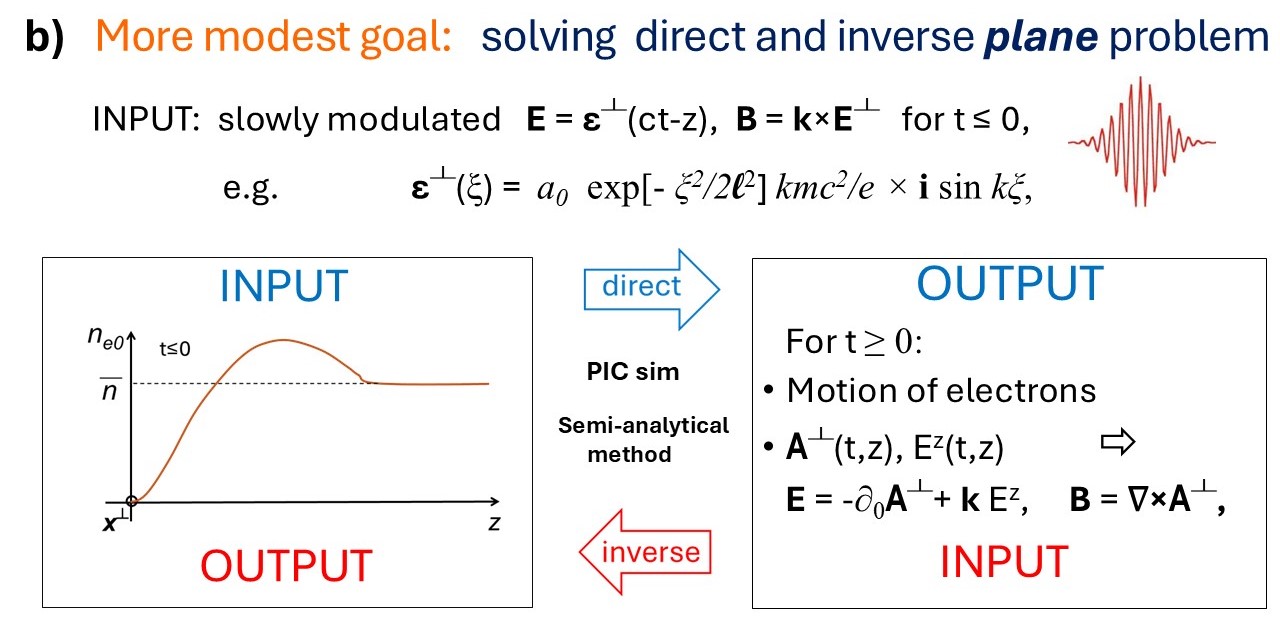}
\caption{Schematic illustration of the direct and inverse problems}
\label{dream}
\end{figure}

Laser wake-field acceleration (LWFA)  
\cite{Tajima-Dawson1979,Sprangle1988,EsaSchLee09,TajNakMou17}
 is the historically first and prototypical mechanism of extreme acceleration of charged particles along short distances: injected electrons are accelerated by the longitudinal electric field of a plasma wave (PW)
driven by a very short laser pulse in a diluted  plasma. In view of the extremely important applications of accelerators in particle physics, materials science, biology, medicine, industry,  etc., nowadays huge investments and research collaborations\footnote{We just mention the EU-funded project {\it EuPRAXIA} \cite{Eupraxia19AIP,Eupraxia19JPCS,Eupraxia20EPJ}, \url{https://www.eupraxia-project.eu/}, \url{https://www.eupraxia-facility.org/}.}  are devoted to develop 
reliable table-top accelerators on the base of such mechanisms.
Dynamics of LWFA is ruled by Maxwell equations coupled to a  kinetic  theory
for plasma  electrons, the ions being usually considered immobile due to their large mass. Solving an associated {\it direct problem} consists in solving these equations equipped with some ``input data", i.e. initial electromagnetic (EM) field,  densities and velocities of the plasma components, to determine the corresponding  evolution of the EM field and motion of the plasma - the ``output"; nowadays this can be done rather accurately via 
particle-in-cell (PIC) codes. Ideally (see fig. \ref{dream}.a), one would like to solve also an {\it inverse problem}: given the desired output (high quality accelerated electron bunches,...), find an input configuration being able to generate it. In general this is an unsolved and  formidable task. 
Approaching such an input by a trial-and-error process is unaffordable by PIC simulations, which involve huge costs for each choice of the input data; it is therefore crucial to run them after a preliminary selection of 
working points that is based on using simpler models. The development of semi-analytic models may lead also to deeper understanding of the underlying complex dynamics and to analytical or semi-analytical results that can be directly applied in  studies of the plasma dynamics. Even better would be to solve the inverse problem, at least approximately and in simplified situations.

This is what we do here, considering first inputs with plane symmetry (see fig. \ref{dream}.b), then with cylindrical symmetry.
Pursuing the research line of Ref.\cite{FioCat18,FioDeAFedGueJov22,FioDeNAkhFedJov23,FioAAC22},  we use conditions enabling a hydrodynamic description (HD) of the interaction 
 of an ultra-short and intense laser pulse 
 with a cold diluted plasma initially at rest, so as to study the propagation 
 and  
wave-breaking (WB) at  density inhomogeneities  \cite{Daw59} of the excited PW. We  subsequently 
derive preliminary conditions for early and moderate self-injection of electrons at a density downramp (self-injection at density downramps has been studied: e.g. in Ref. \cite{BulEtAl98,SukEtAl2001,England2002, BraEtAl08} for plane symmetric conditions, i.e. 1D models; e.g. in Ref.  \cite{BraEtAl08,Tomas03,GonEtAl2011,HanEtAl15,TooEtAl2017,HueWanLevMal23,JaiEtAl2024}  for 3D-models with cylindrical symmetry, in particular in the {\it bubble regime})
and for maximizing the 
acceleration  of the 
trapped particles 
in the first stages of their journey.
To that end we adopt an improved, mainly Lagrangian (as opposed to Eulerian),  fully relativistic plane model (see  section \ref{Setup}, for a review see also \cite{Fio18JPA,FioDeNAkhFedJov23} and references therein) where 
the input data depend only on time $t$ and the laser pulse direction coordinate $z$, and $\xi=ct\!-\!z$ replaces  $t$ as the independent variable. Correspondingly, all the  electrons having the same initial longitudinal coordinate $Z$ for $t\le\! 0$ (we name them the {\it $Z$ electrons})  comove and keep the same $z$ also for $t>0$ (sections \ref{planemodel}-\ref{specialcase}). The model is valid  as long as we can: i) regard ions as  immobile, ii) neglect 2-particle collisions, iii) neglect the effects of the pulse envelope evolution and subluminal propagation 
on the electrons, iv) neglect transverse dynamics effects; it also determines the spacetime regions where these simplifications are justified  (see sections \ref{Back-reaction}, \ref{3Deffects}, \ref{PIC-conclusions}).  The plane model is  first  hydrodynamic  \cite{Fio18JPA}, then {\it multi-stream} \cite{Daw59,FioPostHydro} (i.e. different $Z$ electron sheets cannot cross each other in the former, but may in the latter). 
In the former, the Lorentz-Maxwell and electrons' fluid continuity equations are reduced to the family (parametrized by $Z\!>\!0$) of {\it decoupled pairs} of Hamilton equations 
 (\ref{heq1}) with $\xi$ playing the role of `time'; each pair rules the longitudinal motion of the sheet of $Z$ electrons and becomes autonomous after the pulse has overcome the latter. The  PW emerges as a collective effect of all the sheets. The latter do not intersect as long as the Jacobian $\hat J$ of the map from Lagrangian to Eulerian coordinates keeps positive. We localize WBs in spacetime solving the equation $\hat J= 0$ with the help of the relation  \cite{FioDeNAkhFedJov23}, valid for each $Z$ after the end of the laser-plasma interaction, 
\be
\hat J(\xi,Z)=a(\xi,Z)+\xi \, b(\xi,Z),
 \label{lin-pseudoper}
\ee
where $a,b$, defined after (\ref{pseudoper}), are $\xiH$-periodic in $\xi$, and $b$ has zero mean over the period $\xiH(Z)$
given by (\ref{period}) (section \ref{WB-localization}). 
After the first WB, the hydrodynamic regime (HR) of the plasma breaks. Remarkably,  we can avoid  resorting to a full kinetic theory by the mentioned multi-stream  Lagrangian plane model \cite{FioPostHydro}. 
Using these tools, in  section \ref{WFA} we first discuss wakefield-acceleration of injected electrons in the plane problem; then (section \ref{3Deffects}) we argue that
for a realistic laser pulse with a sufficiently large, but finite, waist $w_0$ 
crucial results keep valid for on-axis electrons self-injected in the PW and travelling inside a suitable causal cone trailing the pulse, as far as the evolution of the latter at the base of the cone is negligible. 
In section \ref{optimizeLWFA} we present our multi-step optimization procedure of the LWFA of self-injected  electrons: assigned a laser pulse, by the first two steps we determine the initial plateau density $\bar{n}$ maximizing the maximal value  $E^z_{\scriptscriptstyle M}$ of the longitudinal electric field generated by plasma oscillations; by the third step we find possible linear density downramps making the first  self-injected electrons (FSIE) into the PW have the right phase to experience the maximal possible acceleration; by the fourth step we choose one of the 
possible upramps preventing WBs that may causally interfere with the above injected electrons; by the fifth step we may improve the results by fine-tuning
of the downramp parameters. This maximization procedure could be implemented via machine learning and by the development of a comprehensive AI based model.
In section \ref{Applic} we present three sample applications of our procedure. In section \ref{PIC-conclusions}:  first we compare the results obtained by our plane model to those obtained by (1D-equivalent) FB-PIC simulations with the same plane input data, in order to validate the former;  then  we present results of quasi-3D FB-PIC simulations with more realistic, Gaussian and cylindrically symmetric laser pulse that is well approximated by the plane wave one
within a finite distance $R\ll w_0$ from the $\vec{z}$-axis, determining a minimum waist $w_{0m}$ above which the on-axis electrons self-injected in the PW by the first WB experience essentially the same early acceleration as in the plane model.
On the contrary,  if  $w_0<w_{0m}$ 3D-effects (including possibly diffraction, self-focusing, wave-front shaping, 
bubble formation, transverse self-injection,...) dominate, and our model looses all predictivity. We conclude section \ref{PIC-conclusions} adding further remarks (in particular on the range of validity of our predictions), as well as our conclusions, and suggesting directions for further investigations.

\section{Setup and plane model}
\label{Setup}

In this section we summarize the improved fully relativistic plane model developed in  \cite{Fio14JPA,Fio18JPA,FioDeAFedGueJov22,FioDeNAkhFedJov23,FioPostHydro} that we use. For further details we address the reader in particular 
to \cite{FioDeNAkhFedJov23,FioPostHydro}.

\subsection{Reformulation of the dynamics of a single charged particle}
\label{1-particle}

\begin{figure}
\includegraphics[height=5.4cm]{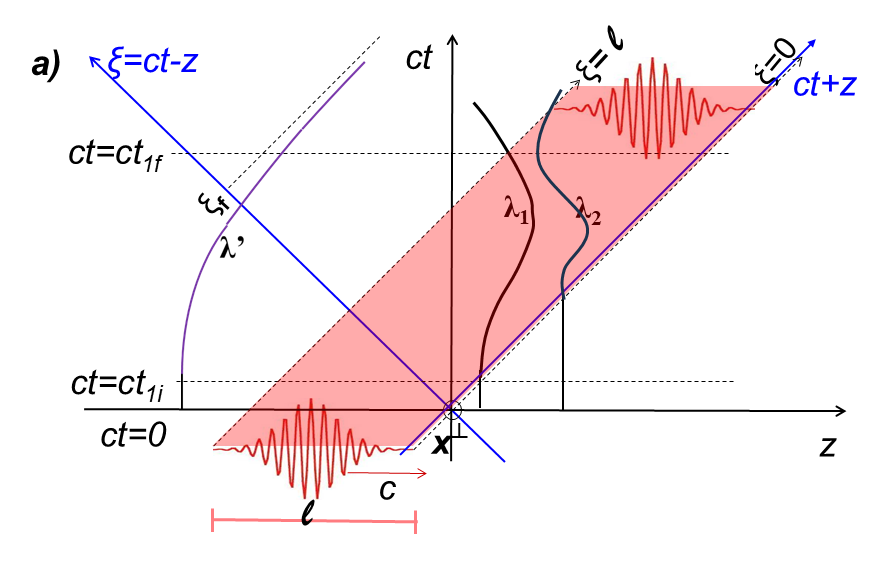}\hfill
\includegraphics[height=5.4cm]{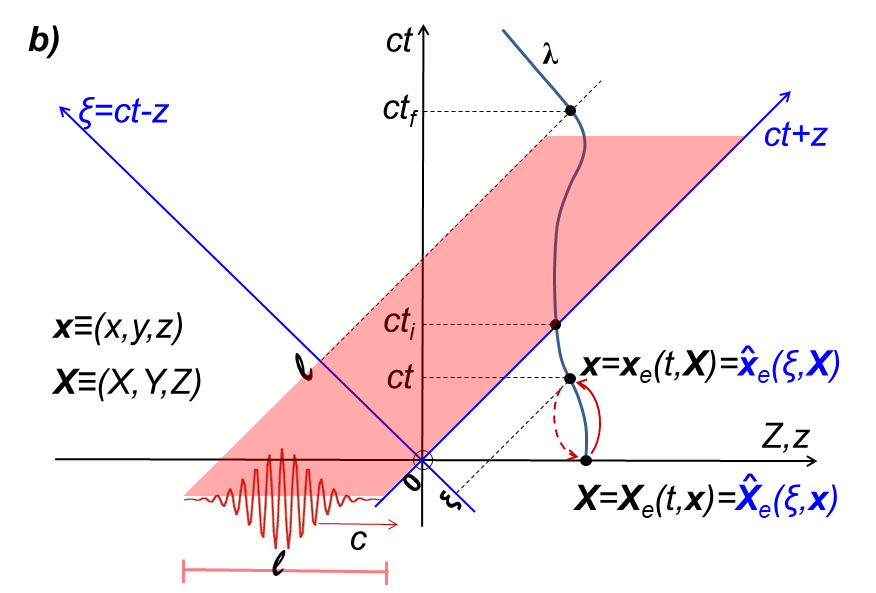}
\caption{a) As every particle travels slower than light, $\tilde \xi(t)=ct\!-\!z(t)$ grows strictly, and 
{\color{blue}{$\xi=ct\!-\!z$}} can replace $t$ as the independent parameter along its worldline (WL) $\lambda$ in Minkowski space  and  in its  eq. of motion 
\cite{Fio18JPA}.   
$\tilde\xi\to\xi_f<\infty$ as $t\to\infty$ implies $ \dot z\to c$, see e.g. WL $\lambda'$. \\
While the instants of intersection with the front and the end of a plane EM wave (\ref{EBfields0}) (whose support we have painted in pink) depend on the particular WL
(we have pinpointed the ones $t_{1i},t_{1f}$ for $\lambda_1$), the corresponding light-like coordinates are the same for all WLs: $\xi_{1i}=\xi_{2i}=0$, $\xi_{1f}=\xi_{2f}=l$: the pulse acts as a ``clock" for ``time" $\xi$. \\ b) We label the ``particles" (i.e. elements) of the electron fluid by their initial positions $\bX$; the hydrodynamic regime (HR) holds as long as  their WLs  do not intersect, i.e. the maps \ 
$\bX\mapsto \bx$ \ are one-to-one.
We denote Eulerian, Lagrangian  observables as follows:
{\color{red}{Eulerian observables}} \ ${\color{red}{f}}(t,\bx)={\color{red}{\check f}}({\color{blue}{\xi}},\bx)={\color{magenta}{\tilde f}}(t,\bX)={\color{magenta}{\hat f}}({\color{blue}{\xi}},\bX)$ \ {\color{magenta}{Lagrangian  observables}}.}
\label{fig2}
\end{figure}

The (Lorentz)  equations of motion of a charged particle in a given  external
electromagnetic  (EM) field are non-autonomous and highly nonlinear in the unknowns \  $\bx(t)$, \ $\Bp(t)=mc\,\bu(t)$:
\bea
\ba{l}
\displaystyle\dot\Bp(t)=q\bE[t,\bx(t)] + \frac{\Bp(t) }{\sqrt{m^2c^2\!+\!\Bp^2(t)}}  \times  q\bB[t,\bx(t)] 
,\\[6pt]  
\displaystyle
\frac{\dot \bx(t)}c =\frac{\Bp(t) }{\sqrt{m^2c^2\!+\!\Bp^2(t)}} ,
\ea
\label{EOM}
\eea
Here \ $m,q,\bx,\Bp$ \ are the  rest mass, electric charge, position
and   relativistic momentum of the particle, $c$ is the speed of light. We use  Gauss CGS units. 
As usual, it is convenient to use dimensionless variables: \
$\bb\!\equiv\!\bv/c\!\equiv\!\dot \bx/c$, \  the Lorentz relativistic factor  
$\gamma\!\equiv\!dt/d\tau\!=\!1/\sqrt{1\!-\! \bb^2}$ ($\tau$ is the proper time of the particle), the 4-velocity \ $u\!=\!(u^0\!,\bu)
\!\equiv\!(\gamma,\!\gamma \bb)\!=\!\left(\!\frac {p^0}{mc^2},\!\frac {{\Bp}}{mc}\!\right)$, i.e. the dimensionless version of the 4-momentum; whence
$\gamma=\sqrt{1\!+\! \bu^2}$. $\bx\!=\!x\bi\!+\!y\bj\!+\!z\bk\!=\bxp\!+\!z\bk$ will be the decomposition of $\bx$ in the Cartesian coordinates of the laboratory frame; for any vector ${\bf V}$ we denote by ${\bf V}^\perp$ its component in the $xy$ plane.
In terms of the EM potential 4-vector  $A\equiv (A^\mu)=(A^0,\bA)$  the electric and magnetic field read
\ $\bE=-\partial_t\bA/c-\nabla A^0$ and $\bB=\nabla\!\times\!\bA$. \
If $\bE,\bB$ depend only on $t,z$, we partially fix the gauge choosing $A=(A^0,\bA)$ to depend only on $t,z$, too.
We consider first an {\it external} EM field of the form
\be
\bE(t ,\bx)=\Bep(ct\!-\!z)+\bk E^z(t,z),\qquad \bB(t ,\bx)=\bk\times\Bep(ct\!-\!z),
\label{EBfields0}
\ee
i.e. the sum of  a (possibly vanishing) 
longitudinal electric field $E^z$ and a plane transverse travelling wave  (the {\it pump}) propagating in the $z$-direction, determined by the function $\Bep(\xi)$. We can model the  EM field of a laser pulse 
by the transverse parts $\bEp,\bBp$ of (\ref{EBfields0});  $\Bep(\xi)$ is in general a rapidly varying function of $\xi$ in its  support.
Since no particle can reach the speed of light $c$ [$|\dot\bx|\!<\!c$ by (\ref{EOM}b)], \ $\tilde \xi(t)\!\equiv\!ct\!-\!z(t)$  grows strictly and admits the inverse $\hat t(\xi)$; hence,
we can make the change  $t\mapsto \xi\!=\!ct\!-\!z$ of independent parameter  along its worldline (WL), see Fig. \ref{fig2}.a, so that the expression  $\Bep[ct\!-\!z(t)]$ in  (\ref{EOM}), where the {\it unknown} $z(t)$ is in the argument of the highly nonlinear and rapidly varying $\Bep$, becomes the {\it known} forcing term $\Bep(\xi)$. Let $\hat \bx(\xi)$ be the position of the particle
 as a function of $\xi$,  $\hat \bx(\xi)=\bx(t)$,  and $c\,\hat t(\xi)\!\equiv\!\xi\!+\!  \hat z(\xi)$. More generally  for any given function $f(t,\bx)$ we  denote $\hat f(\xi, \hat \bx)\equiv f\big[\hat t(\xi),  \hat \bx\big]$, abbreviate $\dot f\!\equiv\! df/dt$, $\hat f'\!\equiv\! d\hat f/d\xi$ (total derivatives). 
It is convenient to make also the change of dependent 
variable $u^z\mapsto s$, where $s$  is the light-like component of $u$  
\be
s\equiv\gamma\!- u^z=u^-=\gamma(1-\beta^z)=\frac{\gamma}c \frac{d\tilde \xi}{dt}>0;         \label{defs0}
\ee
$\gamma,\bu,\bb$  are the rational function of $\bu^{{\scriptscriptstyle\perp}}\!\!,s$ 
\be
\gamma\!=\!\frac {1\!+\!\bu^{{\scriptscriptstyle\perp}}{}^2\!\!+\!s^2}{2s}, 
\qquad  u^z\!=\!\frac {1\!+\!\bu^{{\scriptscriptstyle\perp}}{}^2\!\!-\!s^2}{2s}, 
 \qquad  \bb\!=\! \frac{\bu}{\gamma}                                    \label{u_es_e}
\ee
((\ref{u_es_e}) hold also with the carets); $s\!\to\!0$ implies $\gamma,u^z\!\to\!\infty$. 
Replacing $d/dt\mapsto(c s/ \gamma)d/d\xi$ and putting carets  on all variables makes (\ref{EOM}) rational in the unknowns $\hat\bu^{{\scriptscriptstyle\perp}},\hat s$, in particular  (\ref{EOM}b) becomes $\hat\bx'=\hat\bu/\hat s$. 
For an electron in an EM field of the form (\ref{EBfields0}) the transverse component of (\ref{EOM}a) becomes \ $\dot{\Bp}^{{\scriptscriptstyle\perp}}=\frac ec \frac{d\bAp}{dt}$  \ or equivalently $\hat\bu^{\scriptscriptstyle \perp}{}' (\xi)=-\Bep(\xi)\frac e{mc^2}$, \  equipped with $\hat\bu^{\scriptscriptstyle \perp} (0)=\0$. Replacing the solution 
\be
\hat\bu^{\scriptscriptstyle \perp} (\xi)=\frac e{mc^2}\Bap(\xi) ,\qquad \quad 
\Bap(\xi)\equiv -\!\int^{\xi}_{ -\infty }\!\!\!d\zeta\:\Bep(\zeta)    \label{bup1}
\ee
of the latter [note that $\bAp(t,z)\equiv\Bap(ct\!-\!z)$ can be adopted
as the transverse vector potential of the EM field (\ref{EBfields0})] into the $z$-component  of 
(\ref{EOM}) one obtains  (abbreviating $v\equiv \hat\bu^{\scriptscriptstyle \perp}{}^2$)
\be
\hat z' =\frac {1\!+\!v}{2\hat s^2}\!-\!\frac 12, \qquad 
\hat s'(\xi)  = \frac {e}{mc^2}\check E^z(\xi,\hat z).          \label{reduced}
\ee
Once this 
is solved, one determines  
also $\hat\bx^{\scriptscriptstyle \perp}(\xi)$ 
integrating the rest of $\hat\bx'=\hat\bu/\hat s$ over $\xi$:
\bea
&\hat\bx(\xi)=\bx_0+\hat \BD\!(\xi), \qquad &\mbox{where}\quad \hat \BD\!(\xi)\!\equiv\!\!\displaystyle\int^\xi_{0}\!\!\! d\eta \,\frac{\hat\bu(\eta)}{\hat s(\eta)}.          \label{hatsol}
\eea
Summarizing, (\ref{EOM}) is thus reduced to (\ref{reduced}).
The latter  are also the Hamilton eqs. for a system with 1 degree of freedom;
 $\hat z,-\hat s,\xi$  resp. play the role of $q,p,t$, while the Hamiltonian  (made dimensionless by dividing it by $mc^2$) reads $\hat H(z,s,\xi)=\gamma-  \check A^0(\xi, z)e/mc^2$, where
$\gamma$ depends on $s,\xi$ via (\ref{u_es_e}a),  (\ref{bup1}a).
If $\hat s(\xi)$ vanishes  as $\xi\uparrow \xi_f<\infty$ at least as fast as $\sqrt{\xi_f\!-\!\xi}$, 
then the {\it physical} solution as a function of $\xi$  is defined  only for $\xi<\xi_f$, 
whereas  as a function of $t$ is defined for {\it all} $t<\infty$, because \
$\hat t(\xi_f) =\xi_f +\hat z(\xi_f) =\infty$, \ by (\ref{reduced}a). 

If the pulse is a slowly modulated quasi-monochromatic wave (SMMW), i.e. 
\be
\Bep\!(\xi)\!=\!\underbrace{\epsilon(\xi)}_{\mbox{modulation}}
\underbrace{[\bi \cos\psi\,\sin (k\xi\!+\!\varphi_1)\!+\!\bj \sin\psi\sin (k\xi\!+\!\varphi_2)]}_{\mbox{carrier wave $\Be_o^{{\scriptscriptstyle \perp}}\!(\xi)$}},
 \label{modulate}
\ee
where $\bi=\nabla x$,  $\bj=\nabla y$,   $k=2\pi/\lambda$ is its
 wave number,   the modulating amplitude $\epsilon(\xi)\ge 0$ has support $[0,l]$, and \
$|\epsilon'|\!\ll\! |k\epsilon|$ for $\xi\in[0,l]$, \  one  can easily show (see \cite{Fio18JPA}, appendix A.4)  that 
\be
\Bap(\xi)\!\simeq\!  \frac{\epsilon(\xi)}{k}\,\Bep_o\left(\xi\!+\!\frac{\pi}{2k}\right) 
,
 \label{alphaapprox}
\ee
(whence in particular $\Bap\!(\xi),\bup\!(\xi),v(\xi) \!\simeq\!   0$ if $\xi\!>\!l$). Then the following is valid:

\noindent
{\bf Remark 2.1.}   $\hat s$ is practically insensitive
  to the rapid oscillations of the pump $\Bep$ (as e.g. fig. \ref{graphsb}.b illustrates), 
because it essentially depends only on the average-over-a-cycle \ $v_a(\xi)\equiv \frac 1{\lambda}\int_{\xi}^{\xi+\lambda}d\eta\: v(\eta)$ \ of $v$ (the ponderomotive potential),
i.e. essentially only on $\epsilon/k$;  e.g. the linearly polarized  $\Bep(\xi)=\bi \,\epsilon(\xi)\cos(k\xi)$  yields $v_a(\xi) \simeq \frac 12\left[\frac{q\,\epsilon(\xi)}{kmc^2}\right]^2$.
Consequently, replacing $v$ by $v_a$ in (\ref{reduced}) 
does not change the integrals significantly, but makes $\hat s
$ much easier to compute\footnote{In fact, note that the Cauchy problem (\ref{reduced}) with initial conditions $\big(\hat z(\xi_0),\hat s(\xi_0)\big)=(z_0,s_0)$ is equivalent to
\bea
\hat z(\xi)=z_0\!+\!\int^\xi_{\xi_0}\!\! d\zeta
\left[\frac {1\!+ \!v(\zeta)}{2\hat s^2(\zeta)}\!-\!\frac 12\right]\!, \qquad  
\hat s(\xi)=s_0\!- \!\int^\xi_{\xi_0}\!\!d\zeta\,\frac{qE_s^z[\hat z(\zeta)]}{mc^2}.
\label{heq1rint} 
\eea
The fast oscillations of $ v$  induce
by the  integration in (\ref{heq1rint}a) much smaller relative oscillations of $\hat z$, because $v/\hat s^2\!\ge\!0$ 
and its integral  grows with $\xi$; the  integration in  (\ref{heq1rint}b) averages the residual small oscillations of $E_s^z[\hat z(\xi)]$ to yield an essentially smooth $\hat s(\xi)$.  The results do not change perceptibly if we replace  $v$ by $v_a$.
}.

\subsection{Plane problem eqs: EM pulse interacting with 
a cold plasma at rest}
\label{planemodel}

\begin{figure}
\includegraphics[width=8cm]{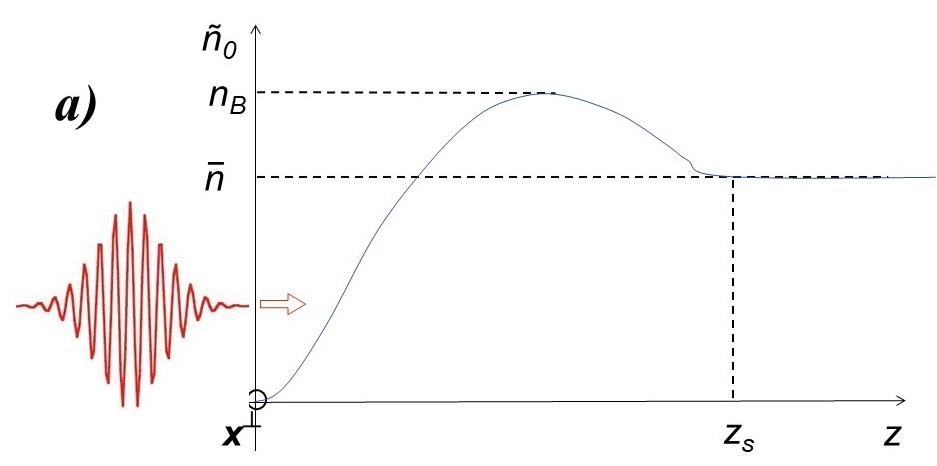}\hfill
\includegraphics[width=7cm]{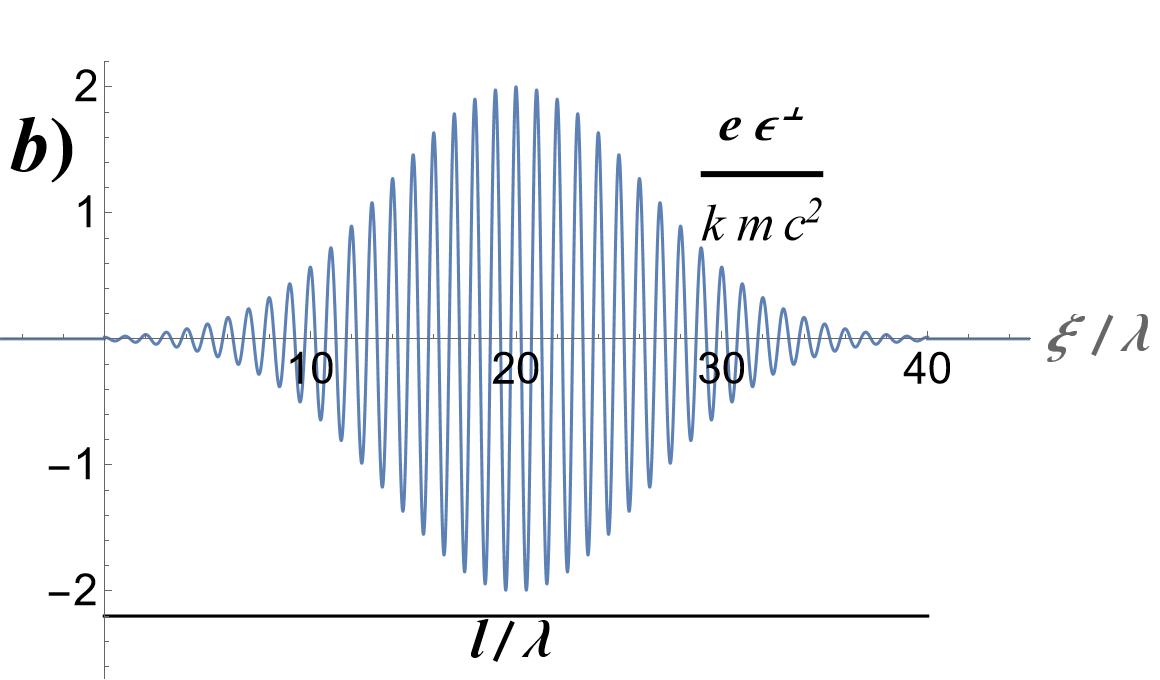} 
\caption{a) An initial plasma density of the type
(\ref{n_0bounds}) with an up-ramp, a down-ramp, and a plateau.
b) An {\it essentially short}, slowly modulated quasi-monochromatic (SMM) laser pulse.}
\label{fig1}
\end{figure}

We apply the changes of independent and dependent variables  considered in section \ref{1-particle} also to the plasma electrons (while we regard ions as immobile). 
We denote as $\bx_e(t,\bX)$  the position at time $t$
of the electrons' fluid element contained in the volume element $d^3\!X$ initially located at $\bX\!\equiv\!(X,\!Y,\!Z)$, as $\hat \bx_e(\xi,\!\bX)$ the same position    as a function of $\xi$.
 For brevity   we refer to the electrons initially contained: in $d^3\!X$, as the  `$\bX$ electrons';  in a 
region $\Omega$, as the `$\Omega$ electrons'; in the layer between $Z,Z\!+\!dZ$,
as  the `$Z$ electrons'. In the HR
the map $\bx_e(t, \cdot):\bX\mapsto \bx$ must be one-to-one for every $t$;
equivalently,   $\hat\bx_e(\xi, \cdot):\bX\mapsto \bx$ must be one-to-one  for every $\xi$.
The inverses $\bX_e(t,  \cdot):\bx\mapsto \bX$, $\hat \bX_e(\xi,  \cdot):\bx\mapsto \bX$ fulfill 
\be
\bX_e(t,  \bx)=\hat \bX_e(ct\!-\!z,  \bx). \label{clear}
\ee
We assume that the Eulerian electron fluid   density $n_e$, velocity $\bv_e$ fulfill the initial conditions
\be 
\bv_e(0 ,\!\bx)\!=\!\0, \qquad n_e(0,\!\bx)\!=\!\widetilde{n_0}(z), 
 \label{asyc}
\ee
where the initial electron (as well as proton) density $\widetilde{n_0}(z)$ satisfies
\be 
 \widetilde{n_0}(z) \le  n_B,\qquad \widetilde{n_0}(z)=\left\{\!\!\ba{ll}0 \:\: &\mbox{if }\: z\!\le\! 0, \\
\bar{n} \:\: &\mbox{if }\: z\!\ge\! z_s \ea\right.
 \label{n_0bounds}
\ee
for some $n_B\!\ge\! \bar{n}\!>\!0$ and $z_s\!>\!0$ (see e.g. fig. \ref{fig1}.a): $\widetilde{n_0}(z)$ is bounded by $n_B$ and for $z\ge z_s$ becomes a plateau of height  $\bar{n}$.
Up to section \ref{Applic} we model the EM fields $\bE,\bB$ before the impact ($t\le 0$) 
as (\ref{EBfields0}) with $E^z=0$, assuming that the support of
 $\Be^{{\scriptscriptstyle\perp}}\!(\xi)$ is an interval $0\le\xi\le l$ of length $l$ fulfilling \ 
$l\lesssim  \sqrt{\!\pi mc^2/n_B e^2}$ \cite{SprEsaTin90PRL,FioFedDeA14}, or more generally \cite{FioDeAFedGueJov22} (\ref{Lncond'}), so as to maximize the  oscillation amplitude, and thus also the energy transfer from the pulse to the PW.
 $\widetilde{n_0}(z),\Be^{{\scriptscriptstyle\perp}}(\xi)$ make up the {\it input data} of our problem.

Since the problem is independent of $x,y$, also the EM field, densities and
velocities obtained solving the Maxwell+plasma equations will depend only on $t,z$; 
similarly, the displacements  \ $\BD_e\equiv \bx_e(t,\!\bX\!) -\!\bX$ will 
actually depend only on $t,Z$ [and their ``hatted" counterparts 
$\hat\BD_e\equiv \hat\bx_e(\xi,\bX)\!-\!\bX$ only on $\xi,Z$].
Hence we can partially fix the gauge choosing
the EM potential $A=(A^0,\bA)$ to depend only on $t,z$ as well, and its
transverse part as the physical observable
\be
\bA\!^{{\scriptscriptstyle\perp}}(t,z) \equiv - c\!\! \int^{t}_{ -\infty }\!\!\!  dt' \: \bEp\!(t',z).
\label{Ap}
\ee
By (\ref{EBfields0}), \ $\bAp(t,z)=\Bap(ct\!-\!z)$ \ for $t\le 0$.
The Eulerian electrons' momentum $\Bp_e(t,z)$ obeys equation
(\ref{EOM}), where one has to replace $\bx(t)\mapsto\bx_e(t,\bX)$,
 $\dot \Bp\mapsto d\Bp_e/dt\equiv$ {\it total} derivative;  by (\ref{Ap}) and 
$\Bp^{{\scriptscriptstyle\perp}}_e(0,\bx)\!=\!\0$,
again the transverse part of (\ref{EOM}a)  becomes
$\frac{d\Bp^{{\scriptscriptstyle\perp}}_e}{dt}=\frac ec \frac{d\bAp}{dt}$ 
and  implies
\be
\Bp^{{\scriptscriptstyle\perp}}_e=\frac ec \bAp \qquad\quad 
\mbox{i.e. }\quad \bup_e=\frac e{mc^2} \bAp         \label{bupExpl}
\ee
(the analogue of (\ref{bup1})), or equivalently its hatted counterpart. This relation allows to trade $\bup_e$  for $\bAp$ as an unknown. 
In \cite{FioPostHydro} it is shown that this plane, fully relativistic Lagrangian approach can be pursued even after the first wave-breaking (WB), provided the plasma remains essentially collisionless; the only novelty is that for $t$ (resp. $\xi$) sufficiently large the map $Z\mapsto z_e(t,Z)$  (resp. $Z\mapsto \hat z_e(\xi,Z)$) may be no longer  one-to-one, i.e. two or more different electron layers can go through each other.
The number  of  electrons  per unit transverse surface  having longitudinal coordinate $z'\!\le\! z$ at time $t$ is 
\be
 N_e(t,z):=\int^{\infty}_0\!\!\! d Z\:\widetilde{n_{0}}(Z)\:\theta\big[z-z_e(t,Z)\big],
\label{DefN_e}
\ee
where $\theta$ is Heaviside step function (clearly, for any fixed $t$ this vanishes for sufficiently small $z$,  because $|v_e^z|< c$ and $z_e(0,Z)=Z>0$).
It is easy to check that $\partial_z N_e=n_e$. 
In \cite{FioPostHydro}  it is shown that, assuming immobile ions, the solution $E^z(t,z)$ of the Maxwell equations
\bea
\nabla\cdot\bE=\partial_z E^z=4\pi j^0 
,\qquad \qquad \partial_0E^z+4\pi j^z /c=(\nabla\times\bB)^z=0
\label{Maxwell'}
\eea
($j^0(t,z)=e[\widetilde{n_0}(z)-n_e(t,z)]$, $\bjc=-en_e \bv_e$ are the electric charge density and current density)
 fulfilling the plane  initial conditions (\ref{asyc}) 
is related to $N_e (t, z)$ by the left equation in
\be
E^z(t, z)\!=\!4\pi e \big[
\widetilde{N}(z)\!-\! N_e (t, z)\big]
\qquad \Leftrightarrow\qquad \check E^z(\xi, z)\!=\!4\pi e  \big[
\widetilde{N}(z)\!-\! \check N_e (\xi, z)\big], \label{elFL}
\ee
where \ $\widetilde{N}(Z):=\int^{Z}_0\!\!\! d Z'\,\widetilde{n_0}(Z')$ \  
is the number  of  protons (due to ions of all kinds)  per unit surface in the layer $0\!\le\!Z'\!\le\! Z$ (as well as the same number for the electrons at $t\le 0$); the right equation yields the Eulerian longitudinal electric field as a function of $\xi, z$.
More generally, one finds   \cite{FioPostHydro} \ $E^{{\scriptscriptstyle z}}(t,z)=4\pi \sum_h q_h N_h(t,z)$, \
where the sum runs over all particle species (ions or electron), $N_h(t,z)$ is the analogue of $N_e(t,z)$ for the $h$-th one, and $q_h$ is the associated electric charge.
These are the same results of  Proposition 2 of \cite{Fio14JPA} (see also formula (67) in  \cite{Fio18JPA})  under the present more general assumptions and
definitions of the $N_h(t,z)$; 
as long as the map $z_e(t,\!\cdot):\, Z\mapsto z$ is one-to-one (HR) the definition (\ref{DefN_e}) reduces to that \ $N_e(t,z)\!:=\!\widetilde{N}[ Z_e(t,z)]$ \
given in \cite{Fio18JPA}.
We point out that $N_e(t,z)$ keeps {\it well-defined and finite}, and (\ref{elFL}) make sense , {\it everywhere}, even at  spacetime points $(t,z)$ where the HR is broken and WB occurs, i.e. where  $\partial z_e/\partial Z=0$, 
what makes  $n_{e}(t,z)$ divergent and (\ref{Maxwell'}) ill-defined.
Therefore we can regard (\ref{elFL}) as a regularization of (\ref{Maxwell'}).

Relations (\ref{DefN_e}), (\ref{elFL})  and $n_e=\partial_z N_e$ allow to express $n_e,E^z$  in terms of  the assigned $\widetilde{n_0}$
and of the still unknown  $ z_e(t,Z)$  (or equivalently, $\hat z_e(\xi,Z)$), i.e. of the longitudinal motion, also after WB; 
thereby  they further reduce the number of unknowns. The remaining ones are $\bA\!^{{\scriptscriptstyle\perp}},\bx_e$ and $u_e^z$, or - alternatively - $s$. 
$\bAp$ is coupled to the current via the Maxwell eq. \ $\Box\bAp=4\pi\bjp$ (in the Landau gauges), where \ $-\bjp\!=e n_e\bb^{\scriptscriptstyle \perp}_e=e n_e\bup_e/ \gamma_e= e^2 n_e\bAp/ mc^2\gamma_e\equiv \bAp f/\pi$, \  by (\ref{bupExpl}). 
Under our assumptions about the supports of $\Bep,\widetilde{n_0}$, we have $\bj\!=\!\0$ if  $ct\!\le\! |z|$, by causality. Let $\xip:= ct\!+\!z$.
Changing independent variables $(ct,z)\mapsto(\xi,\xip)$, noting that $\Box=4\frac{\partial}{\partial \xi}\frac{\partial}{\partial \xip}$  and defining the function $\check g$ via
$\check g(\xi,\xip)=g(t,z)$ (for all functions $g$), we obtain $\check\bj\!=\!\0$ if either $\xi\!\le\! 0$ or $\xip\!\le\! 0$ and transform the previous eq. into
\be
\frac{\partial}{\partial \xi}\frac{\partial}{\partial \xip}\check\bAp(\xi,\xip)=\pi \check\bjp
(\xi,\xip)=-[\check\bAp  \check f](\xi,\xip);
\label{diffeq1p}
\ee
this equation must be equipped with the ``initial conditions" $\bAp(t,z)=\Bap(ct\!-\!z)$ \ for $t\le 0$. 

Using  (\ref{bupExpl})   and abbreviating \
  $v\!\equiv\!\hat\bu^{\scriptscriptstyle \perp}_e{}^2\!=\![ e\hat\bAp/{mc^2}]^2$, \
$\hat\Delta(\xi,Z)\!\equiv\!\hat\Delta^z(\xi,Z)\!=\!\hze(\xi,Z)\!-\!Z$, \
the remaining equations   to solve  are  (\ref{reduced}) for the plasma electrons,
which take the form  \cite{FioDeN16}
\bea
\hat\Delta'(\xi,Z)=\displaystyle\frac {1\!+\!v}{2\hat s^2}\!-\!\frac 12, \qquad 
\hat s'(\xi,Z)= \frac {e}{mc^2} \check E^z[\xi,\hat z_e\!(\xi,\! Z)]=K\left\{\!
\widetilde{N}\left[Z\!+\!\hat\Delta\right] \!-\! \widetilde{N}(Z)\!\right\} 
\label{heq1} 
\eea
($K\equiv\frac{4\pi e^2}{mc^2}$) in the HR. 
More generally, $\check E^z$ is given by (\ref{elFL}b), and (\ref{heq1}b) has to be replaced by
\be
\hat s'(\xi,Z) 
= K  \left\{ 
\widetilde{N}\big[
\hat z_e(\xi,Z) \big]-\!\int^\infty_0\!\!\!\!\! d\zeta\,\widetilde{n_0}(\zeta)\, \theta\big[\hat z_e(\xi,\!Z)\!-\!\hat z_e(\xi,\!\zeta)\big] \right\};  \label{Newheq1b}
\ee
the first, second  term at the right-hand side (rhs) are due to the interaction of the $Z$ electrons resp. with ions,  other electrons.
Eqs. (\ref{heq1}-\ref{Newheq1b}) are equipped with the initial conditions (IC)
\ $\hat\Delta(-Z,\!Z)\!=\!0$, $\hat s(-Z,\!Z)\!=\! 1$.  Since 
$\bAp\!=\!\0$ for  $ct\!\le\!z$, then $v,\hat\Delta,\hat s-1$ remain zero until $\xi=0$, 
and we can shift the IC to
\bea
&&  \: \hat \Delta(0,\!Z)\!=\!0,  \qquad\qquad
 \hat s(0,\!Z)\!=\! 1. \qquad\qquad\qquad\qquad  \label{heq2}
\eea
Since  $\bA\!^{{\scriptscriptstyle\perp}}(t,z)=\Bap(ct\!-\!z) $ for $t\le 0$,  this equality and 
the one  \  $\hat\bu^{{\scriptscriptstyle\perp}}_e\!=\! e\Bap/{mc^2}$ \ approximately hold  at least for small $t>0$.
Replacing $\bA\!^{{\scriptscriptstyle\perp}}(t,z)\mapsto\Bap(ct\!-\!z) $ makes $\hat\bu ^{{\scriptscriptstyle\perp}}_e$ and the forcing term $v$   {\it known} functions of $\xi$ (only), and 
 (\ref{heq1})  a family parametrized by $Z$ of {\it decoupled ODEs}:
correspondingly, in the HR  (only)  the equations of motion of each electron
layer are decoupled from those of the others. In section \ref{Back-reaction} we will justify such a replacement  for a rather long time lapse, arguing that under our assumptions the corresponding change of 
$\bA\!^{{\scriptscriptstyle\perp}}(t,z)$ with respect to (w.r.t.) $\Bap(ct\!-\!z) $ 
does not significantly affect the form of the PW.

As in the single particle case, for every $Z$  (\ref{heq1})  have the form of Hamilton equations \ $q'=\partial \hat H/\partial p$, $p'=-\partial \hat H/\partial q$  of a 1-dim system: \  $\xi,\hat\Delta, -\hat s$  play the role of $t,q,p$, and the Hamiltonian (which gives the total energy of a $Z$ electron) is rational in $\hat s$ and reads \cite{FioDeN16} 
\bea
\hat H( \hat \Delta, \hat s,\xi;Z)\equiv \gamma(\hat s;\xi)+ \U( \hat \Delta;Z), \qquad\qquad
    \label{hamiltonian} \\[6pt]
\gamma(s;\!\xi) \!\equiv\frac{s^2\!+\!1\!+\!v(\xi)}{2s},\qquad
\frac{\U( \Delta;\!z)}K \!\equiv\! \int^{z \!+\!  \Delta}_z\!\!\!\!\!\!\!\!\! d\zeta\,\widetilde{N}(\zeta)
- \widetilde{N}\!(z)  \Delta;
\nonumber                            
\eea
$\gamma\!-\!1$, $\U$ act as kinetic, potential energy in $mc^2$ units.
We  can 
solve (\ref{heq1}-\ref{heq2}) in the unknown $\hat P\equiv(\hat\Delta,\hat s)$ numerically, or also by quadrature for 
$\xi\!\ge\! l$. Finally, \ $\hat\bx^{{\scriptscriptstyle\perp}}_e{}'\!=\!
\hat\bu ^{{\scriptscriptstyle\perp}}_e/\hat s$  is solved by
\vskip-.2cm
\be
\hat\bx^{\scriptscriptstyle \perp}_e(\xi,\bX)-\bXp=\!\int^\xi_0\!\!\! d\eta \,\frac{\hat\bu^{\scriptscriptstyle \perp}_e(\eta)}{\hat s(\eta,Z)},        \label{hatsol'}
\ee
cf. (\ref{hatsol}). 
In the HR,  the energy $H$ is conserved along the solution for $\xi\!\ge\!l$; its
value 
equals 
\be
h(Z) \equiv 1+\!\int^l_0\!\!d\xi v'(\xi)/\hat s(\xi,Z).  \label{defh}
\ee
For $Z\!>\!0$ the associated path in $P\equiv (\Delta,s)$ phase space is a cycle around $C\equiv\big(0,\gamma^{\scriptscriptstyle \perp}_f\big)$, where $\gamma^{\scriptscriptstyle \perp}_f$ is the value of 
$\hat\gamma^{\scriptscriptstyle \perp}\equiv \sqrt{1\!+\!\hat\bu^{\scriptscriptstyle \perp}_e{}^2}=\sqrt{1\!+\!v}$ for  $\xi\ge l$.
As in fig. \ref{graphsb}.a, below we assume for simplicity that the pulse is a slowly modulated quasi-monochromatic wave (SMMW); this makes $v(l)\ll 1$ (see e.g. \cite{FioDeNAkhFedJov23}. 
Approximating $v(l)\!=\!0$, it follows $C\!=\!(0,1)$, $v(\xi)\!=\!0$,  $\bup(\xi)\!=\!\0$, $\gamma^{\scriptscriptstyle \perp}=1$ and by (\ref{hatsol'}) \
$\hat\bx^{\scriptscriptstyle \perp}_e(\xi,\!\bX)\!=$const (purely longitudinal motion)  for $\xi\!\ge\! l$. 
Below we print 1 in boldface when it is the result of
approximating $\hat\gamma^{\scriptscriptstyle \perp}(l)\!\simeq\!1$ (so that the exact
results can be reobtained replacing $\1$ by $\hat\gamma^{\scriptscriptstyle \perp}_f$, or a suitable power thereof). The points $P$ of the cycle with center
$C\!=\!(0,\1)$ solve the equation $\hat H(\Delta,s;Z)\!=\!h(Z)$; 
\be
P_0\!\equiv\!(\Dm,\!\1), \quad
 P_1\!\equiv\!(0,\!\smm), \quad
P_2\!\equiv\!(\DM,\!\1), \quad
 P_3\!\equiv\!(0,\!\sM)  \label{Points_i}
\ee
minimize/maximize $\Delta$ or $s$, see e.g. fig. \ref{graphsb}.c. $\DM\!>\!0,\Dm\!<\!0$ solve the equation $\U(\Delta;Z)\!=\!h(Z)\!-\!\1$.  The periodic motion can be determined by quadrature;  the period is
\be
\xiH(Z)=2\int^{\DM(Z)}_{\Dm(Z)}\!\!\!\!\!\!\!\!\!d\Delta\, \frac{\breve{\gamma}(\Delta;Z)}{\sqrt{ \breve{\gamma}^2(\Delta;Z)-\1}}, \quad  \breve{\gamma}\equiv h\!-\!\U.                        \label{period}
\ee
The points $P$ with given  $\Delta\!\in\![\Dm,\!\DM]$ are
$P=(\Delta,\spm)$, where $\spm\!\equiv\! \breve{\gamma}\pm\sqrt{\breve{\gamma}^2\!-\!\1}$,
and  $\sM(Z)=\sp(0,\!Z)=h(Z)\!+\!\sqrt{h^2(Z)\!-\!\1}$, $\smm(Z)=\sm(0,\!Z)=h(Z)\!-\!\sqrt{h^2(Z)\!-\!\1}$. 
$\hat P(\xi,Z)$ moves anticlockwise along the  cycle and passes in the order through
$P_0,P_1,P_2,P_3$ at each turn; we shall denote as $\uxi^i$ ($i=1,2,3,4$) 
the $\xi$-lapse necessary to go from $P_{i-1}$ to $P_i$, and  as $\xi^i_k$ ($i=0,1,2,3$) the value  of $\xi\ge l$ such that  $P(\xi^i_k,Z)=P_i$ during the $k$-th turn, $k\in\NN_0$. Clearly
$\xiH=\uxi^1\!+\!\uxi^2\!+\!\uxi^3\!+\!\uxi^4$, \ and 
\bea
\xi_k^0=\xi_0^0\!+\!k\xiH,\qquad\xi_k^1=\xi_k^0\!+\!\uxi^1,\qquad\xi_k^2=\xi_k^1\!+\!\uxi^2,
\qquad\xi_k^3=\xi_k^2\!+\!\uxi^3.   \label{Def_xi^0_k}
\eea
The $\xi$-lapses  $\uxi^i$ are given by
\bea
 \uxi^1=\frac14 \xiH+\Dm,\quad
\uxi^2=\frac14 \xiH-\DM,\quad
\uxi^3=\frac14 \xiH+\DM,\quad
\uxi^4=\frac14 \xiH-\Dm .  \label{xilapses}
\eea

The PW emerges as a collective effect, 
see e.g. fig.  \ref{Worldlinescrossings'}. The  map  $\hat \bx_e(\xi,\cdot)\!:\!\bX\!\mapsto\!  \bx$ from Lagrangian to Eulerian coordinates is invertible,  and the HD  justified,  as long as 
\be
\hat J \equiv  \det\left(\frac{\partial\hat \bx_e}{\partial \bX}\right)   \stackrel{(\ref{hatsol'})}{=} \frac{\partial\hat z_e}{\partial Z} > 0 
\ee
for all $Z$; we  recover the Eulerian $\bu_e,\!\gamma_e,\!\bb_e$   replacing $(\xi,\!Z)\!\mapsto\! \big(ct\!-\!z,Z_e(t,\!z)\big)$ in 
 $\hat\bu_e,\!\hat\gamma_e,\!\hat\bb_e$, e.g.
\bea
\bu_e(t,z )=\hat\bu_e\!\left[ct-z, Z_e(t, z)\right].      \label{sol'}
\eea
The electron density diverges where $\hat J=0$, because \cite{FioDeN16}
\be
n_e(t,z)\!=\!\left[ \frac{\hat\gamma\,\widetilde{n_{0}}}{\hat s \,\hat J}\right]_{(\xi,Z)=\big(ct\!-z,\hat Z_e(ct\!-z,z)\big)}=\left.\frac{\widetilde{n_0}(Z)}
{(1\!-\!\hat\beta^z) \hat J}\right\vert_{(\xi,Z)=\big(ct\!-z,Z_e(t,z)\big)}.
\label{expln_e}
\ee
Below we assume  that the pulse  is not only a SMMW, but also {\it essentially short (ES) w.r.t. $\widetilde{n_0}$} \cite{FioDeAFedGueJov22}, in the sense
\be
\hs(\xi,Z)\ge \1,  \qquad \mbox{for all $\xi\!\in\![0,\!l]$, $Z\!\ge\! 0$.}
 \label{Lncond'}
\ee
Physically this implies that the pulse overcomes each plasma electron before the $z$-displace- ment $\Delta$ of the latter  reaches a negative minimum for the first time, as it occurs for WL $\lambda_1$, but not $\lambda_2$, in fig. \ref{fig2}.a.
ES pulses simplify the control of the PW and its WB, 
in particular  guarantee that  $\hat J>0$ for all $Z\ge 0$ and $0\le\xi\le l$, so that no 
wave-breaking during the laser-plasma interaction (WBDLPI) takes place. \
Sufficient conditions on the input data for ES 
pulses are given in \cite{FioDeAFedGueJov22,FioDeNAkhFedJov23}.
With stronger conditions (SC) 
on the input data one could ensure that $\big|\hat J-1\big|\ll 1$ for $0\le\xi\le \xi_1^0$.

\subsection{Special case: \ constant initial density}
\label{specialcase}

If \ $\widetilde{n_0}( Z ) =\bar{n}$ \ then 
the $Z$-dependence disappears completely from
(\ref{heq1}-\ref{heq2}), which  reduces to the  Cauchy problem of a forced, relativistic harmonic oscillator with equilibrium at $\Delta\!=\!0$:
\bea
\hat\Delta'=\displaystyle\frac {1\!+\!v}{2\hat s^2}\!-\!\frac 12,\qquad\quad
\hat s'=M\hat\Delta,\qquad \qquad \hat\Delta(0)\!=\!0, \quad \hat s(0)\!=\! 1,
\label{e1}
\eea
 where $ M \!\equiv\!K\bar{n}\!=\!4\pi e^2\bar{n}/mc^2 \!=\!\omega_p^2/c^2
$ ($\omega_p$ is the {\it plasma angular frequency}).  The solution  of (\ref{e1}), as well as  $h$, $\hat\bu_e,\hat\BD$, depend on $\bar{n}$, but not on $Z$; it describes the motion  of every $Z$ layer of electrons. From $\partial\hat\Delta/\partial Z\!\equiv\!0$ it follows $\hat J\!=\!1$, which guarantees the HR for ever. In the sequel we denote by $\bar P(\xi;\bar{n})\equiv \big(\Delta(\xi 
;\bar{n}),s(\xi;\bar{n})\big)$ or more synthetically by $\bar P(\xi)\equiv \big(\Delta(\xi),s(\xi)\big)$
the solution of (\ref{e1}), by $\bar{h}(\bar{n}),\ospm,\osmm,\osM,\oDM,\oDm,\bar P_i,\bxiH,\ouxi^i,\bar\xi^i_k...$ the associated  $h,\spm,\smm,\sM,\DM,\Dm,P_i,\xiH,\uxi^i,\xi^i_k,...$. 
Since $\U\!\equiv\!M\Delta^2/2$, also $\oDM\!=\!-\oDm\!=\!\sqrt{2(\bar{h}\!-\!1)/M}$ is
$Z$-independent.
The inverse functions $Z_e(t,\!z)$ and $\hat Z_e(\xi,\!z)$ of $z_e(t,\!Z),\hat z_e(\xi,\!Z)$ have the closed form  
\be
 Z_e(t,z)=z-\Delta(ct\!-\!z)\qquad\Leftrightarrow\qquad \hat Z_e(\xi,z)=z\!-\!\Delta(\xi),                                                \label{sol"}
\ee 
which makes all Eulerian fields - such as   (\ref{sol'}) or (\ref{expln_e}) - completely explicit and 
dependent on $t,z$ only through $ct\!-\!z$, i.e. propagating as travelling-waves; in particular 
(\ref{expln_e}) becomes \cite{AkhPol56}
\bea
 &&  n_e(t,z)
=\frac{\bar{n}}2 \,
\left[1\!+\! \frac{1\!+\!v(ct\!-\!z)}{s^2(ct\!-\!z)}\right]=
\frac{\bar{n}}{1-\beta^z(ct\!-\!z)}.       \label{expln_e'}
\eea
{\bf Remark 2.3.} \  If the pulse is a SMMW, then this plasma density would not perceptibly change for $\xi\equiv ct\!-\!z\ge l$ if we  replaced $v$  by $v_a$ in eq. (\ref{reduced}), by Remark 2.1 (recall that $v(\xi)=$const$\simeq 0$ if $\xi\equiv ct\!-\!z\ge l$).

\begin{figure}
\includegraphics[width=8.5cm]{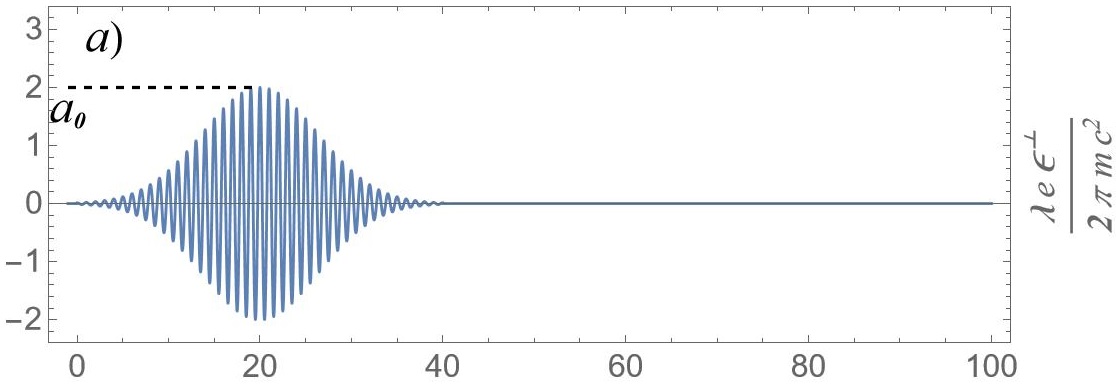} \hfill\includegraphics[width=6.2cm]{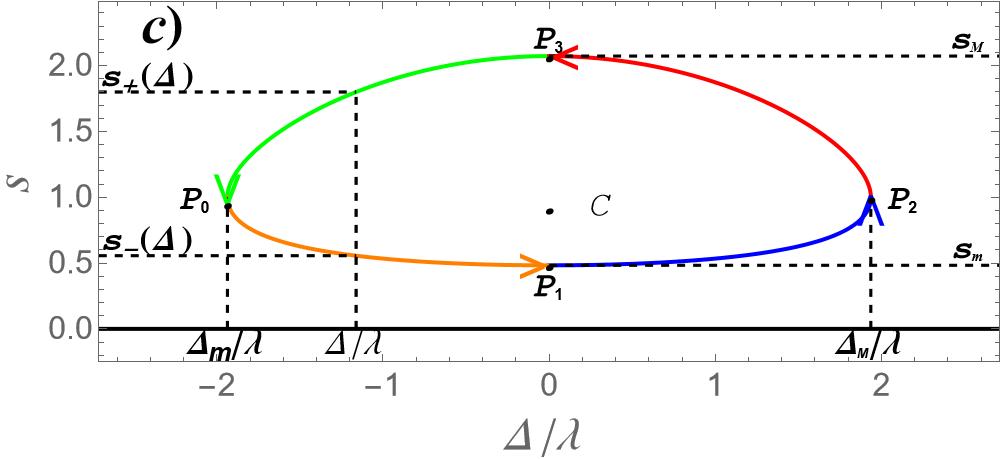}\\
\includegraphics[width=8.6cm]{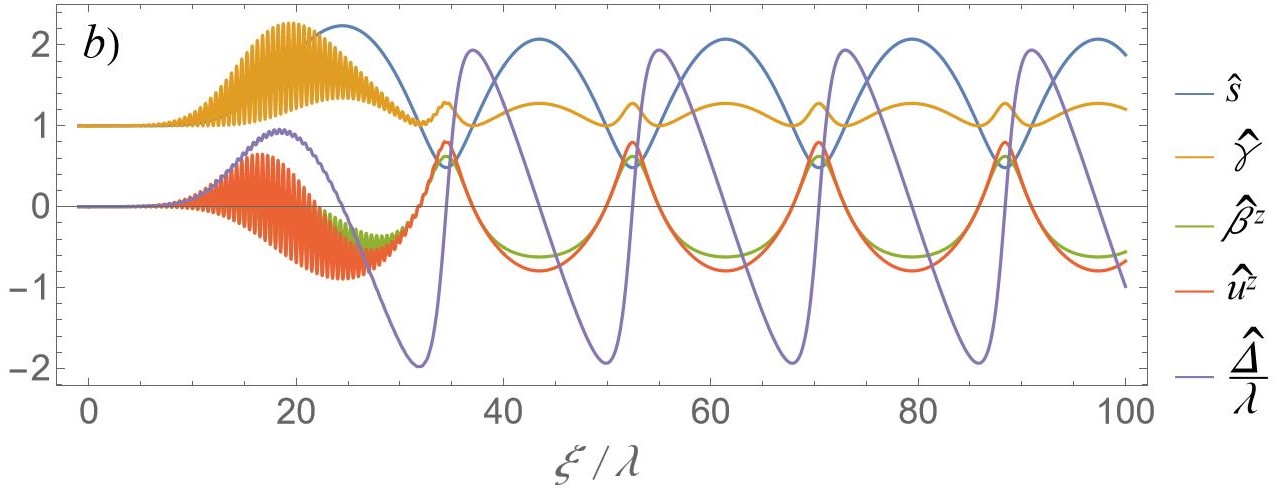}\hfill \includegraphics[width=7.6cm]{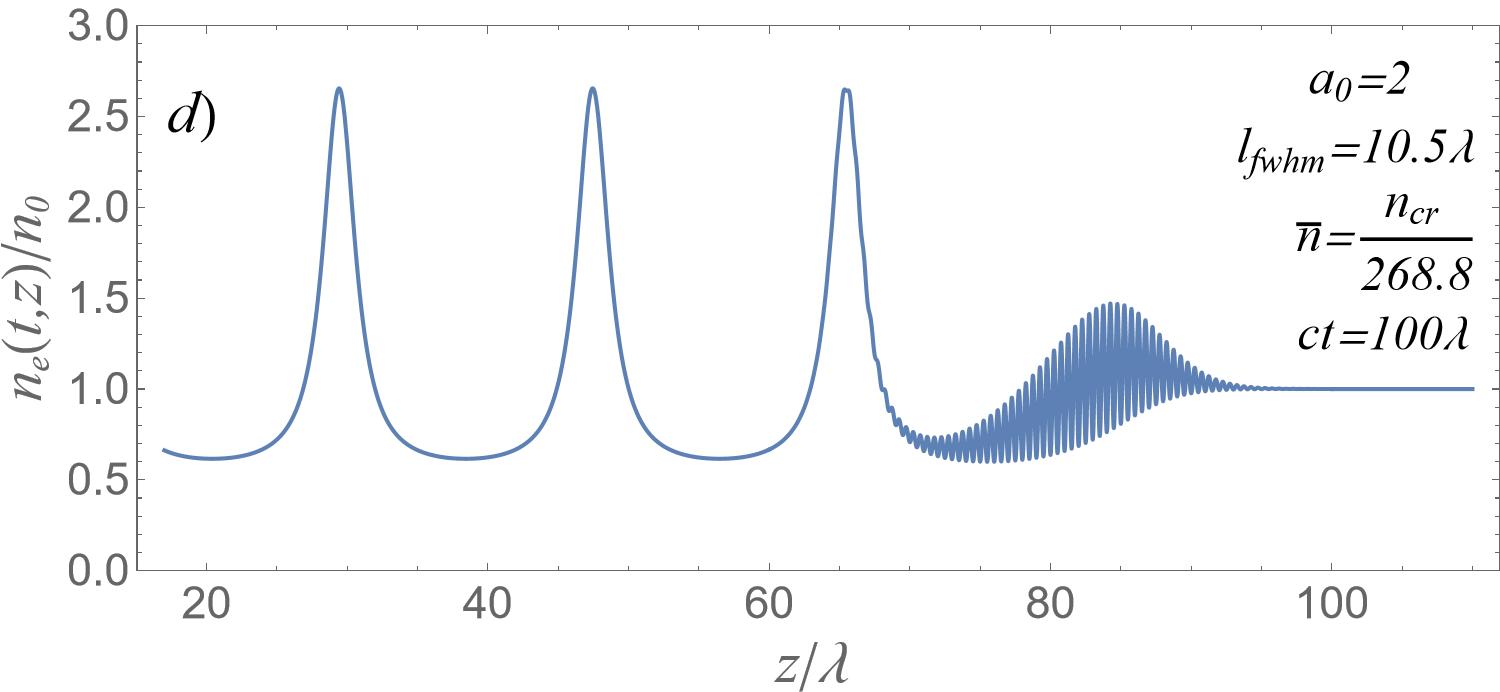}
\caption{
(a)  Normalized Gaussian pulse of FWHM $l'
\!=\!10.5\lambda$, linear polarization, 
peak  amplitude $a_0\!\equiv\!\lambda eE^{\scriptscriptstyle \perp}_{\scriptscriptstyle M}/2\pi mc^2\!=\!2$,
as in section III.B of \cite{BraEtAl08}. We consider $l\!=\!40\lambda$ and cut the tails outside $|\xi\!-\!l/2|\!<\!l/2$. 
\ \ (b) \ Corresponding  solution of (\ref{heq1}-\ref{heq2})  if $\widetilde{n_0}(z)\!=\!\bar{n}^j\!\equiv\! n_{cr}/268.8$ 
($n_{cr}\!=\! \pi mc^2/e^2\lambda^2$ is the critical density); as a result, $E/mc^2\equiv h\!=\!1.28$. Adopting $\bar{n}=\bar{n}^j$ as the density plateau maximizes the LWFA of test electrons, see section \ref{WFA}.{B}. A wavelength $\lambda=0.8\mu$m  leads to a peak  intensity $I\!=\!1.7\!\times\!10^{19}$W/cm$^2$ and
$\bar{n}^j\!=\!6.5\times\! 10^{18}$cm$^{-3}$. 
\ \ (c) \ Corresponding  phase portrait (at $\xi>l$). \ \ (d) \ Corresponding electron density
at $t=100\lambda/c$. }
\label{graphsb}
\end{figure}

The period (\ref{period}) becomes
\be
\bar \xi_{{\scriptscriptstyle H}}\!\big(\bar{h};\bar{n}\big)
=8\sqrt{\!\frac{\bar{h}\!+\!1}{2K\, \bar{n}}}\left[\E(\alpha)-\frac
{\K(\alpha)}{\bar{h}\!+\!1}\right], 
\quad \alpha\equiv\sqrt{\!\frac{\bar{h}\!-\!1}{\bar{h}\!+\!1}} ; \label{period0}
\ee
$\K,\E$ are the complete elliptic integrals of the 1st, 2nd kind.
$\bar \xi_{{\scriptscriptstyle H}}/c$ resp. reduces to the nonrelativistic, ultrarelativistic harmonic oscillator periods $t_{{\scriptscriptstyle H}}^{{\scriptscriptstyle nr}}\!\equiv\!\sqrt{\!\pi m/\bar{n} e^2}$, $t_{{\scriptscriptstyle H}}^{{\scriptscriptstyle ur}}\!\simeq\! \tfrac{15}{8}\sqrt{\!\bar{h}\pi m/2\bar{n} e^2}$ 
 in the nonrelativistic  limit $\bar{h}\!\to\!1$ and the  ultrarelativistic one $\bar{h}\!\to\!\infty$.
 In fig. \ref{graphsb} we plot a Gaussian-modulated $\Bep(\xi)$ and the corresponding solution of (\ref{e1}) 
if $Ml_{fwhm}^2\!\simeq\!11$, $Ml^2\!\simeq\!158$. The abrupt rise of $\hat\Delta$ 
 where $\hat s$ reaches a minimum [see the comment after eq. (\ref{heq2})] is evident; the solution manifestly becomes periodic   for $\xi\!\ge\! l$, i.e.  after the pulse has passed. 

\subsection{Back-reaction of the plasma on the pulse}
\label{Back-reaction}

\begin{figure}
\includegraphics[width=5.4cm]{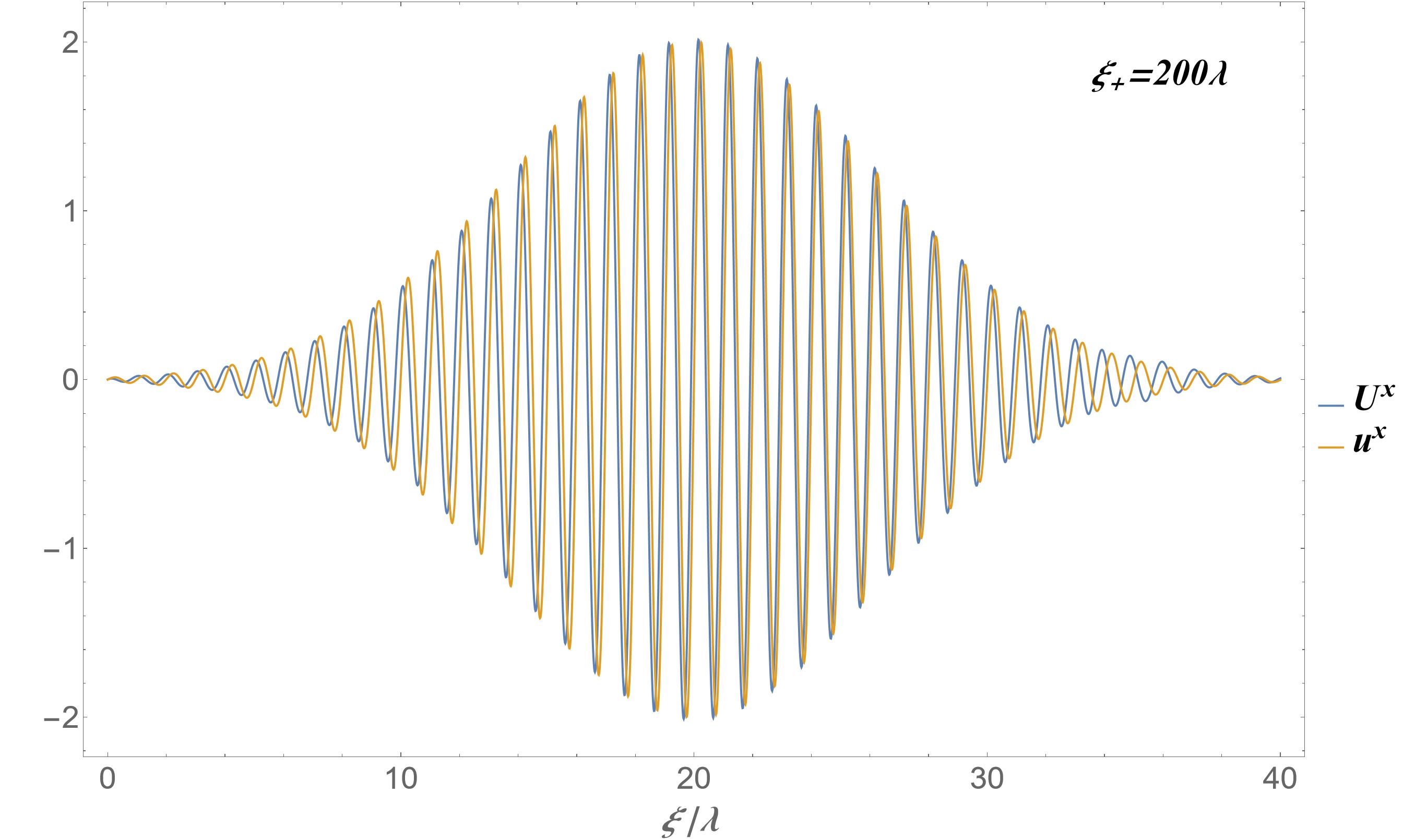} 
\includegraphics[width=5.4cm]{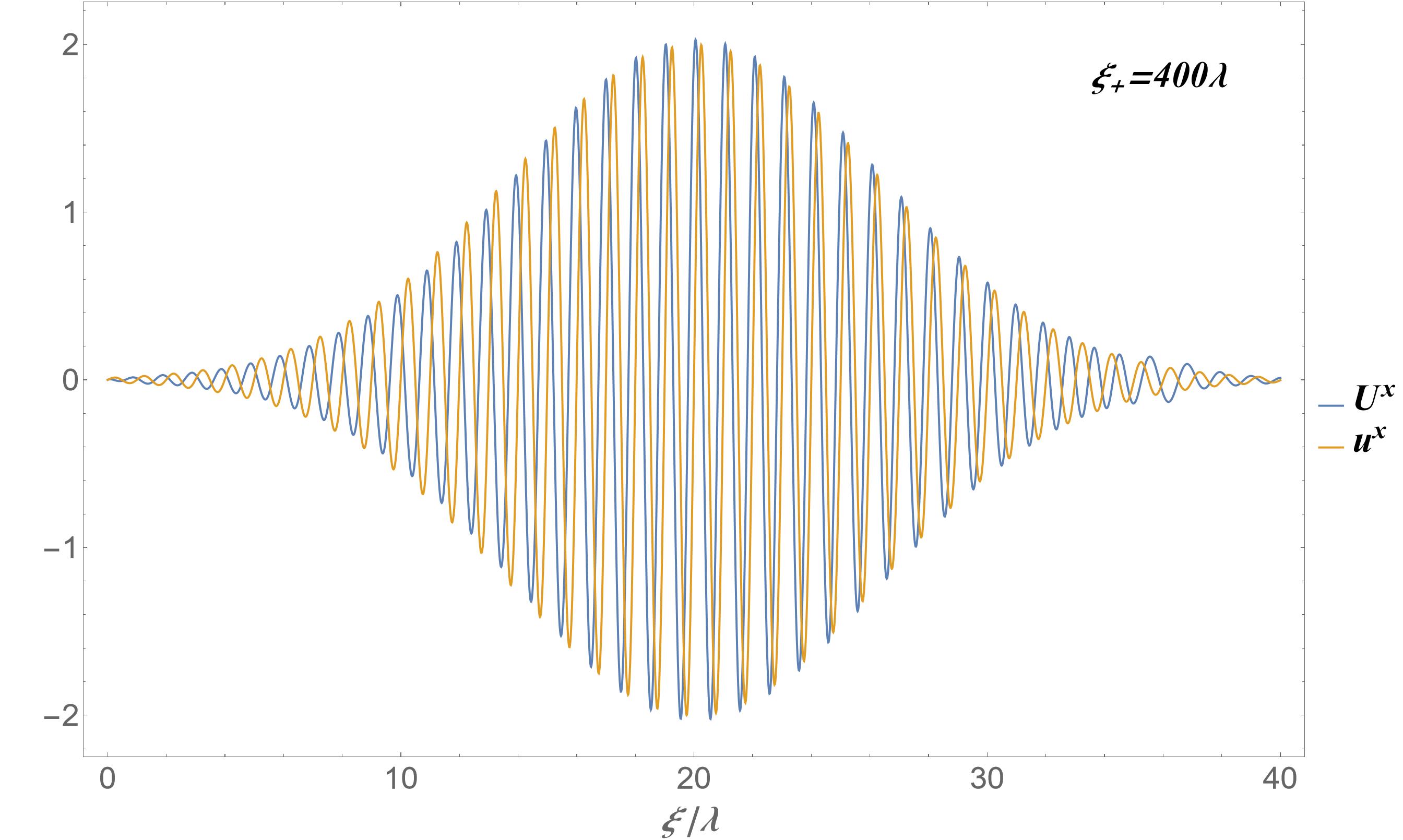}\includegraphics[width=5.4cm]{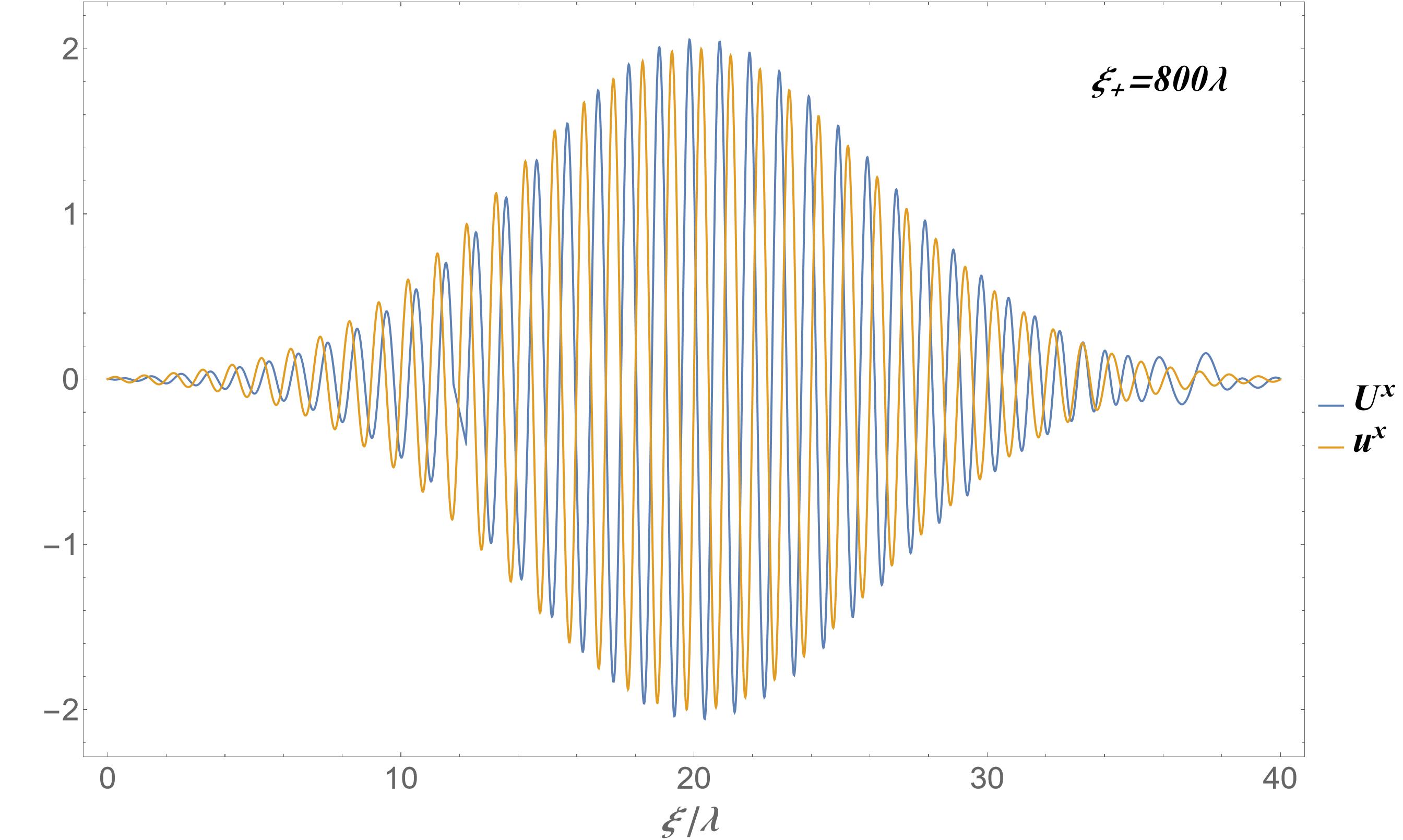}\\
\includegraphics[width=5.4cm]{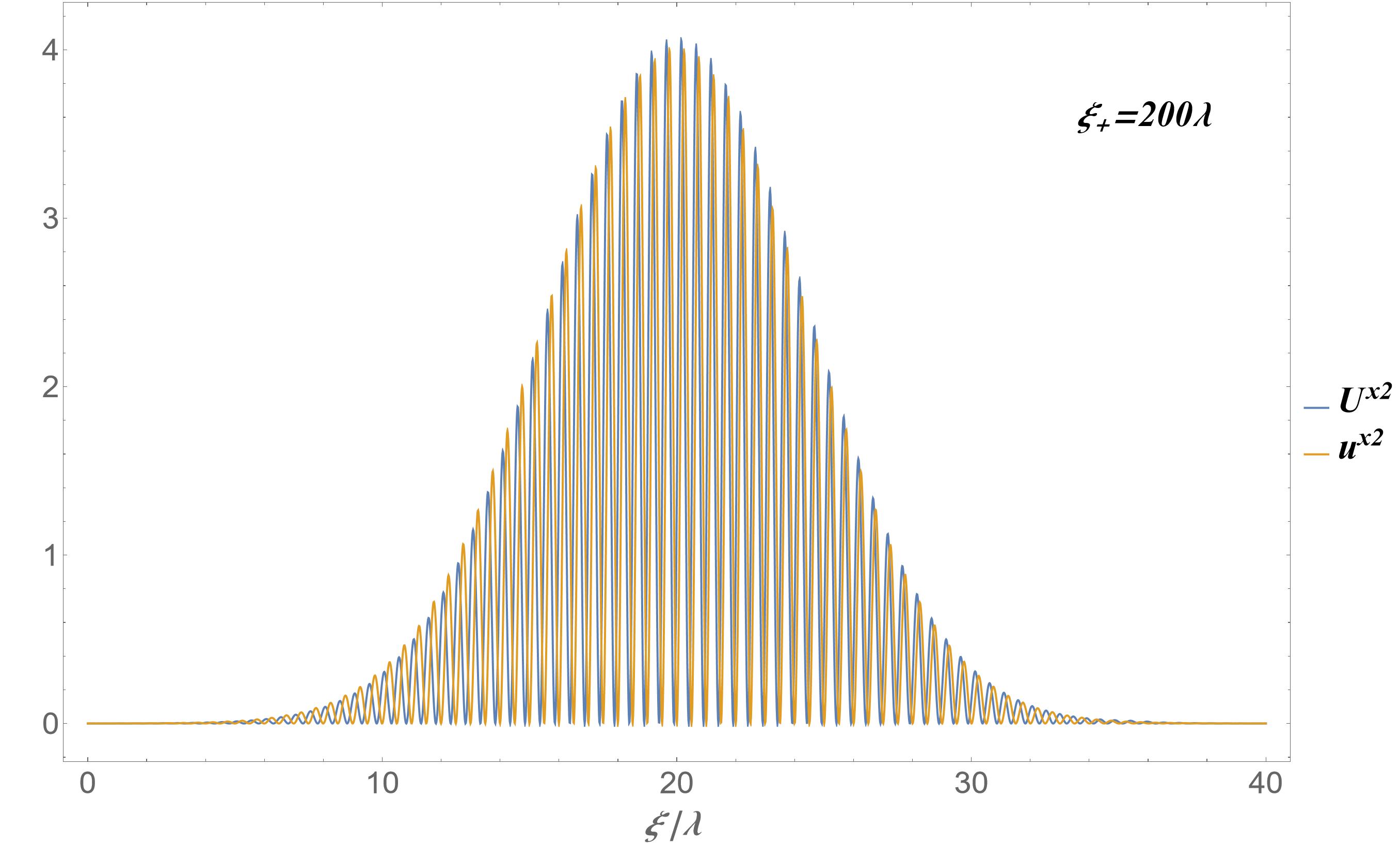} \includegraphics[width=5.4cm]{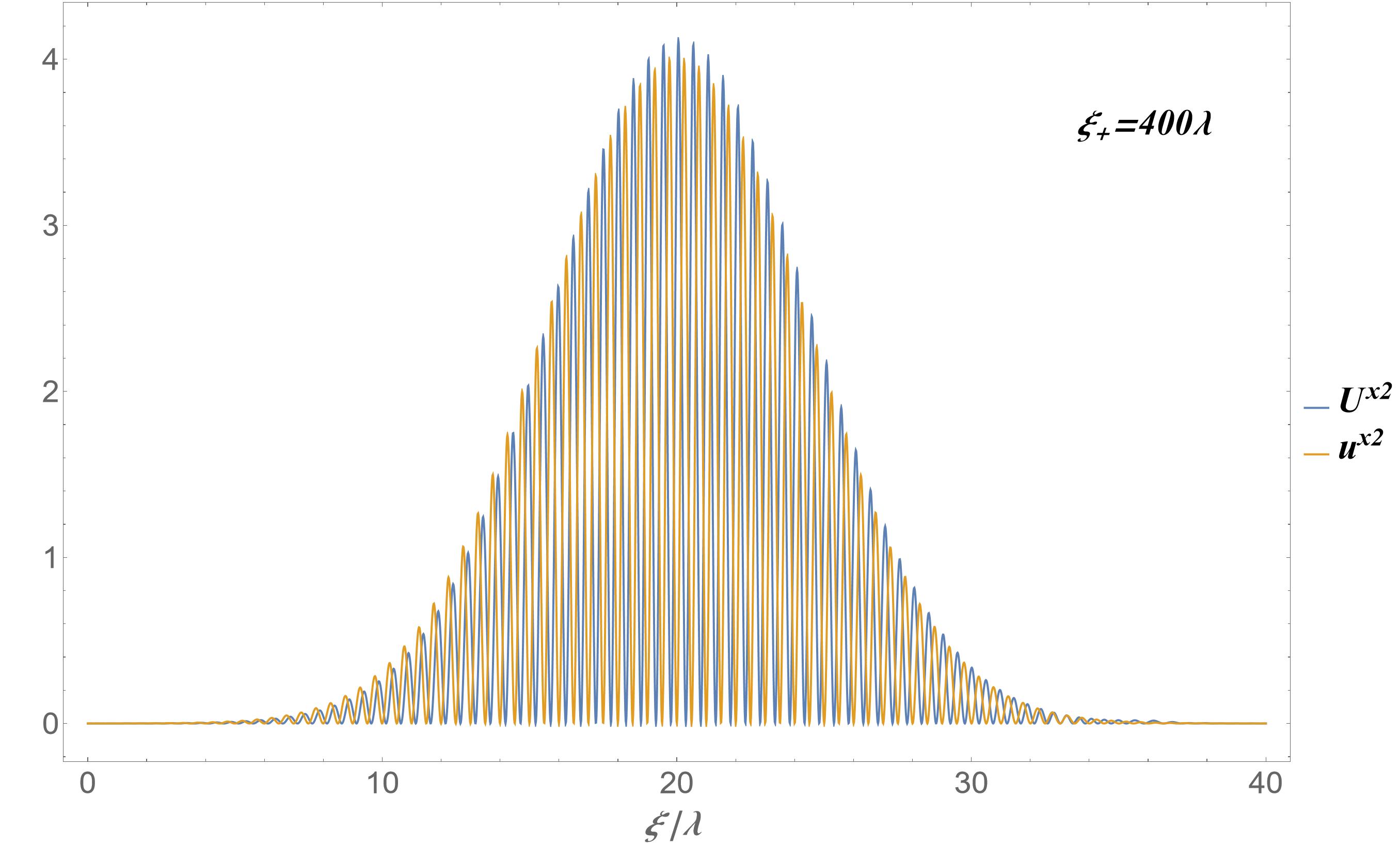} \includegraphics[width=5.4cm]{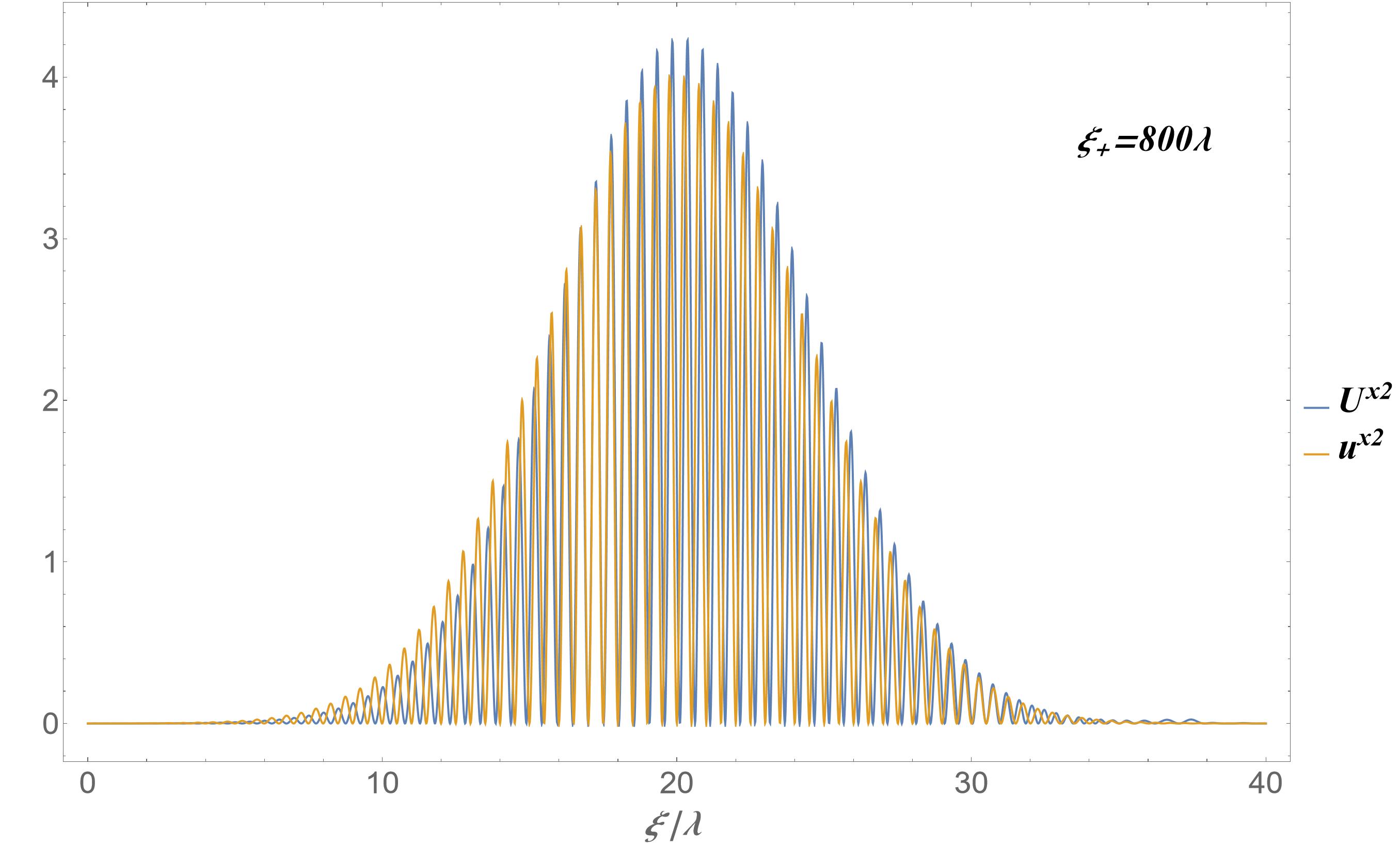}\\
\includegraphics[width=5.4cm]{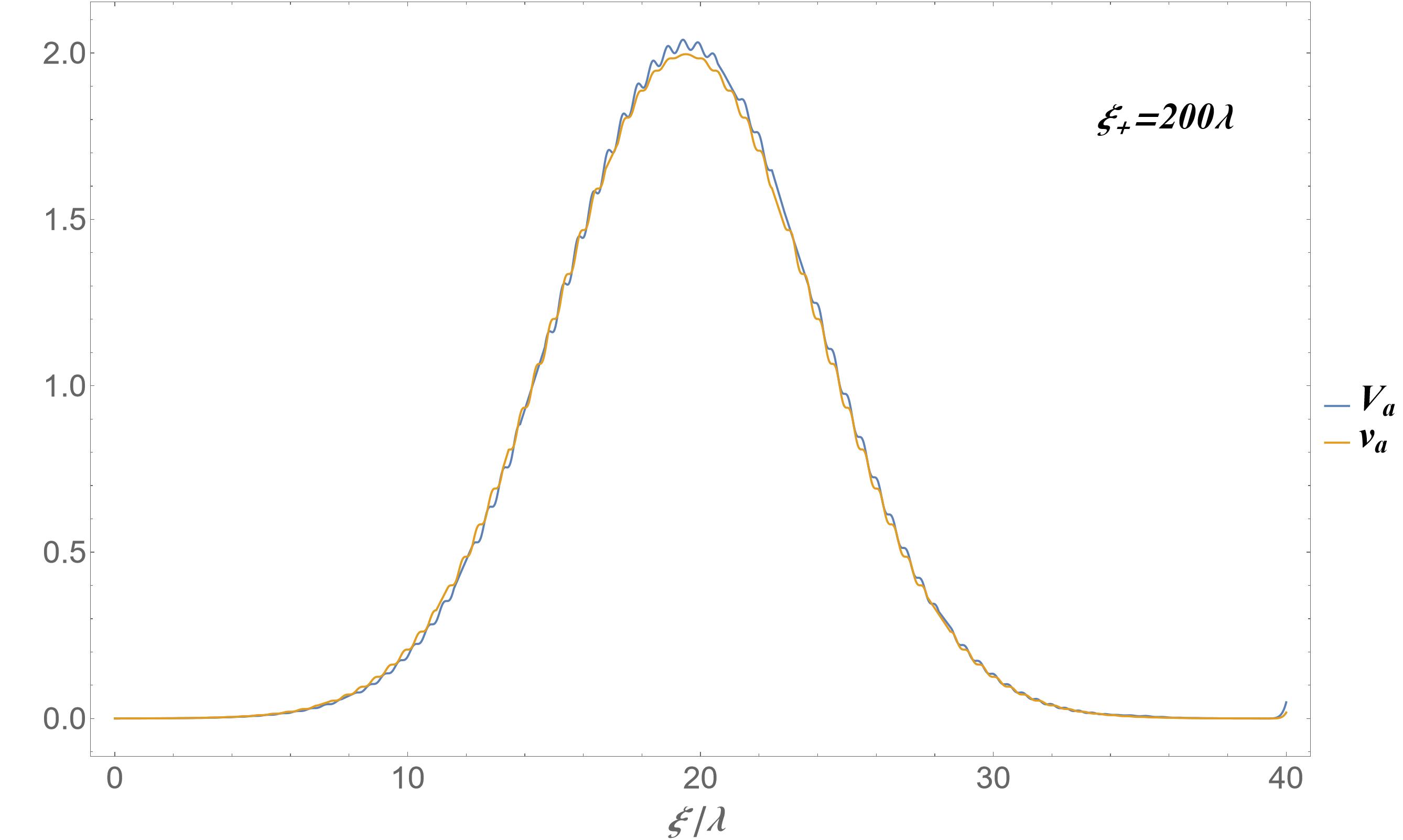} \includegraphics[width=5.4cm]{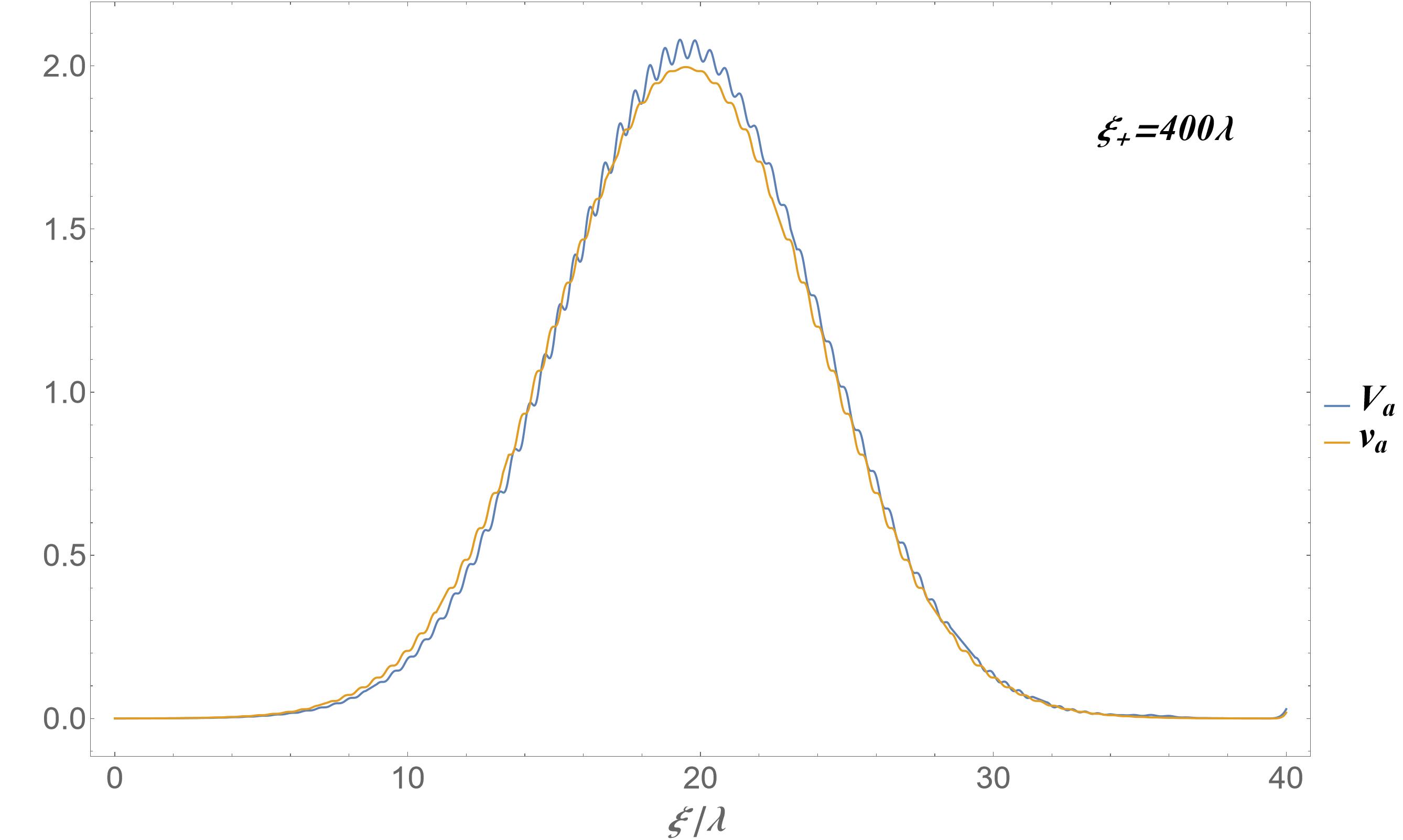} \includegraphics[width=5.4cm]{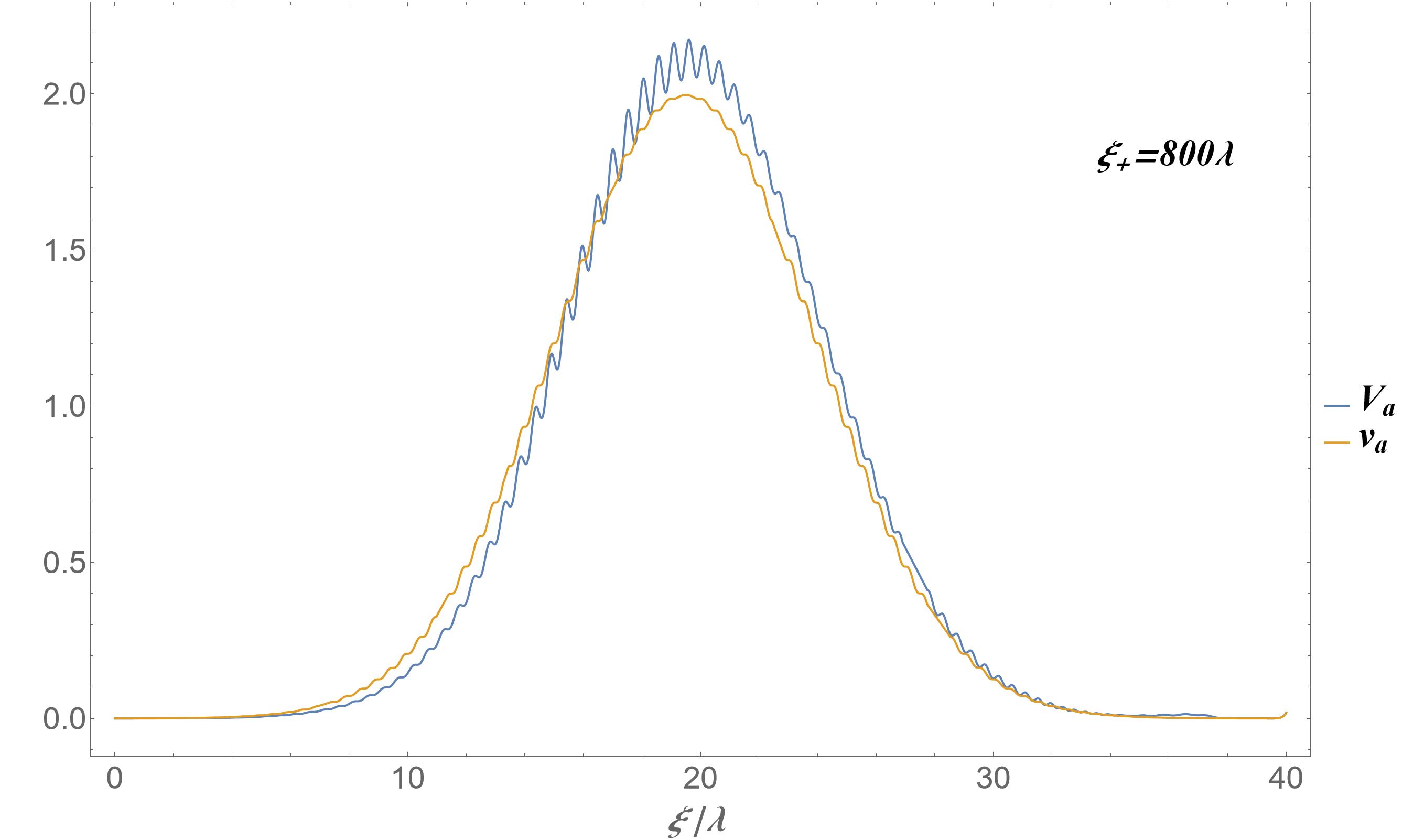}
\caption{Assuming the same input data as in fig. \ref{graphsb}, we plot: in the first row, $\cbUp(\xi,\xip)\equiv e\check \bAp(\xi,\xip)/mc^2$ and $\hbup(\xi)\equiv e\Bap(\xi)/mc^2$
vs. $\xi$ resp. for $\xip/\lambda=200,400,800$;  in the
second row, $V\equiv \cbUp{}^2(\xi,\xip)$ and $v\equiv \hbup{}^2(\xi)$ vs. $\xi$ 
resp. for $\xip/\lambda=200,400,800$; in  in the third row, the averages $V_a(\xi,\xip),v_a(\xi)$ vs. $\xi$  resp. for $\xip/\lambda=200,400,800$.}
\label{graphsUx_vs_ux_Ux2_vs_ux2_Va_vs_va}
\end{figure}

By (\ref{expln_e}), for 
a $\widetilde{n_0}$ of the type (\ref{n_0bounds}) the function $\check f$ at the rhs of Eq. (\ref{diffeq1p}) reduces to
\bea
\check f(\xi,\xip) \equiv \theta(\xi)
\theta(\xip)\frac{\pi e^2}{mc^2}\frac{\bar n}{s(\xi) }
\label{def_f}
\eea
if \ $z\equiv\frac {\xip\!-\!\xi}2> z_q\!\equiv\! z_s\!+\!\DM(\bar{n})$, \ i.e. on the plateau; here $s(\xi)$ is defined and computed as in section (\ref{specialcase}) by setting $\bA\!^{{\scriptscriptstyle\perp}}(t,z)=\Bap(ct\!-\!z) $. In  \cite{FioPostHydro} the solution of (\ref{diffeq1p}) with such a $\check f(\xi,\xip)$ (and ``initial conditions" $\bAp(t,z)=\Bap(ct\!-\!z)$ \ for $t\le 0$)
is found in closed form. For comparison, in fig.s \ref{graphsUx_vs_ux_Ux2_vs_ux2_Va_vs_va} we plot  $\cbUp(\xi,\xip)\equiv e\check \bAp(\xi,\xip)/mc^2,V\equiv \cbUp{}^2(\xi,\xip)$ and the average $V_a(\xi,\xip)$ (which accounts for the evolving ponderomotive potential) together with their unperturbed counterparts  
$\hbup(\xi)\equiv e\Bap(\xi)/mc^2,v\equiv \hbup{}^2(\xi),v_a(\xi)$,
vs. $\xi\in\mbox{supp}(\Bep)=[0,l]$ for few fixed values of $\xip$.
As one can see, for $\xip\le 800\lambda$ the change of $\bAp$ w.r.t. $\Bap$ (mainly a phase shift  of the carrier wave) does not  affect very significantly  $v_a$ (the average-over-a-cycle of $v$) and (by Remarks 2.1, 2.3) the formation of the PW;
in other words $|V_a\!-\!v_a|$, and thus the difference between the evolving and the unperturbed ponderomotive potentials, keep small. Hence for $z\lesssim \zM =400\lambda$ the first longitudinal oscillations of the plasma electrons - and thus the first PW buckets - induced by the passing of the unperturbed laser pulse are self-consistent.
For larger $\xip$ the size of $|V_a\!-\!v_a|$
begins to be more evident. 

\medskip
A good estimate of the total energy depletion of the pulse can be obtained as follows.
Denote by $\E(\bar z)$ the energy per unit transverse surface carried by the pulse across the plane $z=\bar z$. For $\bar z\le 0$, i.e.  before the impact with the plasma, this is
$$
\E(0)=\frac{1}{8\pi}\!\int^0_{-l}\!\!\!\!d\zeta\left[\bE^{{\scriptscriptstyle\perp}2}(0,\zeta)\!\!+\!\bB^{{\scriptscriptstyle\perp}2}(0,\zeta)\right]=\frac{1}{4\pi}\!\int^0_{-l}\!\!\!\!d\zeta\,\Bep{}^2(-\zeta)=\frac{1}{4\pi}\!\int_0^l\!\!\!\!d\xi\,\Bep{}^2(\xi).
$$
If  $z_s$ of (\ref{n_0bounds}) is so small (and $\E(0)$ so large) that
$\E(z_s)\!\simeq\!\E(0)$, we  can approximate $\widetilde{n_0}(z)=\bar{n}\,\theta(z)$. Since the energy gained by each plateau plasma electron overcome by the pulse is 
$mc^2\big[1-\bar h(\bar n)\big]$, by total energy conservation the variation of $\E( z)$
(pulse  depletion) is 
\be
\Delta \E(z)\equiv  \E(z)- \E(0)=mc^2\big[1-\bar h(\bar n)\big]\,\bar n\: z<0
\ee
for (not too large)  $z>z_s$.
For a SMMW with carrier wavelength $\lambda$ and Gaussian modulation with full width at half maximum (FWHM) $l'$  one easily finds\footnote{I.e. $\epsilon(\xi)\!=\! E^{{\scriptscriptstyle\perp}}_{{\scriptscriptstyle M}}\exp\!\left[-\frac{(\xi\!-\!l/2)^2}{2\sigma}\right]$, where  $E^{{\scriptscriptstyle\perp}}_{{\scriptscriptstyle M}}\equiv \max|\bEp|$,  $\sigma\!=\!\frac{l'{}^2}{4\ln2}$. Since  $l$ is several times $l'$ we can  approximate $ \int_0^l\!d\xi\Bep{}^2(\xi)\simeq\int_{-\infty}^{\infty}\!\!\!\!d\xi\Bep{}^2(\xi)$ and thus find (\ref{depletion});
$a_0\equiv \frac{e}{mc^2}\alpha^{{\scriptscriptstyle\perp}}_{{\scriptscriptstyle M}}\simeq E^{{\scriptscriptstyle\perp}}_{{\scriptscriptstyle M}}\frac{e\lambda}{2\pi mc^2}$ and $\alpha^{{\scriptscriptstyle\perp}}_{{\scriptscriptstyle M}}\equiv \max|\Bap|$.}
\bea
\frac{|\Delta \E(z)|}{\E(0)} &\simeq &\frac{z}{l'} \,
\frac{4 \sqrt{\ln 2}}{ a_0^2 \sqrt{\pi}}\,
\frac{\bar n}{n_{cr}}
\,\big[\bar h(\bar n)-1\big].     \label{depletion}
\eea
We can neglect the pulse depletion as long as \ $|\Delta \E(z)|/\E(0)\le\delta$, \ or  $z\le z_{dp}$; \ the size
of $\delta\!\in]0,1[$ expresses the degree of approximation we are ready to accept, $z_{dp}$ is defined by the eq. 
\be
|\Delta \E(z_{dp})|/\E(0)=\delta.          \label{defz_dp}
\ee
For a stepwise density $\widetilde{n_0}(z)=\bar n^j\, \theta(z)$ and a pulse as in fig. \ref{graphsb} one finds: $z_{dp}\simeq 1070\lambda$ for 
 $\delta=1/20$; $z_{dp}=\zM \simeq 400\lambda$ for 
 $\delta\simeq 0.02$.

\subsection{Localizing wave-breakings}
\label{WB-localization}

\begin{figure}
\includegraphics[width=16cm]{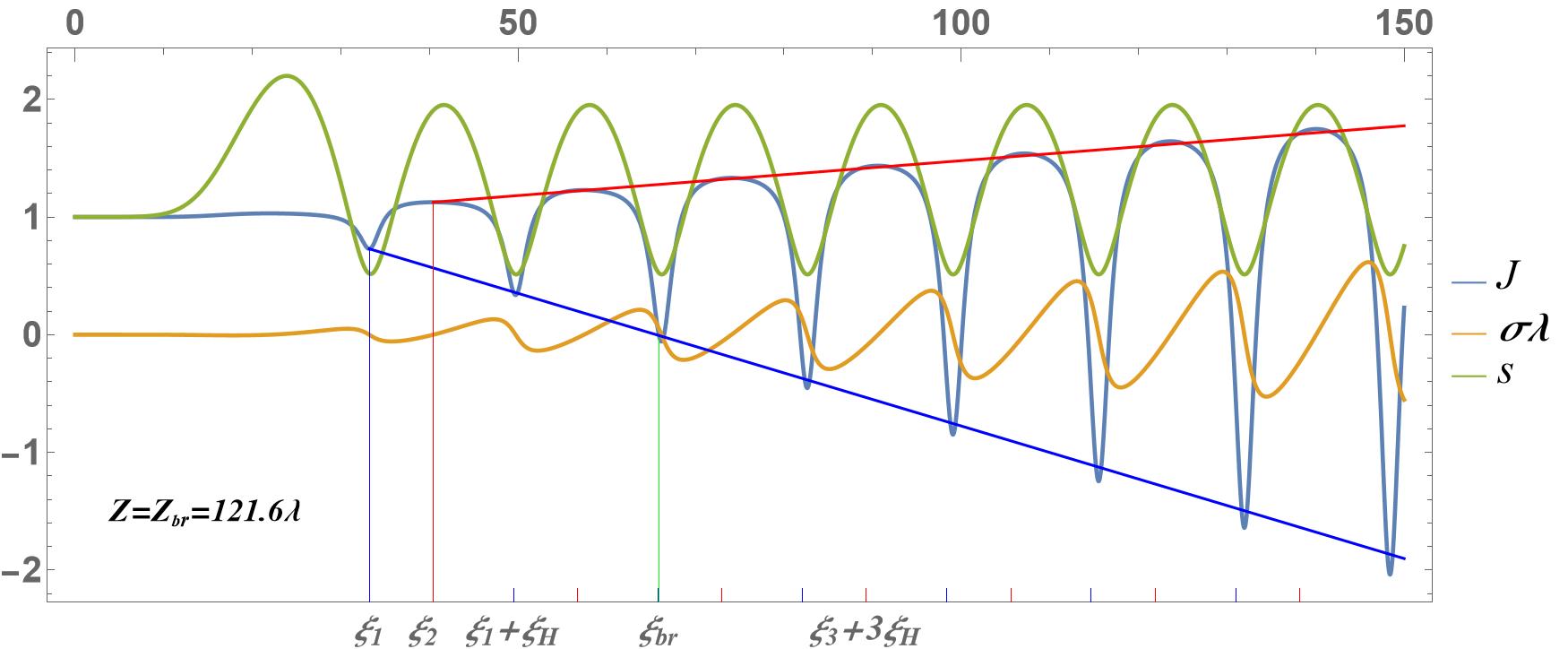} 
\caption{$\hat J,\hat\sigma\equiv \partial \hat s/\partial Z$  vs. $\xi$ for $Z=Z_{b}\simeq121.6\lambda$ and input data as in fig.  \ref{Worldlinescrossings'}.
}
\label{graphs2'}
\end{figure}

The periodicity identity \ $\hat z_e\!\left[\xi+k\,\xiH(Z),\!Z\right]\!=\!\hat z_e(\xi,\!Z)$ \ holds for 
$k\!\in\!\NN$, $\xi\!>\!l$ in the HR; differentiating it w.r.t. $Z$ and setting $\Phi \!\equiv\!
\partial\xiH/\partial Z
$  one  finds \cite{FioDeNAkhFedJov23}
\be
\hat J\!\left(\xi+k\xiH,Z\right)=\hat J(\xi,Z)-k\,\Phi (Z)\,\hat\Delta'\!\left(\xi,Z\right) ,
  \label{pseudoper}
\ee
so that (\ref{lin-pseudoper}) holds with   $b\equiv -\hat\Delta' \frac{\partial \log\xiH}{\partial Z}$, 
$a\equiv \hat J\!-\!\xi b$, see  fig. \ref{graphs2'}. Via (\ref{pseudoper}) we can extend our knowledge of  $ \hat J$ from $[l, l\!+\!\xiH[$ 
to all $\xi\ge l$ preceding the first WB. Unless $\widetilde{n_0}\!=$const, a WB at sufficiently large $\xi$ (and $t$) is unavoidable  (as well known \cite{Daw59}), because $\hat\Delta'$ changes sign twice in each period. Typically,  $\Phi(Z)$ is positive, negative if $\widetilde{n_0}(Z)$ decreases, grows at $Z$; this can be easily shown if $\widetilde{n_0}(Z)$ fulfills (\ref{SlowDensityVar}).
For all $Z$, let
 $\xi_{br}(Z)$ be the smallest $\xi\ge l$ if any,  such that $\hat J(\xi,Z)=0$ (with 
$\hat J$ as determined in the HR), and let $k(Z)\in\NN_0$ be the integer such that
$\xi_{br}(Z)\in[\xi_k^0(Z),\xi_{k+1}^0(Z)[$; if $\xi_{br}(Z)$ does not exist we set $k(Z)=\infty$.
By causality, outside the {\it future Cauchy developments} of all points $P\!_{br}(Z)\equiv \big(\xi_{br}(Z),\hat z_e[\xi_{br}(Z),Z]\big)$ the HD is still valid;
inside them the dynamical description has to be modified solving  (\ref{heq1}a-\ref{Newheq1b}) in place of (\ref{heq1}) and recomputing all the derived variables, including $\hat J$; in particular, (\ref{pseudoper}) will no longer hold.  The `first' (i.e. `earliest' w.r.t. `time' $\xi$) WB takes place at $P\!_{br}(Z_{b})$, where $Z_{b}$ minimizes $\xi_{br}(Z)$; we abbreviate 
$\xi_{b}\equiv\xi_{br}\big(Z_{b}\big)$, $k_{1}\equiv k\big(Z_{b}\big)$.
Assuming a pulse satisfying the SC, we can approximate $\hat J(\xi,Z)=1$ for all   $\xi\in[\xi_0^0(Z),\xi_1^0(Z)]$, and we can determine $k(Z)$ as the smallest integer such that
(\ref{pseudoper}) becomes $\le 0$ at some $\xi$ in such an interval.
On a density downramp $\mbox{sign}\big(\Phi(Z)\big)=1$, and the minimum 
is reached when $\hat P=P_1$, so that $\hat \Delta'$ reaches its maximum 
$(\1/\smm^2-1)/2=h/\sqrt{h^2\!-\!\1}- 1$,  and we find
\be
k(Z)\, \frac{\partial\xiH}{\partial Z}(Z)= \frac {h(Z)}{\sqrt{h^2(Z)\!-\!\1}}- 1;
\label{rate0}
\ee
then $\xi_{br}(Z)$ is determined as the smallest $\xi\in[\xi_k^0(Z),\xi_{k+1}^0(Z)[$
where (\ref{pseudoper})  vanishes; finally, 
$Z_{b}$ is determined as the minimum point for 
$\xi_{br}(Z)$,
 and $k_{1}$ by
$k_{1}=k\big(Z_{b}\big)$. If $k(Z)\gg 1$ it is $\xi_{br}(Z)\simeq \xi_k^1(Z)$,
whence $\hat P\big(\xi_{br},Z\big)\simeq P_1(Z)$.

\medskip
%
One finds that $\hat P(\xi,\!Z)\simeq \bar P\big[\xi;\!\widetilde{n_0}(Z)\big]$, where
 $\bar P(\xi; \bar{n})$ is the constant density solution  of section \ref{specialcase},  if
the relative variation of $\widetilde{n_0}$ over  $\I\!\equiv\![Z\!+\!\Dm(Z),Z\!+\!\DM(Z)]$ is small, i.e. if
\be
\left|1-\frac{\widetilde{n_0}(Z')}{\widetilde{n_0}(Z)}\right|\ll 1\qquad \mbox{for all $Z'\in\I(Z)$}.
\label{SlowDensityVar}
\ee
This implies also $\DM(Z)\simeq\oDM[\widetilde{n_0}(Z)\big]\simeq \big|\Dm(Z)\big|$, and replacing  $\DM(Z),-\Dm(Z)$ by $\oDM[\widetilde{n_0}(Z)\big]\!=\!\sqrt{2\{\bar{h}[\widetilde{n_0}(Z)\big]\!-\!1\}/M}$   in the definition of $\I$ (see section \ref{specialcase}) makes  (\ref{SlowDensityVar}) a condition that can be easily checked. 
If (\ref{SlowDensityVar}) is fulfilled, 
one can  check the no-WBDLPI sufficient conditions given in \cite{FioDeAFedGueJov22,FioDeNAkhFedJov23}  more easily replacing everywhere $\Delta_u(Z)$ by $\oDM[\widetilde{n_0}(Z)\big]$.

\section{WFA of injected electrons 
}
\label{WFA}

Solving (\ref{heq1}.a)-(\ref{Newheq1b}) with the IC (\ref{heq2}) one can in principle determine the motion of all plasma electrons (and thereby also phenomena like WB, longitudinal self-injection, WFA and beam loading)  under the above assumptions of plane symmetry, immobile ions, no 2-particle collisions, negligible effects (dephasing, etc) of the pulse evolution on the electrons' motion; the range of validity of the latter has been discussed at the end of section \ref{Back-reaction}. In next subsection we stick to study the motion of test electrons injected in the PW. In section \ref{WFA.c} we will argue that the results hold also for the first electrons self-injected electrons (FSIE) in the PW.

\subsection{WFA of injected electrons in the plane model}
\label{planeWFA}

If a test electron is injected  in the PW at $\xi_0\!>\!l$ with $\hbup_i(\xi_0)=0$,
then   its $\hat z_i,\hat s_i$ evolve after equations (\ref{reduced}), which become \cite{FioDeNAkhFedJov23} 
\bea
\hat z_i'= \displaystyle\frac {1\!-\!\hat s_i^2}{2\hat s_i^2}, \quad 
\hat s_i'(\xi)=K\!\left\{\!
\widetilde{N}\!\left[\hat z_i(\xi) \right] \!-\! \widetilde{N}\!\left[\hat Z_e\!\left(\xi,\!\hat z_i(\xi) \right) \right] \!\right\}\!.\qquad                 \label{heq-test}
\eea
By (\ref{sol"}), as $\widetilde{N}(z)=\widetilde{N}(z_s)\!+\!\bar{n} (z\!-\! z_s)$ for $z\ge z_s$,  (\ref{heq-test}b) along the density plateau  reduces to $\hat s_i'=M\Delta$, 
i.e. the same equation (\ref{e1}b) fulfilled by $s(\xi)\equiv s(\xi;\bar{n})$.  Imposing initial conditions 
$\big(\hat z_i,\hat s_i\big)_{\xi=\xi_0}\!=\!(z_{i0}, s_{i0})$, with $s_{i0}\!>\!0$,  $z_{i0}\!\ge\! z_q\!\equiv\! z_s\!+\!\oDM(\bar{n})$, then the solution of (\ref{heq-test}) is
\be
\hat s_i(\xi)
=\delta s+s(\xi),\quad\hat z_i(\xi)=z_{i0}+\!\!\int^\xi_{\xi_0}\!\!  \frac{dy}2\left[\frac {1}{\hat s_i^2(y)}\!-\!1\right],  \label{test-motion}
\ee
as long as $\hat z_i\!\ge\! z_q$.  Here 
 \ $\delta s\equiv s_{i0}\!-\!s(\xi_0)$. \ If 
\be
s_i^m\equiv \osmm\!+\!\delta s < 0, \label{trap}
\ee
then  $\hat s_i(\xi_f)=0$, $\hat s_i'(\xi_f)=s'(\xi_f)<0$ for some $\xi_f\!>\!\xi_0$ (corresponding to $t_f=\infty$):  the electron  remains trapped in the bucket of the PW located around $\xi=\xi_f$ while its velocity grows and goes to $c$. 
{\bf Inequality (\ref{trap})  is the trapping condition} \cite{FioDeNAkhFedJov23} that $\delta s$ must fulfill.
For $\xi\simeq \xi_f$ we have $\hat s_i(\xi)\simeq \! \left|s' (\xi_f)\right|(\xi_f\!-\!\xi)=\! M\left|\Delta (\xi_f)\right|(\xi_f\!-\!\xi)$, and in the limit $\xi\to \xi_f$
\be
\hat z_i(\xi)\, \simeq   \,
\frac {1}{2\left[ M\Delta(\xi_f)\right]^2 (\xi_f\!-\!\xi)}
\: \to 
\: \infty. 
\label{approx}
\ee
Meanwhile,  $c\hat t(\xi)-\hat z_i(\xi)-\xi_f=\xi-\xi_f\stackrel{\xi\to \xi_f}{-\!\!\!-\!\!\!\longrightarrow}0$, implying $\big(t,z_i(t)\big)-\big(t,ct\!-\!\xi_f\big)\stackrel{t\to  \infty}{-\!\!\!-\!\!\!\longrightarrow}(0,0)$: as $t\to  \infty$ the test electron tends to the travelling point of  the bucket having spacetime coordinates $(t,ct\!-\!\xi_f)$. On the latter the longitudinal electric field has the constant value $\bar E^z=s'(\xi_f)mc^2/e=4\pi e\bar{n}\Delta(\xi_f)$; these equalities are based on (\ref{reduced}b), (\ref{e1}).
In this simplified model trapped test electrons {\it cannot dephase},
as the phase velocity of the PW is $c$.
Solving (\ref{approx}) for $\xi_f\!-\!\xi$ we can express first $\hat s_i$, then $\hat \gamma_i$ in terms of $ z_i$:
\be
\gamma_i=\frac{1}{2 s_i}+\frac{s_i}2
\:\simeq \: F \: \frac{z_i}{\lambda}\: \stackrel{ z_i\to \infty}{-\!\!\!-\!\!\!-\!\!\!\longrightarrow} \: \infty, \label{s_i^m<0|s_iz_i}
\ee
$F\!\equiv\! M\lambda \left|\Delta(\xi_f)\right|$.
Hence, in this simplified model  {\bf the  energy grows proportionally to the travelled distance}. 
Of course, (\ref{s_i^m<0|s_iz_i}) is reliable in the interval $0\le z_i\le \zM$ (see section \ref{Back-reaction}) where the change of the ponderomotive force by the pulse 
can be neglected.

We point out that the onset of ion dynamics,  collisions and heating  become significant at any position $\bx$ after many plasma oscillations around $\bx$ induced by the passing of the pulse there; hence, by causality, they cannot significantly affect the trapped electrons, which go away from $\bx$ with almost the speed of light,  trailing the pulse.

Fixed  $z_i, \bar{n}$, the maximal rate $F$ of acceleration 
is obtained maximizing $|\bar E^z|=4\pi e|\Delta(\xi_f)|$; this is achieved if
$\xi_f=\bar\xi_k^0$ for some $k\in\NN_0$, so that
 $|\Delta(\xi_f)|$ takes  its maximum value $|\Dm|\!=\!\sqrt{2\big(\bar{h}\!-\!1\big)/M}$ (see section \ref{specialcase}). Correspondingly we obtain 
\be
\gamma_i(z_i; \bar{n}) \: \simeq \: \sqrt{ j(\nu)}\:\:  
\frac{z_i}{\lambda}, \qquad j(\nu)\equiv 8\pi^2\nu\big[\bar h(\nu)\!-\!1\big];         \label{gamma^M(z_i)}
\ee
here $\bar h(\nu)$ is the final electron energy transferred by the pulse if $\widetilde{n_0}(z)\!=\!\bar{n}$, as a function of $\nu\!\equiv\!\bar{n}/n_{cr}$. In this case $F(\nu)=\sqrt{ j(\nu)}= eE^z_{\scriptscriptstyle M}/mc^2$, where $E^z_{\scriptscriptstyle M}$   is  the maximal value reached by the longitudinal electric field (along the plateau of height $\bar{n}$) during its oscillations. By (\ref{test-motion}a), (\ref{Points_i}), \ $\xi_f=\bar\xi_k^0$ and  $\hat s_i(\xi_f)=0$ lead to $0=\hat s_i(\bar\xi_k^0)=\delta s+s(\bar\xi_k^0)=\delta s+\1$, i.e.
\be
\delta s\equiv s_{i0}\!-\!s(\xi_0)=-\1.
\label{PhaseCond}
\ee
This is the {\bf phase condition} \cite{FioAAC22} that $\delta s$ must fulfill in order
to maximize $F$ (and the WFA) for fixed $\bar{n}$.

\subsection{3D-effects}
\label{3Deffects}

%
Now assume that:
 i) the pulse is not a plane wave, but a Gaussian beam 
 with symmetry axis $\vec{z}$, same longitudinal phase and full-width-at-half-maximum $l'$ as in Fig. \ref{graphsb}.a,
waist $w_0$ located at $z=0$, so that 
the EM field does not significantly differ from the plane wave case considered so far
inside a cylinder $\C$ 
of axis $\vec{z}$, radius  $R\ll w_0$,
base at $z=0$, height $\zM \le z_{dp}$, $\zM \ll z_R\equiv \pi w_{0}^{2}/\lambda$ (the Rayleigh length), 
for $ct\le \zM + 4 l'$;
ii) $\widetilde{n_0}$ does not significantly differ from the plane case (in particular, does not depend on $x,y$) inside $\C$. 

 Then, by causality,  our results hold strictly inside the causal cone (of axis $\vec{z}$, base radius $R$) trailing the pulse, and approximately in a neighbourhood thereof. In particular, if the pulse $\Bep(\xi)$ has maximum
at $\xi=l/2=2l'$, and
\be
w_0\:\gg\: R\:>\: \xi_{b}-2l',\qquad R\:\gg\:
 |\Delta\bx^{{\scriptscriptstyle \perp}}_{e{\scriptscriptstyle M}}| \simeq
\frac{a_0\lambda}{2\pi}\left[\bar h\!+\!\sqrt{\bar h^2\!-\!1}\right]
\label{Rcond}
\ee
($\simeq$ holds by formula (50) in \cite{FioDeNAkhFedJov23}; these inequalities exclude the {\it bubble regime} \cite{RosBreKat91,MorAnt96,PukMey2002,LuEtAl07
}), then
 the $\bX\simeq(0,\!0,\!Z_{b})$ electrons  are injected in the PW as above and trail that cone with the same maximal WFA for  $z\le \zM $, i.e. as far as the pulse does not diverge nor is depleted significantly. 

\bigskip
Now we can formulate our multi-steps optimization procedure.

\section{LWFA optimization procedure}
\label{optimizeLWFA}

\begin{figure}
\includegraphics[width=16.5cm]{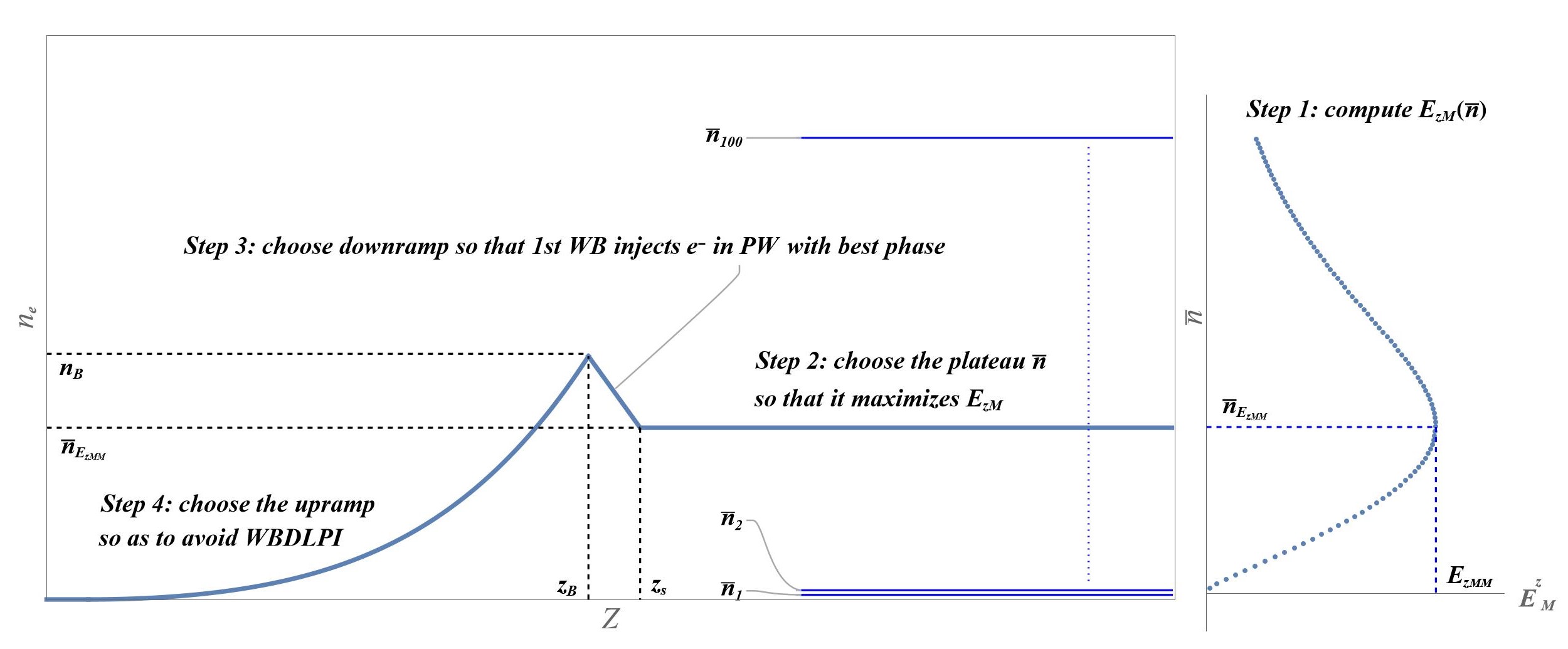} 
\caption{First 4 steps in the determination of the 3rd plasma density studied in section \ref{Applic}}
\label{4Steps}
\end{figure}

Henceforth we focus on the class of  $\widetilde{n_0}(z)$'s
having a maximum $n_B$ at some $z=z_B$, decreasing in an interval $\I_d\equiv[z_B,z_s]$ (downramp) with $z_s\ll z_{dp}$, cf. (\ref{depletion}), and equal to some $\bar{n}<n_B$ for $z>z_s$ (plateau), see 
fig. \ref{fig1}.a. We also assume that $\widetilde{n_0}$ fulfills  (\ref{SlowDensityVar})  for $Z$ in an interval  $\I'$ such that $\I'\cap\I_d\neq\emptyset$, so that (\ref{rate0}) holds, as well as  the SC. In the previous sections we have described how to solve a {\it direct problem}, i.e. given the initial electron (and proton) density $\widetilde{n_0}(z)$ and laser pulse $\Bep$, how to 
determine the motion of the plasma electrons.
Now, fixed $\Bep$, we wish to {\it tailor} $\widetilde{n_0}$  so that the first WB injects a bunch of $Z\in\I_d$ electron layers
in the PW with the right phase and the best plateau level in order to maximize their WFA along the density plateau.
We solve this {\it inverse problem}  by  4 steps; the latter entail some approximations used to make the inversion formulae not too complicated. The effectiveness of the $\widetilde{n_0}$ resulting from the inversion formulae can  be checked, and then further improved by fine-tuning (5th step), solving again the direct problem. In fig. \ref{4Steps} we illustrate the first 4 steps for a particular density.

\subsection{Step 1: Computing  \texorpdfstring{$\bar h(\bar n)$, $j(\bar n)$, $\bxiH(\bar n)$, $\bar\xi_k^i(\bar n)$}{h(n), j(n), xiH(n), xi	extasciicircum{}i	extsubscript{k}(n)} for the given pulse}
\label{WFA.a}

\begin{figure}[b]
\includegraphics[width=16.5cm]{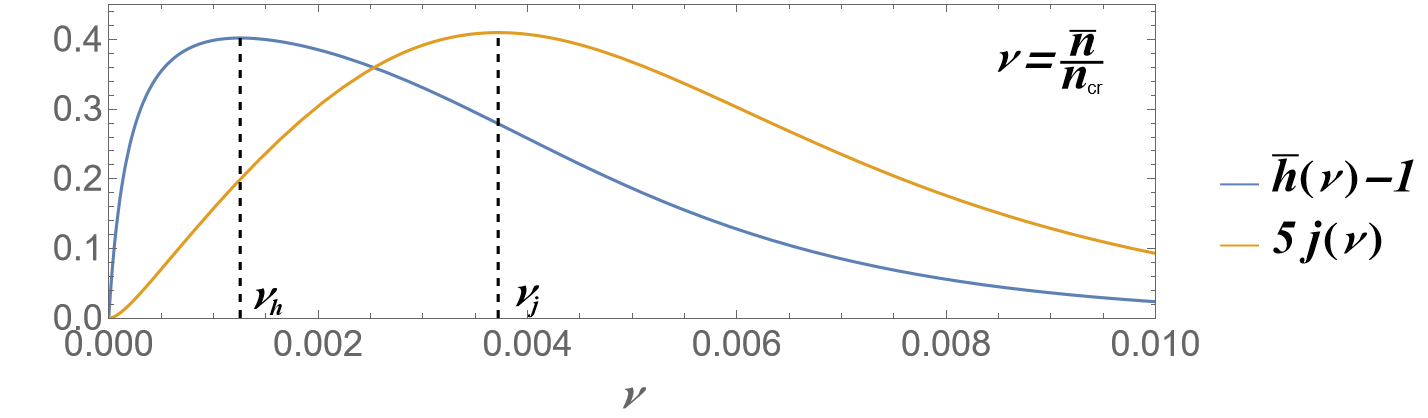} 
\caption{The energy gain per electron $\bar h\!-\!1$ and $j$ vs. the  
density $\bar{n}$.}
\label{graph_bh-j_vs_nu}
\end{figure}

This is done solving (\ref{e1}) for few hundreds values of $\bar n$, computing the corresponding values of $\bar h,j,\bxiH,\bar\xi_k^i$ via (\ref{defh}), (\ref{gamma^M(z_i)}), (\ref{period}), (\ref{Def_xi^0_k}-\ref{xilapses}) and interpolating the corresponding listplots. Fixed the laser pulse, doing these operations takes few seconds through our {\it Mathematica} code. 
In fig. \ref{graph_bh-j_vs_nu} we plot  $\bar h(\nu),j(\nu)$ and their 
maxima  $\nu_h,\nu_j$ for the pulse of fig. \ref{graphsb}.a.

\subsection{Step 2: Optimal choice for the plateau density \texorpdfstring{$\bar{n}$}{}}
\label{WFA.b}

If the plasma longitudinal length $z_i$ available for WFA fulfills $z_i\le \znd(\nu_j)$
 [see (\ref{defz_dp})], by (\ref{gamma^M(z_i)}) the choice maximizing the energy growth for injected electrons along the plateau is $\nu=\nu_j$:
\be
\gammaM_i(z_i)\simeq\sqrt{j(\nu_j)}\, z_i/\lambda.   \label{gamma^MM(z_i)}
\ee

\subsection{Step 3: Optimal linear down-ramp \texorpdfstring{$\widetilde{n_0}$}{} for self-injection, LWFA}
\label{WFA.c}

Let $(\xi_{b},Z_{b})$ be the pair $(\xi,\!Z)$ with $Z\!\in\!\I_d$ and the smallest $\xi$ 
such that  $\hat J(\xi,\!Z)\!=\!0$.  Since $d\xiH/dZ>0$ along the downramp, we find $\hat\Delta'(\xi_{b},\!Z_{b})>0$ by (\ref{pseudoper}), and 
$P_{b}\equiv\hat P(\xi_{b},\!Z_{b})$ belongs to the lower part $P_0P_2$ of the cycle of the $Z_{b}$ electrons. This means that 
\ $\xi^0_{k_{1}}(Z_{b})<\xi_{b}<\xi^2_{k_{1}}(Z_{b})$, \
with the $k_{1}\in\NN$ introduced in section \ref{WB-localization}. For technical simplicity
we shall approximate
\be
\xi_{b}\simeq\xi^1_{k_{1}}(Z_{b}).
\label{approx_xib}
\ee
This is justified at least
if we can tailor $\widetilde{n_0}$  so that $k_{1}\gg 1$; in fact, in each cycle in the $z-s$ plane $\hat J(\xi,Z_{b})$ reaches its minimum approximately when $\hat P(\xi,Z_{b})=P_1$, and, if $k_{1}\gg 1$, in the $k_{1}$-cycle it will be $\hat J(\xi,Z_{b})\le 0$ only  in a small interval around $\xi^1_{k_{1}}(Z_{b})$.
 For $\xi\!>\!\xi_{b}$ a bunch of $Z\!\le\! Z_{b}$ electron layers start to overshoot the ones lying ahead; thereby they break the part of the PW where they were located and are injected in a part ahead.
The $Z_{b}$  layer is the earliest to do, i.e. what so far we have called the FSIE. At each  $\xi\!>\!\xi_{b}$, at least if $\xi\!-\!\xi_{b}$ is small, it crosses a new electron layer ahead that up to $\xi$ has evolved via (\ref{heq1}) and contributed to the (locally still unbroken) PW; hence it moves as a layer of  test particles, and its motion 
$\hat P(\xi,\!Z_{b})$ fulfills (\ref{heq-test}).
Fixed some $z_{i0}\!\ge\!z_q$, let $\xi_0\!>\!\xi_{b}$ be the `instant'  (if any) when $\hat z_e(\xi_0,\!Z_{b})\!=\! z_{i0}$. 
For $\xi\ge\xi_0$, at least if  $\xi\!-\!\xi_{b}$ is small,
$(\hat z_i,\hat s_i)\!\equiv\!\big(\hat z_e(\cdot,\!Z_{b}),\hat s(\cdot,\!Z_{b})\big)$ is given by
(\ref{test-motion}).
In fact, this holds for all $\xi\ge\xi_0$ if the trapping condition (\ref{trap}) is fulfilled, as we shall assume: then the $Z_{b}$ layer overshoots each $Z>Z_{b}$
layer at some $\xi(Z_{b},Z)>\xi_{b}$, and it does for ever, because the 
repulsive forces  push the $Z_{b}$  layer forward and $Z$ one backward. Thus the $Z_{b}$ electrons  become the fastest 
ones  trapped in a bucket of the PW, after being injected there by the first WB.  
In order to satisfy (\ref{PhaseCond}),
$\hat P(\xi_0,z_q)=\bar P(\xi_0;\bar{n})$ must belong to the upper part $\bar P_2\bar P_0$ of the cycle, where $s(\xi_0)>\1$.
Since $\xiH(Z_b)<\bar \xi_{{\scriptscriptstyle H}}(\bar{n})$,    at $\xi=\xi_{b}$ the $Z_{b}$  electrons have completed a number $k_{1}> k_2\equiv$ number of cycles completed by the plateau electrons. We require $k_2=k_{1}-1$ and 
$\bar P(\xi_0;\bar{n})\in \bar P_3\bar P_0$ in order to make $\xi_0\!-\!\xi_{b},z_s\!-\!Z_{b}$  be as small as possible.

In the HR, for all $Z>0$ the system (\ref{heq1}), and hence also its solution, depend on the behavior of $\widetilde{n_0}(z)$ only in the interval $Z+\Dm(Z)\le z\le Z+\DM(Z)$\footnote{In fact, the rhs (\ref{heq1}b) can be written in the form $K\int^{Z+\hat\Delta}_Z d\zeta\,\widetilde{n_0}(\zeta)$, and $\hat\Delta$ changes in the range $[\Dm,\DM]$.}.  This implies that the location $(\xi_{b},Z_{b})$ of the first WB, the injection and the WFA of the $Z_b$ electrons depend on the behaviour of $\widetilde{n_0}(z)$ only for $z\ge Z_{b}+\Dm(Z_{b})$; hence  the beginning $z_B$  of the downramp has to fulfill $z_B\!\le\!Z_{b}\!+\!\Dm\!(Z_{b})$. To minimize the beam-loading we will choose $z_B\!=\!Z_{b}\!+\!\Dm\!(Z_{b})$ and
the maximum in $z_B$ sufficiently sharp from the left.
Assuming that  $\widetilde{n_0}(z)$ is continuous, differentiable in $]z_B,z_s[$, 
 and fulfills (\ref{SlowDensityVar}), 
the first requirement approximately becomes
\be
z_B =Z_{b}+\oDm(n_b), \qquad\mbox{with} \quad \oDm(n_b)=-\sqrt{2\big[\bar{h}(n_b)\!-\!1\big]/Kn_b}.
\ee
Neglecting the back-reaction of the plasma on $\bAp$, a change $\widetilde{n_0}(Z)\mapsto \widetilde{n_{0a}}(Z)\equiv\widetilde{n_0}(Z\!-\!a)$  changes the dynamics for $Z\ge a$ only by a shift  of all longitudinal coordinates  $z,Z$ by $a$ and of the `time' $\xi$ by $a/c$.
Hence the features of WB and WFA depend on $z_s,Z_{b}$ only through their difference
$\delta Z\!\equiv\!z_s\!-\!Z_{b}$.
Abbreviating \   $n_{b}\!\equiv\!\widetilde{n_0}(Z_{b})$, $\Upsilon\!\equiv\!\widetilde{n_0}'(Z_{b})$, \ and applying the Taylor formula   we find 
\be
\widetilde{n_0}(z)=n_b+\Upsilon\, (z-Z_{b}), 
\qquad z_B\le z\le z_s \label{Downramp}
\ee
at order 
$O[(z\!-\!Z_{b})^2]$. 
If the downramp profile is linear, the equality is exact. 
Formula (\ref{Downramp})
is determined by parameters $n_{b},\Upsilon$; alternatively,  we can (and will) adopt $n_{b},\delta Z$ as parameters, because by (\ref{Downramp})
$\Upsilon\!=\!(\bar{n}\!-\!n_{b})/\delta Z\!=\!(\bar{n}\!-\!n_B)/\delta Z_{dr}$ where $\delta Z_{dr}\equiv z_s\!-\!z_B$. We will bound their range requiring that:
 there is no WBDLPI;
$\bar P(\xi_0)$ belongs to the arch $\bar P_3\bar P_0$ (the green part of the cycle of fig. \ref{graphsb}.c), i.e.
 at $\xi\!=\!\xi_0$ the $Z_{b}$ electrons cross plateau ones having negative displacement $\Delta$ and velocity $\Delta'$.

\begin{prop} 
$\bar P(\xi_0)$ belongs to \ $\bar P_3\bar P_0$ \ if  $\xi_0\!-\!\xi_{b}\ll\bxiH$ and, abbreviating $r(\bar{n})\equiv\tfrac{\bar h(\bar{n})}{\sqrt{\bar h^2(\bar{n})-\1}}-1$,
\bea
 \frac 14 \bxiH(n_{b})-\DM(n_{b})
\: \le \: \delta Z \,r(n_{b}) \: \le \:  \frac12 \bxiH(\bar{n}) \:.
\label{deltaZbounds}
\eea
\label{prop1}
\end{prop}

\bp{}
The requirement that  $\bar P(\xi_0)\in \bar P_3\bar P_0$  amounts to the existence of a $k_2\in\NN$ such that 
\be
\bar\xi^3_{k_2}(\bar{n})<\xi_0<\bar\xi^0_{k_2+1}(\bar{n}).
\label{ineq1}
\ee
As explained above, we require   $k_2=k_{1}\!-\!1$, with the $k_{1}\in\NN$ introduced in section \ref{planemodel}, to make $\xi_0\!-\!\xi_{b}$ as small as possible; this
is compatible with the inequalities
 $
\xi^0_{k_{1}}(Z_{b})<\xi_{b}<\xi_0<\xi^0_{k_2}(z_q)
$, since $\xiH(Z_{b})<\xiH(z_q)$,  $\uxi^i(Z_{b})<\uxi^i(z_q)$  for $i=1,2,3,4$. 
Inequalities (\ref{ineq1}) can be equivalently reformulated as
\be
\Xi':=\bar\xi^0_{k_{1}}-\xi_0>0,\qquad\qquad\Xi:=\xi_0-\bar\xi^3_{k_2}
=\ouxi^4-\Xi'>0
\label{ineq1b}
\ee
(the last equality holds by (\ref{Def_xi^0_k}-\ref{xilapses})), and approximately as (\ref{deltaZbounds}), using the approximations
\bea
\Xi'\simeq r(n_{b})\,\delta Z\!-\! \frac14 \bxiH(n_{b})\!-\!\Dm(n_{b}),\qquad
\Xi\simeq\frac12 \bxiH(\bar{n})\!-\! \delta Z\,r(n_{b}). \label{approxXiXi'}
\eea
The latter can be proved
using: i)  $\xi_0\simeq \xi_{b}\simeq \xi^1_{k_{1}}(Z_{b})$, which holds by (\ref{approx_xib}) if  $\xi_0\!-\!\xi_{b}\ll\bxiH$;  ii)  $ \bar\xi^0_{k_{1}}(\bar{n})=\bar\xi^0_0(\bar{n})\!+\!k_{1}\bxiH(\bar{n})$,  $\xi^1_{k_{1}}(Z_{b})=\xi^0_0(Z_{b})\!+\!k_{1}\xiH(Z_{b})\!+\!\uxi^1(Z_{b})\simeq \bar\xi^0_0(\bar{n})\!+\!k_{1}\xiH(Z_{b})\!+\!\uxi^1(Z_{b})$, which hold by (\ref{Def_xi^0_k}) and 
the approximate equality \  $\xi^0_0(Z)\simeq \bar\xi^0_0(\bar{n})
$; iii) the approximate equalities \  $h(Z)\simeq \bar h[\widetilde{n_0}(Z)]$, $\uxi^i(Z)\simeq\ouxi^i[\widetilde{n_0}(Z)]$, whence in particular $h(Z_{b})\simeq \bar h(n_{b})$ and $\uxi^i(Z_{b})\simeq\ouxi^i(n_{b})$:
\bea
\Xi'  &\simeq& \bar\xi^0_{k_{1}}(\bar{n})\!-\!\xi^1_{k_{1}}\!(Z_{b})\stackrel{ (\ref{Def_xi^0_k})}{\simeq} k_{1}  \big[\bxiH(\bar{n}) \!-\! 
\xiH(Z_{b})\big]\!-\!\uxi^1(Z_{b})\simeq k_{1} \frac{d\xiH}{dZ}(Z_{b})\, \delta Z\!-\!\ouxi^1(n_{b})\nn[6pt]
  &\stackrel{ (\ref{rate0})}{\simeq}& \left(\!\frac{h(Z_{b})}{\sqrt{h^2(Z_{b})\!-\!\1}}-1\!\right)\delta Z\!-\!\ouxi^1(n_{b}) \simeq 
r(n_b)\,\delta Z\!-\!\ouxi^1(n_{b})
\stackrel{ (\ref{xilapses}a)}{=}\mbox{rhs(\ref{approxXiXi'}a)},
\nonumber
\eea
while $\Xi\!\stackrel{ (\ref{xilapses}d)}{=}\!\tfrac14 \bxiH(\bar{n})\!-\!\Dm(\bar{n})\!-\!\Xi'\!\simeq$rhs(\ref{approxXiXi'}b),
approximating $\bxiH(n_{b})\simeq \bxiH(\bar{n})$, $\Dm(n_{b})\simeq \Dm(\bar{n})$. 
\ep

By the further approximation $\bar h(n_{b})\simeq\bar h(\bar{n})$,
inequalities (\ref{deltaZbounds}) take the less accurate form
$$
 \frac 14 \bxiH(\bar{n})-\DM(\bar{n})
\: \le \: \delta Z \,r(\bar{n}) \: \le \:  \frac12 \bxiH(\bar{n}) \:,
\eqno{(\ref{deltaZbounds}')}
$$
which determines the range $\D$ of $\delta Z$ independently of the unknown $n_{b}$. 
Since $\bar h(\bar{n})>\bar h(n_{b})$, 
we find  $r(\bar{n})<r(n_{b})$, and the left (resp. right) inequality implies (resp. is implied by) the left (resp. right) inequality  (\ref{deltaZbounds}). From the practical viewpoint
one may start exploring $\D$ around $\delta Z\simeq \big[\frac 14 \bxiH(\bar{n})\!-\!\DM(\bar{n})\big]/r(\bar{n})$.

Inequalities (\ref{deltaZbounds}), (\ref{deltaZbounds}') are slightly different from  inequalities (33) of \cite{FioAAC22}, which had been obtained by slightly different approximations.

\begin{prop}
Under the same assumptions of Proposition \ref{prop1}, we can approximately compute $\delta s\equiv s_{i0} - s(\xi_0)$  \ as a function of \ $n_{b},\delta Z$ \ 
using the approximate relations
\bea
\ba{l}
 s(\xi_0)=\sM\!-\!\frac{M\Xi^2}4\! \left(\!1\!-\!\frac {\1}{\sM^2}\!\right),
\quad\sM\!=\!\bar h(\bar{n})\!+\!\sqrt{\bar h^2(\bar{n})\!-\!1},
\quad\Xi=\frac12 \bxiH(\bar{n})\!-\! \delta Z\,r(\bar{n}),\\[6pt]
s_{i0}=C-\sqrt{C^2\!-\!1}, \quad 
C\!=\! \big[\delta Z\!+\!\oDM(\bar{n})\big]Kn_{b}\zeta\!+\!\bar h(n_{b}), \quad
 \zeta=\frac{\Xi}2\!\left(\!1\!-\! \frac {M\Xi^2}{3\sM^3}\!\right)\!\left(\!1\!-\!\frac {1}{\sM^2}\!\right).
\quad 
\ea 
\label{Calcolo_delta_s}
\eea
\label{prop2}
\end{prop}

\bp{}
A good, simple approximation of $s(\xi)$ in the interval $[\bar \xi^3_{k_2},\bar \xi^0_{k_{1}}[$ is  its Taylor expansion at 2nd order around $\bar \xi^3_{k_2}(\bar{n})=
\bar \xi^0_{k_{1}}(\bar{n})\!-\!\ouxi^4(\bar{n})$,
because both $s',s'''$ vanish there:
\bea
s(\xi)\simeq \osM+\frac{\delta\xi^2}2\,
s''\!\left[\bar\xi^3_j(\bar{n})\right]= \osM+\frac{M\delta\xi^2}4\,
\left(\frac {\1}{\osM^2}-1\right);
\eea
here  $\delta\xi\equiv \xi\!-\!\bar\xi^3_{k_2}(\bar{n})$.
For $\xi=\xi_0$  it  is $\delta\xi=\Xi$, and we get $s(\xi_0)$ as in eq. (\ref{Calcolo_delta_s}).

In order to find a first, manageable approximation in closed form of the solution of (\ref{heq-test}) in the short interval $[\xi_{b},\xi_0]$ we approximate: i) $\widetilde{n_0}= n_{b}$ at the rhs(\ref{heq-test}b) [(\ref{Downramp}) would yield a better but more complicated approximation], whence $ \widetilde{N}(z)=\widetilde{N}(Z_b)+n_{b}(z\!-\!Z_b)$,  and 
$$
\hat s_i'(\xi) =- K n_{b}\:\zeta(\xi), \qquad
\zeta(\xi):=\hat Z_e\left[\xi,\hat z_i(\xi)\right]\!-\!\hat z_i(\xi).
\eqno{(\ref{heq-test}b')}
$$
ii) $\zeta(\xi)$ by its value at $\xi=\xi_0\simeq \xi^1_{k_{1}}(Z_{b})$\footnote{We can check the validity of such an approximation a posteriori, solving the direct problem; it can be heuristically justified if $\delta Z$ is several times $\oDM$, because then the variation  for  $\xi\in[\xi_{b},\xi_0]$ of the rhs(\ref{heq-test}b), and hence of $\zeta(\xi)$, is small, since the electric charges of the overshot ions, electrons in $[\xi_{b},\xi_0]$ approximately coincide.}. Choosing $\hat z_i(\xi_0)= z_q$, $\hat z_i(\xi)$ takes values $z$ in the range where $\hat Z_e(\xi_0,z)-z=-\Delta(\xi_0; \bar{n})$ [see (\ref{sol"})]; hence,
$\zeta(\xi)\simeq \zeta(\xi_0)=-\Delta(\xi_0; \bar{n})$. To compute the latter we use
the Taylor expansion of $\Delta(\xi; \bar{n})$ at the third order around  
$\bar\xi^3_{k_2}(\bar{n})$, 
\bea
\Delta(\xi; \bar{n})\simeq\Delta'\big[\bar\xi^3_{k_2}(\bar{n}); \bar{n}\big] \delta\xi+\Delta'''\big[\bar\xi^3_{k_2}(\bar{n}); \bar{n}\big]\frac{\delta\xi^3}6   = 
\left(\!\frac {\1}{\osM^2}-1\!\right)\frac{\delta\xi}2\left[1- \1 M\frac {\delta\xi^2}{3\osM^3}\right], \label{interm} 
\eea
where  we have used   (\ref{e1}), (\ref{Points_i}) to evaluate $\Delta$ and its derivatives 
at $\bar\xi^3_{k_2}(\bar{n})$ (in particular, $\Delta,\Delta''$ vanish there)\footnote{In fact, differentiating (\ref{e1}a)  and using (\ref{e1}b) we obtain 
$$
\Delta''=-\1\frac {s'}{s^3}=-\1 M\frac {\Delta}{s^3}, \qquad\Delta'''=\1 M\left[\frac {3s'\Delta}{s^4}- \frac {\Delta'}{s^3}\right]=\1 M\left[\frac {3M\Delta^2}{s^4}- \frac {1}{2s^3}\left(\frac {\1}{s^2}\!-\!1\right)\right].
$$
By (\ref{Points_i}), at $\xi=\bar\xi^3_{k_2}(\bar{n})$ it is $\Delta=0$; hence also $\Delta''$ and the 1st term in the square bracket vanish, whence (\ref{interm}).}. For $\xi\!=\!\xi_0$ we get $\delta\xi\!=\!\Xi$ and the last eq. in (\ref{Calcolo_delta_s}). Eq. (\ref{heq-test}$b'$) becomes
$$
\hat s_i'(\xi) =- B, \qquad
B:=K n_{b}\zeta\simeq K n_{b}\left(\!1-\frac {\1}{\osM^2}\!\right)\frac{\Xi}2\left[1- 
\1 M\frac {\Xi^2}{3\osM^3}\right],
\eqno{(\ref{heq-test}b'')}
$$
which 
is solved by $\hat s_i(\xi) =s_{ib}- B(\xi-\xi_{b})$. Correspondingly, (\ref{heq-test}a)
is solved by
\bea
\hat z_i(\xi) -z_{ib}=  \frac{\1}{2B}\left[\frac 1{\hat s_i(\xi)}\!-\!\frac 1{s_{ib}}\right]- \frac{\xi \!-\! \xi_{b}}2=\frac{\xi \!-\! \xi_{b}}2
\left[\frac {\1}{s_{ib}\hat s_i(\xi)}\!-\!1\right].
\label{soltest'} 
\eea
Here and below we abbreviate  $(z_{ib},s_{ib})\equiv \big(\hat z_e(\xi_{b},Z_{b}),\hat s(\xi_{b},Z_{b})\big)$, $\delta z_i\equiv z_{i0}- z_{ib}$, $\gamma_{ib}\equiv (s_{ib}\!+\!\1/s_{ib})/2$.
Approximating $P(\xi_{b}, Z_{b})\simeq P_1(Z_{b})$ yields $z_{ib}\simeq Z_{b}$ and, since $z_{i0}=z_q$, 
\be
\delta z_i\simeq z_q\!-\!Z_{b}=\delta Z\!+\!\oDM(\bar{n}),        \label{Tempz} 
\ee
as well as 
$\U\simeq 0$ and in turn $\gamma_{ib}\simeq h\big(Z_{b}\big)\simeq \bar h(n_{b})$.
Evaluating $\hat s_i,\hat z_i$ at $\xi=\xi_0$ and getting rid of $\xi_0\!-\! \xi_{b}$ we obtain
$s_{i0}\!-\!s_{ib}=B(\xi_{b}\!-\!\xi_0)$ and
\bea
\delta z_i=\frac{\xi_0 \!-\! \xi_{b}}2
\!\left[\frac {\1}{s_{ib}s_{i0}}\!-\!1\right]=
\frac{s_{ib} \!-\! s_{i0}}{2Bs_{i0}}\!\left[\frac {\1}{s_{ib}}\!-\!s_{i0}\right]
=\frac{1}{2B}\!\left[s_{i0}\!+\!\frac {\1}{s_{i0}}\!-\!2\gamma_{ib}\right]
\simeq\frac{1}{2B}\!\left[s_{i0}\!+\!\frac {\1}{s_{i0}}\!-\!2\bar h(n_{b})\right]
\nonumber
\eea
Setting $C\equiv B\delta z_i\!+\!\bar h(n_{b})$, this amounts to the  second degree equation $s_{i0}^2\!-\!2Cs_{i0}\!+\!\1 =0$ in the unknown $s_{i0}$,  which 
is solved by
\bea
s_{i0}=C-\sqrt{C^2-\1}
\label{inter3} 
\eea
(the solution $C\!+\!\sqrt{C^2\!-\!\1}$ must be discarded because it is $>1$).
Replacing (\ref{heq-test}b''),  (\ref{Tempz}) in the definition of $C$, we find that eq.
(\ref{inter3}) gives $s_{i0}$ with $C,\zeta$ as defined in  formula (\ref{Calcolo_delta_s}).
\ep


The assumption $\xi_0-\xi_{b}\ll\bxiH$ made in Proposition \ref{prop1} has to be checked a posteriori; we can show\footnote{In fact, the function $f(x)\equiv x-\sqrt{x^2\!-\!1}$ is convex and decreases for $x\ge 1$, hence  it satisfies 
\be
f(x)-f(y)<(y-x)\, |f'(x)|=(y-x)\left[\frac{x}{\sqrt{x^2\!-\!1}}-1\right]= (y-x)\, r(x)
\label{convex}
\ee
for $y\!>\!x\!\ge\! 1$. Eqs. $s_{i0}\!=\!f(C)$, $s_{ib}\!=\!f(\gamma_{ib})\!\simeq\! f[\bar h(n _{b})]
$,   $s_{i0}\!-\!s_{ib}=B(\xi_{b}\!-\!\xi_0)$, 
 (\ref{convex}), (\ref{deltaZbounds}) imply, as claimed,
\bea
\xi_0\!-\!\xi_{b}=\frac{s_{ib}\!-\!s_{i0}}B=\frac{f\big[\bar h(n _{b})\big]-f(C)}B< \delta z_i\,  r(n_{b})\simeq \big[\delta Z\!+\!\oDM(\bar{n})\big]\,  r(n_{b})\stackrel{(\ref{deltaZbounds}b)}{\le} \frac 12 \bxiH\nn
\!+\!\oDM(\bar{n})\,  r(n_{b}) 
=\tfrac 12 \bxiH\!+\!\sqrt{\!\frac{2\,[\bar h(\bar{n})\!-\!\1]}{M\big[\bar h^2(n_{b})\!-\!\1\big]}}f\big[\bar h(n_{b})\big]\simeq \tfrac 12 \bxiH+\frac{f\big[\bar h(n_{b})\big]}{\sqrt{M}}<\frac 12 \bxiH+\frac{f\big[\bar h(\bar{n})\big]}{\sqrt{M}}.
\eea
} at least that it is compatible with Proposition \ref{prop2}, which implies  \ $\xi_0-\xi_{b}<\tfrac 12 \bxiH+\frac{f[\bar h(\bar{n})]}{\sqrt{M}}$.

\medskip
Let $\Omega$ be  the set of  $(n_{b},\delta Z)$ respecting the bounds (\ref{deltaZbounds}).
The third step consists in determining pairs $(n_{b},\delta Z)\in\Omega$ of the  downramp parameters leading to $\delta s=-\1$, so that  (\ref{gamma^M(z_i)}) applies. In general these pairs make up a whole line $\Lambda\subset\Omega$; changing the
point $(n_{b},\delta Z)\in \Lambda$ one can change $k_{1}$. Under our assumptions, the latter  is determined by the formula 
\be
k_{1}\equiv k(Z_b)= \left[K_{1}\right],\qquad\qquad
K_{1}\equiv \frac{\delta Z\:\: r(n_{b})}{\bxiH(\bar{n})-\bxiH(n_{b})},
\label{rate0inv}
\ee
(here $[a]$ stands for the integer part of $a\ge 0$), which is obtained inverting eq. (\ref{rate0}) and approximating $\frac{d\xiH}{dZ}(Z_b)\simeq\frac{\xiH(Z_b)-\bxiH(\bar{n})}{Z_b-z_s}\simeq\frac{\bxiH(\bar{n})-\bxiH(n_{b})}{\delta Z}$. Numerically, $\Lambda$ can be approximately found adopting a mesh in $\Omega$,  computing the corresponding values of $\delta s$ by formulae (\ref{Calcolo_delta_s}) and choosing the points of the mesh leading to  $\delta s$ closest to  $-\1$.

\begin{figure}[ht]
\includegraphics[width=15cm]{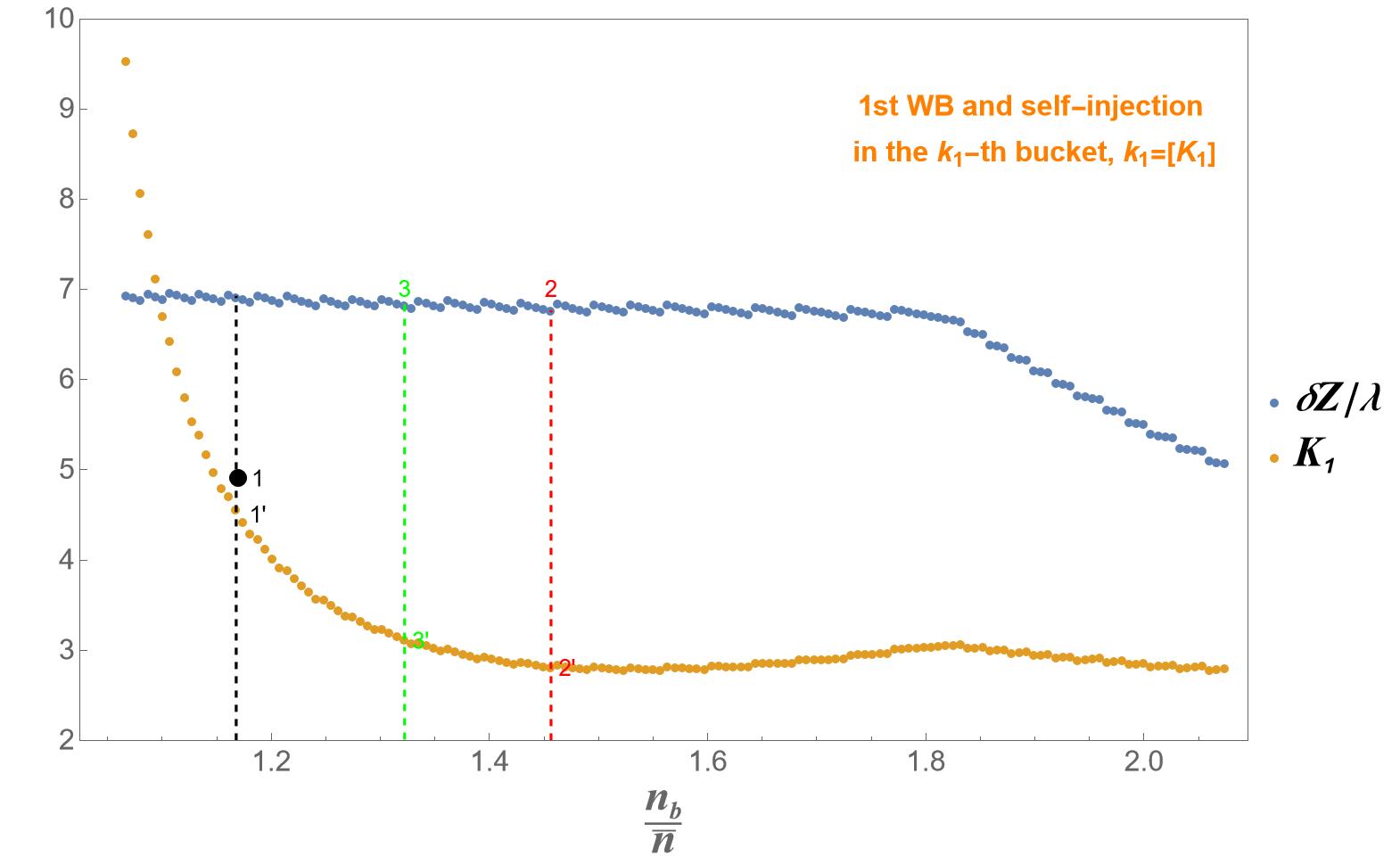}
\caption{The listplot 
of the line $\Lambda$ in the rescaled plane $(n_{b}/\bar n,\delta Z/\lambda)$ is plot  blue; its intersections with the  red, green vertical dashed line give the $(n_{b}/\bar n,\delta Z/\lambda)$ of the  $2^{nd}$ and $3^{rd}$ density profile in section \ref{Applic}. In the same frame we also plot yellow the corresponding listplot 
of $K_1$ vs. $\delta Z/\lambda$; its intersections with the red, green vertical dashed line give the $(K_1,\delta Z/\lambda)$ of the same $2^{nd}$ and $3^{rd}$ density profiles. We have plot the  $(n_{b}/\bar n,\delta Z/\lambda)$ of the $1^{st}$ density profile  of section \ref{Applic} [eq. (\ref{1stInitialDensity}), considered also in \cite{FioAAC22}] as a black thick dot; it lies outside $\Lambda$.}
\label{LambdaPlot}       
\end{figure}

In fig. \ref{LambdaPlot} we plot blue the points of $\Lambda$ [more precisely,
the rescaled ones $(n_{b}/\bar n,\delta Z/\Lambda)$] associated to such a mesh, and the corresponding values of $K_{1}$, for the pulse of fig. \ref{graphsb}.a. An intriguing quasi-periodic structure appears; however, the high $n_b$ part of the plot cannot be considered reliable, because  the assumption of a slowly varying $\widetilde{n_0}(z)$ is no longer valid.

\subsection{Step 4: Choosing the \texorpdfstring{$\widetilde{n_0}$}{}  up-ramp so as to prevent earlier WB}
\label{WFA.d}

We pick one of the $\infty$-ly many \cite{FioDeAFedGueJov22,FioDeNAkhFedJov23} 
$\widetilde{n_0}(z)$ growing from 0 to $n_B$ 
in an (as short as possible)  interval $0\!\le\!z\!\le\!z_B$  and preventing not only WBDLPI, but also WB for $\xi\!<\!\xi_{b}$; to that end a behaviour $\widetilde{n_0}(z)\simeq O(z^2)$  for small $z$ (which is compatible e.g. with the transverse density profile just outside the nozzle of a supersonic gas jet) helps. In our model the specific choice does not affect the first WB at the downramp and associated WFA of the self-injected electrons.


\subsection{Step 5: Checking and  fine-tuning \texorpdfstring{$\widetilde{n_0}$}{} to make it more effective.}
\label{WFA.e}

Adopting the  determined $\widetilde{n_0}$ one finds the motion of the wake-field accelerated electrons solving eq. (\ref{heq-test}) 
numerically in $[\xi_0,\xi_f[$  (direct problem); this reduces to the simpler equations (\ref{heq1}-\ref{heq2}), (\ref{test-motion}) resp. in $[0,\xi_{b}]$, $[\xi_0,\xi_f[$. One finds in particular $Z_{b},\xi_{b},k_{1},F,...$ and can thus check how much they deviate from the values   predicted using the approximations of our procedure to solve the inverse problem. By solving again the direct problem after small variations of  $\widetilde{n_0}$ around the first one can fine-tune the latter to further improve its effectiveness.

Finally, one checks the results via 1D PIC simulations with the determined input data.

\smallskip
Fixed any other plateau density $\bar{n}$, we can apply just steps 1,3,4,5 to 
maximize the WFA of the electrons injected in the PW by the first WB with that plateau level 
 $\bar{n}$.

\medskip

Let us conclude this section noting that the above procedure could be a tool for machine learning and development of a comprehensive AI based model to maximize early electron beam acceleration via laser-plasma interaction with required parameters.

\section{Sample applications of our procedure}
\label{Applic}


As examples, we apply our above multi-steps procedure  to tailor the plasma density   to  the laser pulse of fig. \ref{graphsb}.a, as done in \cite{FioAAC22}; more precisely, we consider the following three profiles, with $\widetilde{n_0}(z)=0$ for $z\le 0$, the same plateau\footnote{The slightly different value $\bar{n}\!=\!n_{cr}/267$ found in \cite{FioAAC22} was due to a coarser mesh. We also point out a slight change of notation: the subscripts $b,br$ of  \cite{FioAAC22} are resp. replaced here  by $B,b$, e.g. $z_b,Z_{br}$ become $z_B,Z_{b}$.} $\bar{n}\!=\!n_{cr}/268.8$ obtained via Step 2, three different
pairs $(n_B,\delta Z_{dr})$ [and correspondingly three different pairs $(n_{b},\delta Z)$]  obtained via Step 3, see   fig. \ref{LambdaPlot}.  
Then, to do a first consistency check, we solve the direct problem [i.e. eqs. (\ref{heq1}-\ref{heq2} of our plane model] with these input data.

\bigskip
{\bf $1^{st}$ Initial density profile.}
We fix the beginning of the downramp  at $z_B=120\lambda $ (as in \cite{BraEtAl08}). We choose the pair $(n_{b},\delta Z)=(1.17 \:\bar{n},4.9\lambda)$ (point 1 of fig. \ref{LambdaPlot})  as done in  \cite{FioAAC22}
\footnote{Actually, this is not in the set $\Lambda$, but has been obtained by a refinement from the point $(n_{b},\delta Z)=(1.17\bar{n},6.9\lambda)\in\Lambda$. The latter is the intersection of the blue listplot of fig. \ref{LambdaPlot} with the black  vertical dashed line, while the intersection of the yellow listplot with the same line shows the corresponding $k_1=[K_1]=4$. Assuming a linear downramp this yields $\delta Z_{dr}=\delta Z\!+\!\oDM(n_b)=8.5\lambda$. Solving the corresponding direct problem gives $Z_b\!-\!z_B=3.6\lambda
$, which is larger than the wished value $\oDM(n_b)=1.6\lambda$. For this reason, in the $1^{st}$ initial density we have reduced the length of the downramp to reduce $k_1$ and make $Z_b-z_B\simeq \oDM(n_b)$, while keeping the same slope. This also reduces $\xi_b\!-\!\xi_0$, improving the validity of the approximations made in section \ref{WFA.c}}. Assuming for simplicity the linear downramp (\ref{Downramp})  for $z\in[z_B,z_s]$, and determining the corresponding plasma electron motion (i.e., solving the corresponding direct problem) as described in sections \ref{Setup}, \ref{WFA}, we find $k_1=[K_1]=3$,  $\delta Z_{dr}=\delta Z\!+\!\oDM(n_b)=6.5\lambda$ and $n_B=\bar{n}17/14=1.21 \:\bar{n}$.
We choose the upramp in one of the  $\infty$-ly many ways mentioned in Step 4. Abbreviating $g_1(Z)\equiv (0.005 Z/\lambda)^2+(0.015 Z/\lambda)^4$, the adopted $\nu\equiv\widetilde{n_0}/n_{cr}$  reads
\be
\nu(Z)=\left\{\ba{ll}
\displaystyle \frac {1}{268.8}\frac {17}{14}\:\frac {1+g_1(120\lambda)}{g_1(120\lambda)}
\:\frac {g_1(Z)}{1+g_1(Z)}
\qquad & 0\le Z\le 120\lambda \:\mbox{ (upramp),}\\[16pt]
\displaystyle \frac {1}{268.8}\left[1+\frac {3}{14}\frac {126.5\lambda-Z}{6.5\lambda}\right]\qquad & 120\lambda\le Z\le 126.5\lambda  \:\mbox{ (downramp),}\\[16pt]
\displaystyle \frac {1}{268.8} \qquad & Z\ge 126.5\lambda  \:\mbox{ (plateau),}
\ea \right.
\label{1stInitialDensity} 
\ee
as in \cite{FioAAC22}. \
In fig. \ref{Worldlinescrossings'} we depict it and
the electron WLs found solving the corresponding direct problem (i.e. (\ref{heq1}-\ref{heq2}), first for $Z=0,\lambda,...,156\lambda$, then again  for  $Z=120\lambda,120.1\lambda,...,140\lambda$, to obtain a better resolution in the downramp region). The first WB
(rotated magenta box in fig. \ref{Worldlinescrossings'}.c) occurs at $\xi_{b}= 65.98
\lambda$ and involves the $Z\!\sim\! Z_{b}=121.6\lambda$ down-ramp electrons; this implies $k_{1}=3$, and also $\delta Z=4.9\lambda$, $n_{b}=1.17\bar{n}$, 
in very good agreement with the values determined by Step 3. Moreover 
$\xi_0=67.34\lambda$ (confirming that $\xi_0\!-\!\xi_{b}\ll\bxiH(\bar{n})\simeq 19.83
\lambda$) and  $\xi_f=67.90\lambda$.
The $Z\!=\!Z_{b}$ electrons  nearly fulfill (\ref{gamma^M(z_i)})
(with $\delta s\!=\!-1.01$) and  (\ref{gamma^MM(z_i)}), hence 
have almost the largest possible WFA factor  (\ref{s_i^m<0|s_iz_i}),  $F\!=\!\sqrt{ j(\nu_j)}\!=\!0.286$ (this is $1.36$ times the highest one found in \cite{FioDeNAkhFedJov23} out of the three input data considered by \cite{BraEtAl08}). 
As wished, the earliest WB (w.r.t. `time' $\xi$) involves the electrons in the downramp and therefore is causally disconnected from the  WBs (encircled in fig. \ref{Worldlinescrossings'}.b) involving up-ramp electrons.  In the figure
we have plot the WL of the $Z_{b}$ electrons in black. 
We find $|\Delta \E(z)|/\E(0)\simeq z/6111 \lambda$ and, choosing $\delta=1/10$ in (\ref{depletion}), that the pulse depletion can be neglected for $z\le z_{dp}\simeq 611 \lambda$. 
If $\lambda=0.8\mu$m,  $F=0.28$ leads to the remarkable energy gain 
$0.357 mc^2\simeq 0.182$MeV per $\mu$m, and $z_{dp}\simeq 490\mu$m.

\bigskip
{\bf $2^{nd}$ Initial density profile.}
Next: i) We reduce $z_B$ 
to $z_B=60\lambda$ to reduce the depletion and deformation of the pulse before the WB at the downramp. ii) We choose  $(n_{b},\delta Z)=(1.46 \:\bar{n},6.75\lambda)\in \Lambda$
 (point 2, blue listplot of fig. \ref{LambdaPlot}), so that  $k_1=[K_1]=2$ (see point  2', yellow listplot in fig. \ref{LambdaPlot}), i.e. $k_1, \xi_{b}$ are reduced as well, and (\ref{Rcond}) can be fulfilled with a smaller $R$; correspondingly, we find $\delta Z_{dr}=\delta Z\!+\!\oDM(n_b)=7.97\lambda$ and $n_B=1.54 \: \bar{n} $, 
if we assume for simplicity again the linear downramp (\ref{Downramp})  for $z\in[z_B,z_s]$. Moreover, we choose a purely convex up-ramp in one of the  $\infty$-ly many ways mentioned in Step 4.
Abbreviating $g_2(Z)\equiv (0.005 Z/\lambda)^2+(0.01 Z/\lambda)^4$, 
 the adopted $\nu\equiv\widetilde{n_0}/n_{cr}$  reads
\be
\nu(Z)=\left\{\ba{ll}
\displaystyle \frac {1}{174.5}\:\frac {g_2(Z)}{g_2(60\lambda)}
\qquad & 0\le Z\le 60\lambda \:\mbox{ (upramp),}\\[16pt]
\displaystyle \frac {1}{174.5}\left[1+\frac {94.3}{268.8}\frac {60\lambda-Z}{8\lambda}\right]\qquad & 60\lambda\le Z\le 68\lambda  \:\mbox{ (downramp),}\\[16pt]
\displaystyle \frac {1}{268.8} \qquad & Z\ge 68\lambda  \:\mbox{ (plateau).}
\ea \right.
\label{2ndInitialDensity}
\ee
Solving the corresponding direct problem (\ref{heq1}-\ref{heq2})  for $Z=0,\lambda,..., 96\lambda$ we have determined the  corresponding  electron WLs and verified that the earliest WB (w.r.t. `time' $\xi$) involves the electrons in the downramp; this is confirmed by solving also eq. (73) in \cite {FioDeNAkhFedJov23} and eq. (\ref{pseudoper}) for the Jacobian $\hat J$.  \ Then we have solved the direct problem for the down-ramp $Z$ electrons more in detail, determining their  WLs for  $Z=60\lambda,60.1\lambda,...,80\lambda$. In particular,
we have found  $(Z_b,\xi_b)\simeq (63.2\lambda, 47.86\lambda)$, $k_1=2$, and the acceleration rate $F=0.285$, in very good agreement with the values determined by Step 3. Moreover 
$\xi_0=49.59\lambda$ whence $\xi_0\!-\!\xi_{b}\ll\bxiH(\bar{n})\simeq 19.83
\lambda$ (confirming that applying Propositions \ref{prop1}, \ref{prop2} is justified) and  $\xi_f=50.18\lambda$.
The $Z\!=\!Z_{b}$ electrons  nearly fulfill (\ref{gamma^M(z_i)})
(with $\delta s\!=\!-0.94$) and  (\ref{gamma^MM(z_i)}), hence 
have almost the largest possible WFA factor  (\ref{s_i^m<0|s_iz_i}).


\bigskip
{\bf $3^{rd}$ Initial density profile.}
In the previous case the application of Step 3 has given $Z_b\!-\!z_B=3.2\lambda$, which is larger than the wished value $\oDM(n_b)=1.22\lambda$. As an example of fine-tuning, we reduce the length of the downramp to make $Z_b-z_B\simeq \oDM(n_b)$ while keeping the same slope $\Upsilon\simeq 0.0677 {\bar n}/\lambda$; this also reduces $\xi_b\!-\!\xi_0$ and thus further improves the validity of the approximations made in section \ref{WFA.c} (cf. Prop. \ref{prop1}, \ref{prop1}):
We pick now $z_B=60\lambda$, the pair $(n_{b},\delta Z)=(1.32\,\bar{n},4.8\lambda)$,
 but the smaller $\delta Z_{dr}=\delta Z\!+\!\oDM(n_b)=6.2\lambda$, whence $n_B=1.42\,\bar{n}$, if we assume for simplicity the linear downramp (\ref{Downramp})  for $z\in[z_B,z_s]$. Moreover, we choose a purely convex up-ramp:
\be
\nu(Z)=\left\{\ba{ll}
\displaystyle \frac {1}{189.4}\:\frac {g_2(Z)}{g_2(60\lambda)}
\qquad & 0\le Z\le 60\lambda \:\mbox{ (upramp),}\\[16pt]
\displaystyle \frac {1}{189.4}\left[1+\frac {79.4}{268.8}\frac {60\lambda-Z}{6.2\lambda}\right]\qquad & 60\lambda\le Z\le 66.2\lambda  \:\mbox{ (downramp),}\\[16pt]
\displaystyle \frac {1}{268.8} \qquad & Z\ge 66.2\lambda  \:\mbox{ (plateau).}
\ea \right.
\label{3rdInitialDensity}
\ee
In Fig. \ref{Worldlinescrossings-new_densita3} we depict it and 
the electron WLs found solving the corresponding direct problem  (\ref{heq1}-\ref{heq2})  for $Z=0,\lambda,...,96\lambda$.
We have verified that the earliest WB (w.r.t. `time' $\xi$) involves the electrons in the downramp; this is confirmed by solving also eq. (73) in \cite {FioDeNAkhFedJov23} and eq. (\ref{pseudoper}) for the Jacobian $\hat J$.   \ Then we have solved the direct problem for the down-ramp $Z$ electrons more in detail, determining their  WLs for  $Z=60\lambda,60.1\lambda,...,80\lambda$. In particular,
we have found  $(Z_b,\xi_b)\simeq (61.5\lambda, 47.77\lambda)$, $k_1=2$, and again the acceleration rate $F=0.286$, in very good agreement with the values determined by Step 3. Moreover 
$\xi_0=49.44\lambda$ (confirming that $\xi_0\!-\!\xi_{b}\ll\bxiH(\bar{n})\simeq 19.83
\lambda$) and  $\xi_f=50.04\lambda$.
The $Z\!=\!Z_{b}$ electrons  nearly fulfill (\ref{gamma^M(z_i)})
(with $\delta s\!=\!-0.984$) and  (\ref{gamma^MM(z_i)}), hence 
have almost the largest possible WFA factor  (\ref{s_i^m<0|s_iz_i}), with $F=0.286$. 
If $\lambda=0.8\mu$m, this leads to a remarkable energy gain of
$
0.182$MeV per $\mu$m.  \\
With the above results eq. (\ref{Rcond}b) is automatically satisfied, while eq.  (\ref{Rcond}a) takes the form \ $w_0\gg R>\xi_b-2l'\simeq 27\lambda\simeq 21\mu$m. Hence, if the pulse is a Gaussian beam as described in section \ref{3Deffects} with e.g. $w_0\ge w_{0m}\equiv 75\mu$m, the motion of plasma electrons inside the causal cone of base radius $R=25\mu$m trailing the pulse, including the WFA of the FSIE, is practically the same as determined with the plane pulse, over a longitudinal distance $z\le \zM =400\mu$m. This estimate of $w_{0m}$ agrees with that obtained by PIC simulations (cf. next section).

\section{
PIC simulations,  
discussion, conclusions and outlook}
\label{PIC-conclusions}

To check the validity of the plane model
and derive the  details of the corresponding WFAs, we first ran Particle In Cell (PIC) simulations with an equivalent 1D geometry (and the same input data) with the FBPIC \cite{FBPIC} code. Then we ran quasi 3D simulations with  the plane wave  replaced by more realistic Gaussian beams with various waists $w_0$, in order to assess the usefulness of the model also with transverse effects included in the dynamics. Although FBPIC is embedded in a quasi-3D geometry \cite{CalderCirc}, the first simulations were designed in order to make transverse effects completely negligible. This was achieved by setting a plane-wave distribution of the injected laser pulse, inside a simulated cylinder moving at the speed of light, having radius $R_{cyl}=440 \mu m$ and length $L_{cyl}=90 \mu m$. This ensured that boundary effects  did not modify the plasma dynamics around the simulation cylinder axis, where the fields and the particles distribution were extracted. The longitudinal grid resolution was of $dz=\lambda/32$, with cell aspect-ratio radial-to longitudinal  of $dr:dz=800:1$. The time step was $dt=dz/c$, and the number of simulated macroparticles per cell was of 192~ppc.

\begin{figure}
\includegraphics[width=16.5cm]{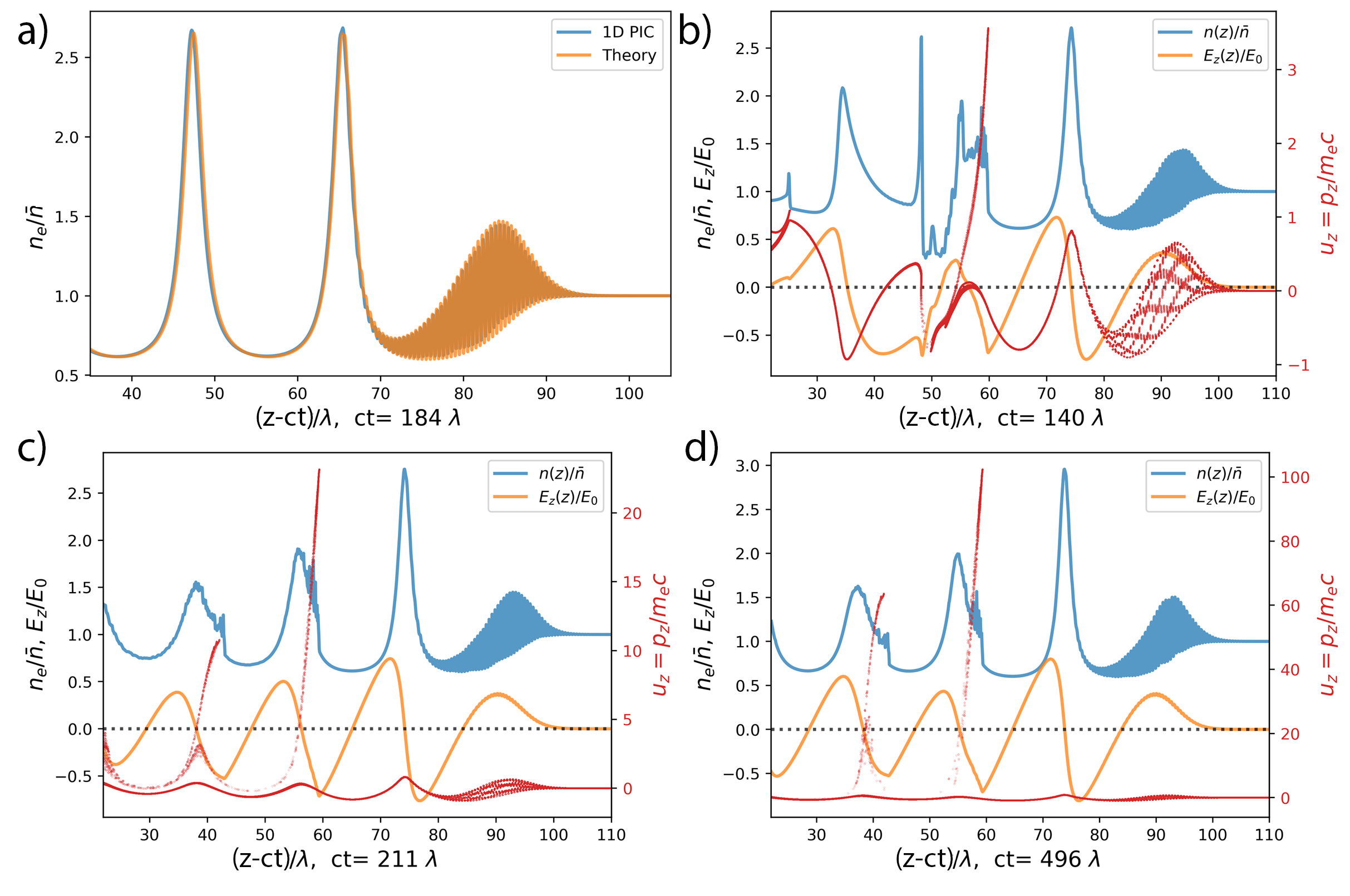}
\caption{ FB-PIC (1D equivalent) simulations run with  the pulse of Fig. \ref{graphsb}a and: a flat initial plasma density profile having  $\bar{n} = n_{cr}/268$ (panel a) at $ct=184\lambda$;  the $3^{rd}$ initial density (\ref{3rdInitialDensity}), taken at 3 different instants (panels b-d). In panel a) the comparison between the semi-analytical results and the PIC simulation is reported, showing the good matching between them. In panels b)-d) the blue line shows the normalized electron density, while the orange line shows the normalized accelerating field. Moreover, we have plot red the longitudinal phase-space distribution of the simulated particles (the $u^z$ scale is reported along the right vertical line). The erosion of the 2nd, 3rd (leftwards) blue peaks is due to plasma electrons self-injected  in the 2nd, 3rd bucket; in longitudinal phase-space the latter are distributed along the two  very steep discontinuous red lines. The latter manifestly have maxima at maximum points of $-E^z$, as foreseen by the theory.\\   
}
\label{1DPIC}       
\end{figure}

\begin{figure}
\includegraphics[height=5.2cm]{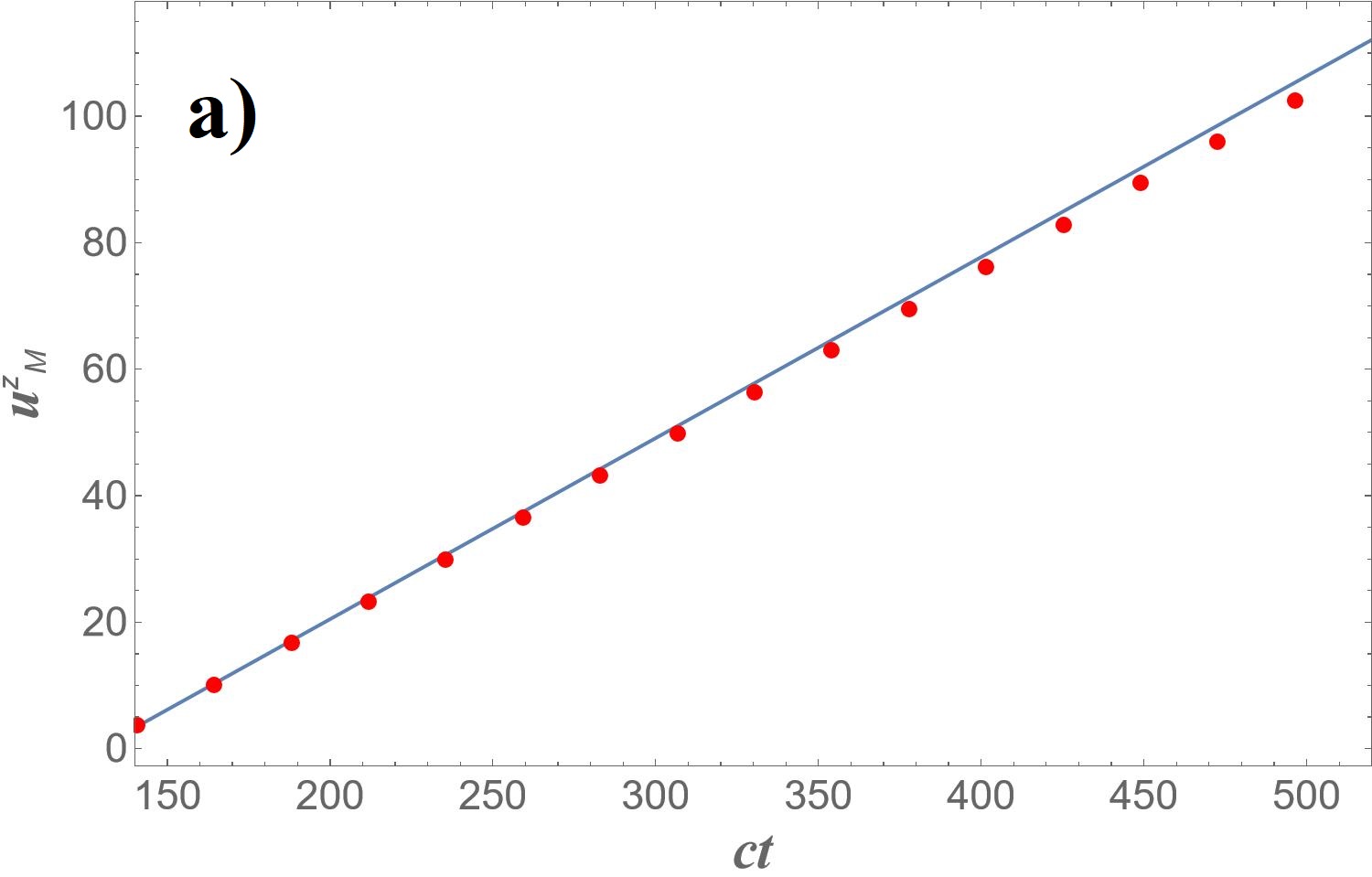} \hfill \includegraphics[height=5.8cm]{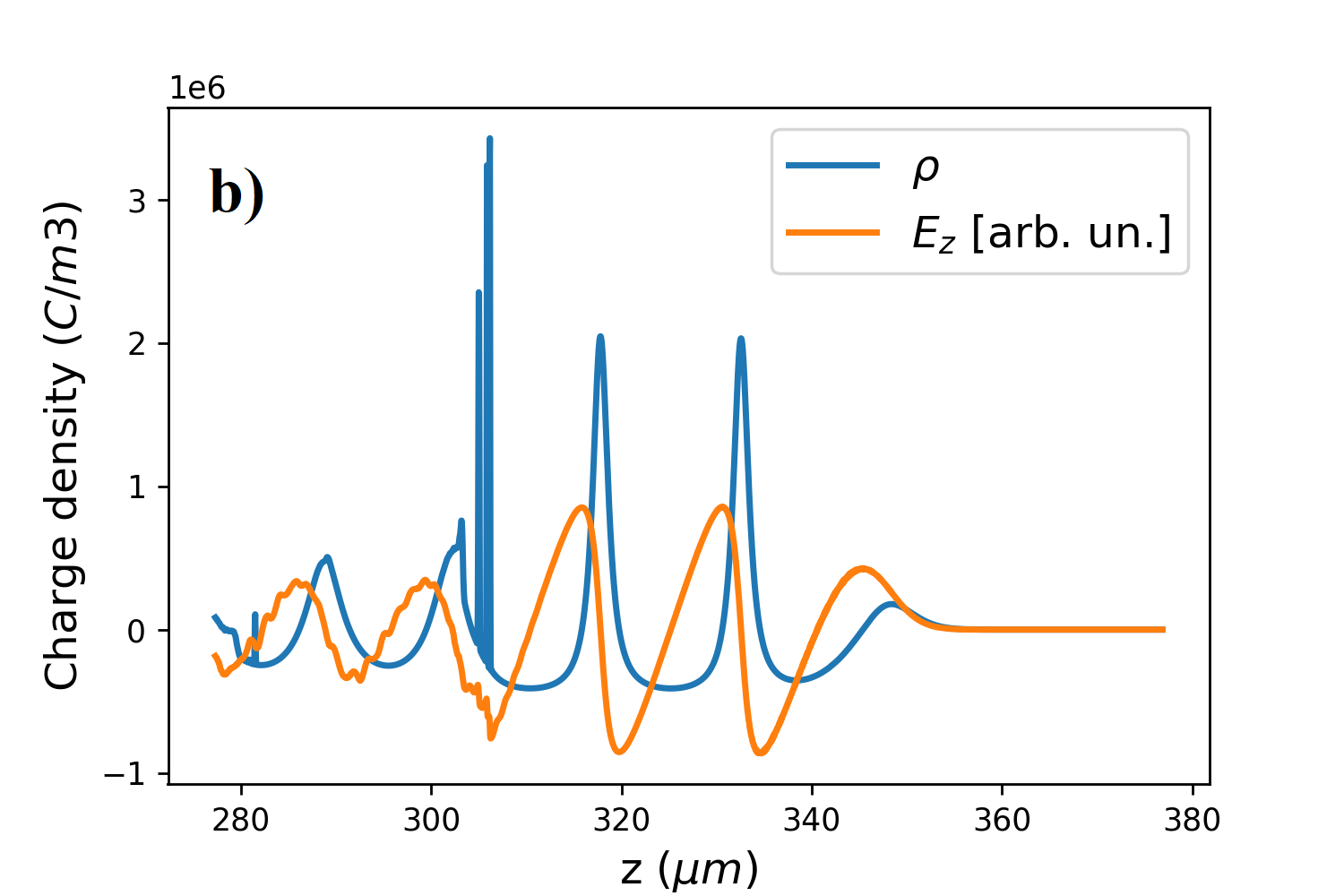}
\caption{(a) \ Comparison between semi-analytical model and FB-PIC (1D equivalent) simulations run with  the $3^{rd}$ initial density \ref{3rdInitialDensity}: maximum  longitudinal momentum $u^\zM $ obtained by the PIC simulation (red circles) and prediction from the theory (blue line). \\
(b) \ Some results of the FB-PIC simulations run with the input data of Fig. \ref{Worldlinescrossings'}  [in particular, the initial density (\ref{1stInitialDensity})] and $\lambda=0.81\mu$m: plots of the electron density and longitudinal electric field after 1~ps; WB has injected electrons in the 3rd bucket of the PW.}
\label{uzmax}       
\end{figure}

A simulation with the same laser parameters described above and a flat plasma density profile having $\bar{n} = n_{cr}/268$, gives us results very close to the semi-analytical ones described in the text (see fig.\ref{1DPIC}-a)). As the semi-analytical results assume that the group laser pulse velocity in the plasma is $c$ and that no pulse shape modification occurs, the matching between 1D PIC simulations and the semi-analytical results can be accurate only for short propagation distances and in this case at least for about 
400~$\lambda$ of pulse propagation in the plasma. In figures \ref{1DPIC}-b)-d)), snapshots of fields and particle phase spaces for the case having a downramp described by the $3^{rd}$ initial density profile, eq. (\ref{3rdInitialDensity}), are shown.  The normalized electron density $n/\bar{n}$ (blue line) shows a 
WB of the PW in the second bucket and at $z-ct\simeq 60\lambda$. The normalized accelerating field $E_z/E_0$ (orange line), where $E_0 = mc^2k_p/e$ is the nonrelativistic wave-breaking limit ($k_p=2\pi/\lambda_p=2\pi/\bxiH(\bar{n})$), confirms the wave breaking in the second bucket and suggests the presence of a severe beam-loading effect  due to the large current of the trapped particles. The longitudinal phase space of the simulated particles is shown in red and is related to the right axis. At $ct= 140\lambda$, {\it i.e.} right after the propagation in the downramp region (see panel b, right axis), a fraction of the plasma electrons leave the wave motion and is going to be trapped in the PW. As in the  simulated  1D geometry the laser pulse center of mass moves with a speed $\beta_g =\sqrt{1-{\bar n}/n_{cr}}$, {\it i.e.} with a relativistic Lorentz gamma factor $\gamma_g = \sqrt{n_{cr}/{\bar n}}\simeq 16.39$, electrons reaching a minimum longitudinal momentum $u_{tr}=\beta_g\gamma_g\simeq 16.36$ are trapped in the PW. This condition is not yet satisfied at  $ct= 140\lambda$ (panel b), while it is satisfied for a large fraction of the accelerated particles in later times (see panels c) and d)).

\begin{table*}
\begin{tabular}{|l|}
\hline
\ \ \ $a_0=2$, \ wavelength $\lambda=0.8\mu$m, \
pulse FWHM $l'=10.5 \lambda$, \  initial density (\ref{3rdInitialDensity})\\[4pt]
\hline
\begin{tabular}{|l|c|c|c|c|c|c|c|c|}
\hline
$ct \, (\mu$m)  & 141 & 188 & 236 & 283 & 331 & 378 & 425 & 473 \\[2pt]
\hline
\hline
$u^z_{{\scriptscriptstyle M}}$,  plane model             & 3.6  & 16.5& 29.8 & 43.0 & 56.4&  69.4& 82.5 &  95.9  \\[2pt]
$u^z_{{\scriptscriptstyle M}}$,  1D PIC                  & 3.61 & 16.5& 29.7 & 42.8 & 55.9&  68.2& 80.1 &  92.6       \\[2pt]
$u^z_{{\scriptscriptstyle M}}$,  q3D-PIC, $w_0=100\mu$m  & 3.71 & 17.3 & 31.5 & 45.8 & 59.7 & 71 & 79.3  & 89.6 \\[2pt]
$u^z_{{\scriptscriptstyle M}}$, q3D-PIC, $w_0=75\mu$m    & 2.22 & 16.3 & 30.4 & 43.6 & 55.5 & 63.8 & 71.5 & 91.2  \\[2pt]
$u^z_{{\scriptscriptstyle M}}$, q3D-PIC, $w_0=50\mu$m    & 2.29 & 16.4 & 30.3 & 42.8 & 60.9 & 74.5 & 86.8 & 98.0  \\[2pt]
\hline
$\Delta \E/\E$ in \%,   plane model          & -0.49   &  -0.76&  - 1.03& -1.30 & -1.57 & -1.84 & -2.11 & -2.38\\[2pt]
$\Delta \E/\E$ in \%, 1D PIC                 &  -0.48 & -0.71 & -0.92 & -1.13 & -1.32 & -1.52 & -1.78 & -2.09   \\[2pt]
$\Delta \E/\E$ in \%, q3D-PIC, $w_0=100\mu$m & -0.50 & -0.68 & -0.88 & -1.03 & -1.22 &     -1.39 & -1.58 & -1.78 \\[2pt]
$\Delta \E/\E$ in \%, q3D-PIC, $w_0=75\mu$m & -0.54 & -0.71 & -0.89 & -1.06 & -1.24 &  -1.40 & -1.59 & -1.79 \\[2pt]
$\Delta \E/\E$ in \%, q3D-PIC, $w_0=50\mu$m & -0.55 & -0.70 & -0.89 & -1.05 & -1.23 &-1.41 & -1.61 & -1.81\\[2pt]
\hline
\end{tabular}
\end{tabular}
\caption {Comparison  of the maximal $u^z$ of self-injected electrons and of the relative pump energy 
depletion $\Delta \E/\E$ corresponding to few values  of $ct$, as computed by the plane model, 1D PIC, and quasi-3D PIC (q3D-PIC) simulations (with 3 different values of the waist); in the latter case $u^z_{{\scriptscriptstyle M}}$ is computed for electrons moving along the symmetry $\vec{z}$-axis. The input data are the laser pulse of fig. \ref{graphsb}.a and the initial density (\ref{3rdInitialDensity}).
}
 \label{tab1}  
\end{table*}
A further analysis of the energy gain rate of the accelerated electrons reveals an excellent agreement between the analytical method and the equivalent 1D simulations. In fig. \ref{uzmax}.a 
and table \ref{tab1} the maximum longitudinal momentum of the beam's particles is shown for selected time-steps of the PIC simulation (red circles). The blue line in the same figure shows the predicted momentum gain by using Eq.\ref{gamma^M(z_i)}, for the case with the $3^{rd}$ initial density profile, which resulted in a steady momentum gain of $du^z/d(ct/\lambda)=0.28$. 
As it is apparent in the figure, the matching between the analytical model and the PIC simulation is perfect at the early stages of the beam acceleration, while the small and negative curvature of the PIC related curve can be attributed to the dephasing of the electrons in the PW, which is moving at a speed $v_g=c\sqrt{1-1/268.8}$. This mechanism, which in later times will rise to the dephasing process, is not included in the model, which assumes the laser group velocity and therefore the wakefield phase velocity equals $c$. Nevertheless, the model's predictions are very accurate at the early stages of the acceleration phase (at least until $ct\approx \zM =400\lambda$).  

\begin{figure}[hbtp]
\includegraphics[width=16cm]{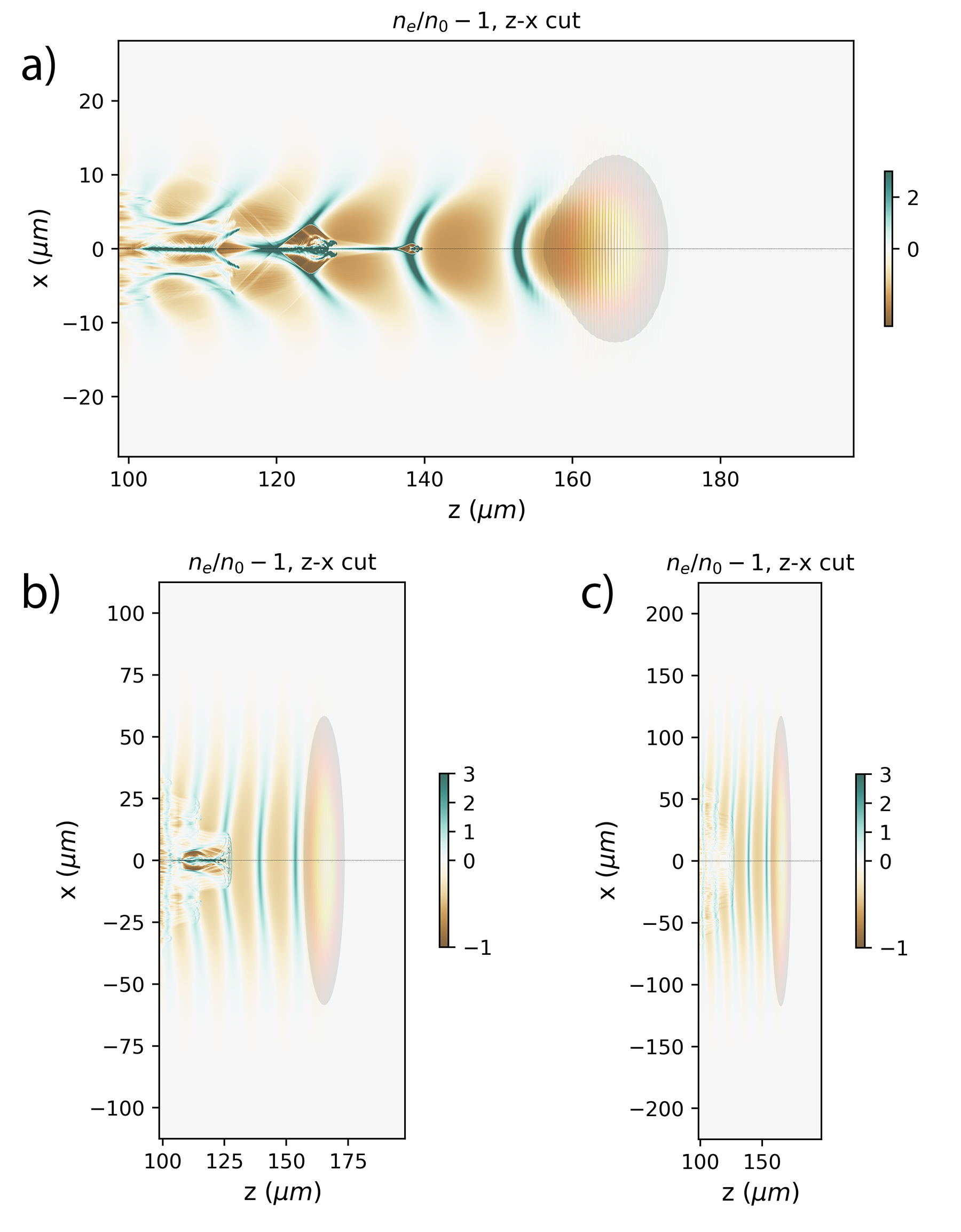}
\caption{Quasi 3D FBPIC simulations with pulse waist in the range $12.5-100 \mu m$, corresponding to $k_p w_0 \simeq  6-50$. Snapshot at $ct= 200 \mu$m of the densities in the z-x plane right after the density downramp for the cases of $w_0=12.5 \mu$m (panel a), $w_0=50 \mu$m (panel b) and $w_0=100 \mu$m (panel c). As it is apparent in panel c), a quasi-1D structure is obtained for the wider pulse case.
}
\label{fig:3DPIC}       
\end{figure}

  Several results of FB-PIC simulations with the same laser parameters described above and the $1^{st}$ initial density profile, eq. (\ref{1stInitialDensity}), are available at the repository \url{https://people.na.infn.it/~gfiore/FB-PIC_simulations} . The results show\footnote{The time lapse between the last frame (at $t=1242$fs) and the last frame before self-injection  (at $t=502$fs) is $\Delta t=740$fs, corresponding to a traveled distance $\Delta z=c\Delta t\simeq 222\mu$m$\simeq 274\lambda$, where $\lambda=0.81\mu$m; correspondingly, we read off from these two plots that $\Delta u^z\simeq 75\simeq\Delta \gamma$. As a result
  $F\simeq \Delta \gamma\lambda/\Delta z\simeq 0.27$.
  }
the presence of a bunch of electrons with  acceleration rate  $F\simeq 0.27$, in good agreement with the prediction in the previous section.
In fig. \ref{uzmax}.b we just plot
the electron density and longitudinal electric field  at the last time of the simulation, i.e.  after 1 ps.

\begin{figure}
\includegraphics[width=16cm]{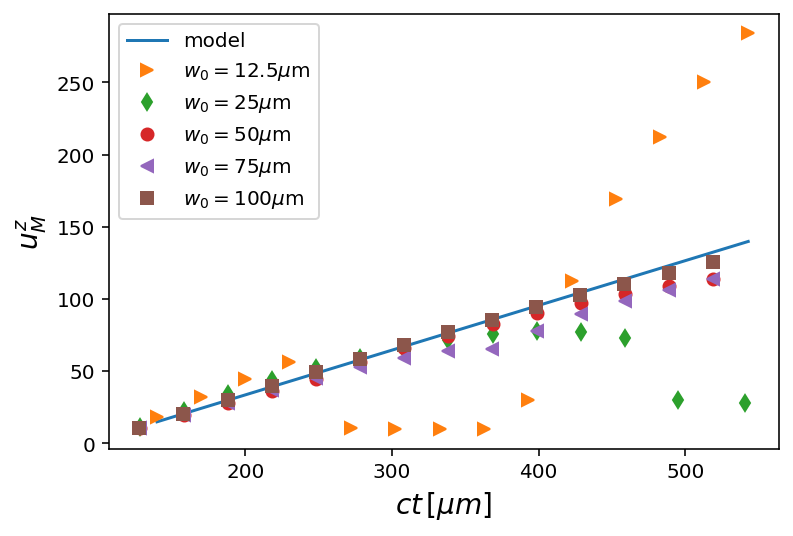}
\caption{Maximal  longitudinal momentum $u^\zM $, obtained by quasi 3D FBPIC simulations with pulse waist in the range $12.5-100 \mu m$,
and the $3^{rd}$ initial density, see eq. (\ref{3rdInitialDensity}), vs. $ct\sim z$, 
and prediction from the theory (blue line).
The results show that the 1D approximation closely mimics the 3D geometry simulation results for $w_0=100\mu$m, corresponding to a normalized waist of $k_p w_0\simeq 50$, and is already acceptable for $w_0=50\mu$m.}
\label{fig:uzM_vs_ct-various w_0}   
\end{figure}

\begin{figure}[ht]
\includegraphics[width=14.5cm]{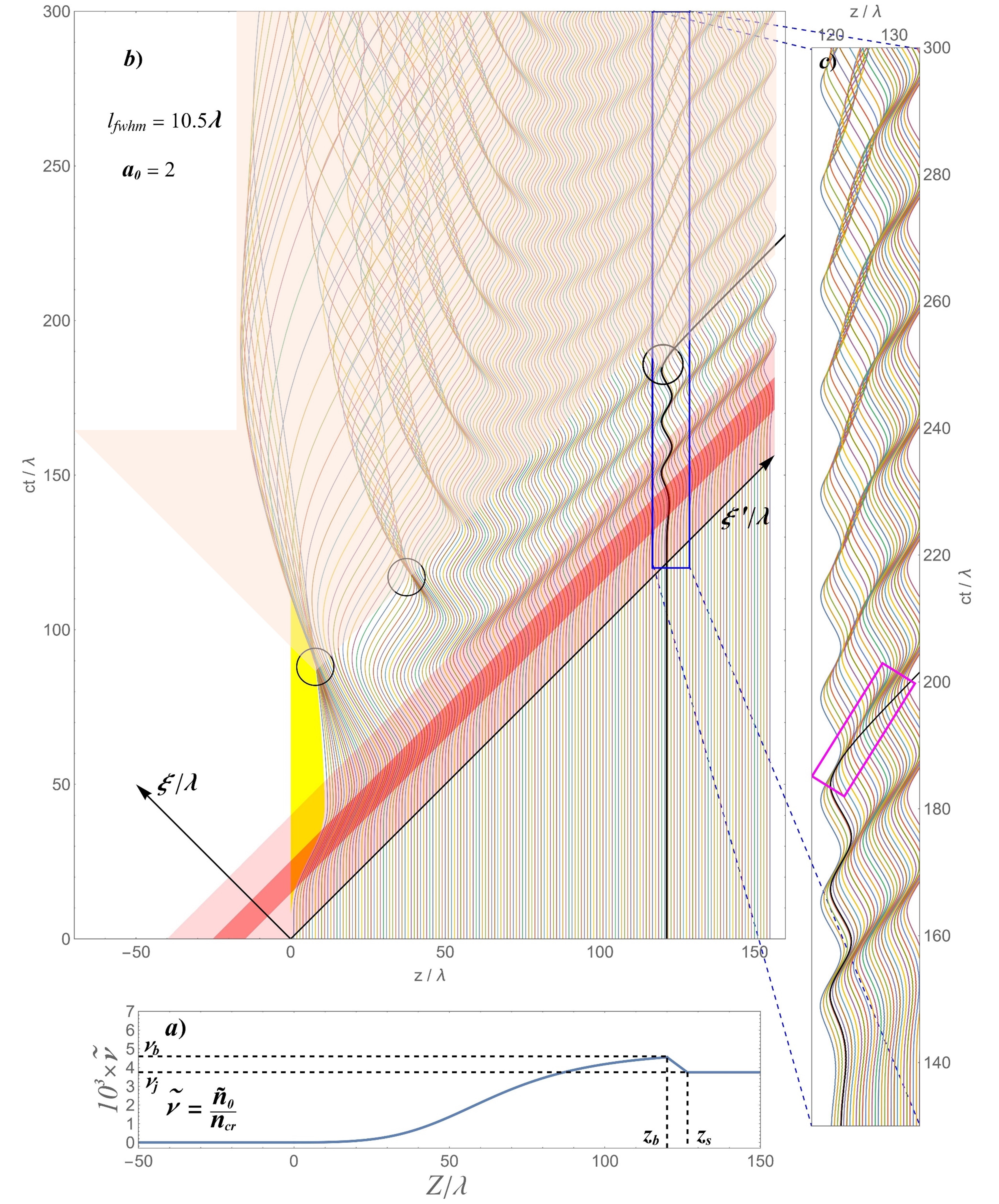}
\caption{ \ a) \  Plot of the initial plasma density (\ref{1stInitialDensity}) - an `optimal' one for the pulse of fig. \ref{graphsb}.a: $n \!=\!n_0^j\!=\!n_{cr}/268.8$,  \
$n_{b}\!=\!1.21\times n_0^j$, \ $z_B \!=\! 120\lambda$, \
$z_s\!-\!z_B \!=\! 6.5\lambda$.  \\
b) \ Corresponding solutions of (\ref{heq1}-\ref{heq2}) 
 for $Z=0,\lambda,...,156\lambda$; \ we have studied the down-ramp electrons more in detail, solving (\ref{heq1}-\ref{heq2}) also   for  $Z=120\lambda,120.1\lambda,...,140\lambda$.
 These curves are the (projections from Minkowski space onto the $z,ct$ plane  of the) 
WLs of the corresponding $Z$ electrons, at least {\it outside} the light-orange shaded wedges; inside the latter, which are the causal wedges
having apex in the earliest WBs (i.e. the first intersections of WLs, here encircled), the WLs have to be found as solutions of the more general eqs. (\ref{Newheq1b}-\ref{heq2}), and hence might differ from the drawn curves. So far we have completely  solved (\ref{Newheq1b}-\ref{heq2})  only for the $Z_{b}$ electrons, and painted  the corresponding complete  WL in black. 
Here: $\xi'\equiv ct\!+\!z$; in the dark yellow region only ions are present; we have painted   in pink the support of $\Bep(ct\!-\!z)$  (considering $\Bep(\xi)\!=\!0$ outside $0\!<\!\xi\!<\!40\lambda$), in red the region where the modulating intensity 
is above half maximum, i.e. $-l'/2\!<\!\xi\!-\!20\lambda\!<\!l'/2$, with $l'
=10.5\lambda$; the pulse  can be considered ES [cf. (\ref{Lncond'})] if we consider  some $l_r\le 27\lambda$ as the pulse length, instead of $l=40\lambda$. \ \ \ 
In \ c)   \ we zoom the blue box of b). \\
}
\label{Worldlinescrossings'}       
\end{figure}

\begin{figure}
\includegraphics[width=\textwidth]{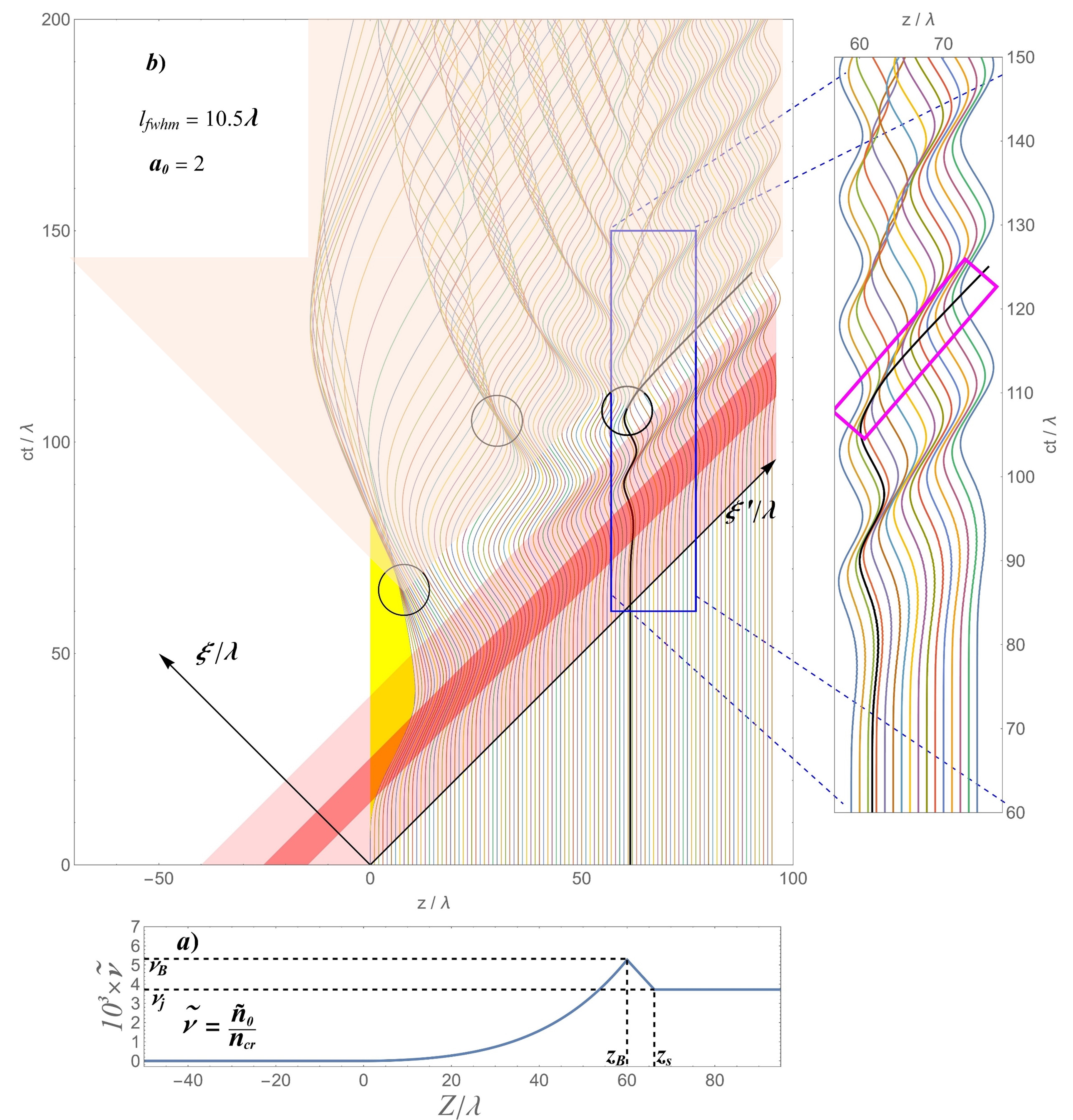}
\caption{ {\bf a}) \ Plot of the initial plasma density (\ref{3rdInitialDensity}) - another `optimal' one for the pulse of fig. \ref{graphsb}.a:  $\bar{n} \!=\!\bar{n}^j\!=\!n_{cr}/268.8$,  
$n_{b}\!=\!1.32\times \bar{n}^j$, \ $z_B \!=\! 60\lambda$, \ $z_s\!-\!z_B \!=\! 6.2\lambda$,  $n_B\!=\!1.42\times \bar{n}^j$.  \\
{\bf b}) \   Corresponding  solutions of (\ref{heq1}-\ref{heq2}) for $Z\!=\!0,\lambda,\!...,95\lambda$; \ we have studied the down-ramp electrons more in detail, solving (\ref{heq1}-\ref{heq2}) also   for  $Z=60\lambda,60.1\lambda,...,80\lambda$.  
These curves are (the projections from Minkowski space onto the $z,ct$ plane  of) the 
WLs of the corresponding $Z$ electrons, at least {\it outside} the light-orange shaded wedges; inside the latter, which are the causal wedges
having apexes in the earliest WBs (i.e. the first intersections of WLs, here encircled), the WLs have to be found as solutions of the more general eqs. (\ref{Newheq1b}-\ref{heq2}), and hence might differ from the drawn curves. So far we have completely  solved (\ref{Newheq1b}-\ref{heq2})  only for the $Z_{b}$ electrons, and painted  the corresponding complete WL in black. The other symbols and colors are as explained in fig. 
\ref{Worldlinescrossings'}. \\
   {\bf c})    \ Zoom of the blue box of b). \\}
\label{Worldlinescrossings-new_densita3}       
\end{figure}

Three dimensional effects are mutually linked and act on the pulse propagation, on the wakefield excitation, and on the electron beam dynamics. As the 1D limit is reached when $k_p w_0 \rightarrow \infty$, in order to assess a lower limit of the laser pulse waist $w_0$ below which 3D effects start to be dominating, we performed a series of quasi-3D simulations with waist ranging from $w_{0,min}=12.5 \mu m$ up to $w_{0,max}=100 \mu m$, which span the dimensionless radial size $k_p w_0$ in the range 6-50. In fig. \ref{fig:3DPIC} we have reported snapshots of the obtained relative density for three different waists at the same snapshot placed right after the density downramp, showing the expected emergence of a quasi-1D structure for the larger waists. In fig.\ref{fig:uzM_vs_ct-various w_0} and table \ref{tab1} the comparison between the theoretical predictions for the dependence of $u^\zM $ on $ct\sim z$ and the corresponding results of quasi-3D simulations with different waists is shown. As is apparent, the 1D analytical model maintains its level of accuracy in a 3D geometry with larger waist, thus setting a minimum normalized waist $k_p w_0\approx 50$ for the case analyzed here.  In table \ref{tab1} we have compared also the pulse energy depletion rates computed by the semi-analytic plane model (see section \ref{Back-reaction}) and the FB-PIC simulations, finding a very good agreement.

\medskip

To summarize, here we have first considered the problem with plane symmetry with a fixed pulse and constructed a procedure by which  we can maximize the LWFA of the first  self-injected electrons (FSIE) in the PW 
at the downramp during the early stages of their journey (i.e. for $z\le \zM $, see section \ref{Back-reaction}): i) by choosing the plateau level $\bar{n}$ as the unique value $\bar{n}_j=\nu_j\, n_{cr}$ associated to the pulse (section \ref{WFA.b}); ii) with the present level of approximation, and assuming a linear downramp for simplicity, by choosing the pair of 
parameters ($n_b,\delta Z$) featuring the downramp in the set (broken line) $\Lambda$  (section \ref{WFA.c}), and then - if one wishes -  improving it by fine-tuning, as done e.g. for the downramp (\ref{3rdInitialDensity}); iii) by choosing the upramp so that WBs involving upramp electrons cannot causally affect the WFA of the FSIE  (this can be done with a very large freedom, see section \ref{WFA.d}). Although the plane model at the basis of our procedure assumes a collisionless plasma, initially cold and at rest, with immobile ions all the time, so that it cannot predict long-term plasma evolution or multi-stage acceleration processes, by causality its predictions for the motion of the FSIE in the early acceleration stages are very accurate (see section \ref{planeWFA}). Secondly, in section \ref{3Deffects} we have shown how to estimate conditions (e.g., if the pulse is a Gaussian-beam then its waist $w_0$ must fulfill a minimum condition $w_0\ge w_{0m}$,...) and range of validity of the above predictions for more realistic 3D input data.  We can thus conclude that our analytical model can reliably predict outcomes in a specific range, offering a valuable tool for preliminary optimization of LWFA before costly simulations or experiments, thus saving time and resources. Of course, the model looses predictivity after longer
distances covered by the FSIE, or more generally for long-term plasma and pulse evolution.

The procedure is very promising and deserves further investigations  in view of improvements and possible developments. First of all, it would be important to develop a better code to solve the system of ODEs (\ref{heq1}a-\ref{Newheq1b}) and thus determine the motion in the plane model also of the other plasma electrons self-injected in the PW, beside the FSIE; this would hopefully give a handle to further tailor the downramp so as to control the energy distribution of these electrons (and the beam loading) already at the early stages of their LWFA. Further analytical investigations about the back-reaction of the plasma on the pulse would also be important. On the other hand, replacing the density plateau by a slowly $z$-dependent $\widetilde{n_0}(z)$ adapted to the pulse evolution could postpone electron dephasing and thus longitudinally extend the acceleration region.

\section*{Acknowledgments}      
P.T. acknowledges the support of the Romanian Government and the European Union through the European Regional Development Fund - the Competitiveness
Operational Programme (1/07.07.2016, COP, ID 1334) Phases II, and the Romanian Ministry of Research, Innovation and Digitalization: Program Nucleu PN23210105. Accessing the ELI-NP facility is supported by the
IOSIN funds for research infrastructures of national interest funded by the Romanian Ministry of Research, Innovation and Digitalization. P.T. draws support also from the contract ELI-RO/RDI/2024 14 SPARC funded by PN III/P5/Subprogram 5.1.

\end{document}